\documentclass[11pt,a4paper]{article}
\usepackage{jheppub}
\usepackage[utf8]{inputenc}
\usepackage{enumitem}
\usepackage{bm}
\usepackage{bbm}
\usepackage{xcolor}
\usepackage{scalerel}
\usepackage{enumitem}  
\allowdisplaybreaks

\makeatletter
\def\namedlabel#1#2{\begingroup
    \normalfont #2\!%
    \def\@currentlabel{#2}%
    \phantomsection\label{#1}\endgroup
}
\makeatother

\def\be{\begin{equation}}
\def\ee{\end{equation}}
\def\beal{\begin{equation}\begin{aligned}}
\def\eeal{\end{aligned}\end{equation}}
\def\nn{\nonumber}
\def\bra#1{\langle #1|}
\def\ket#1{|#1 \rangle}
\def\braket#1{\langle #1 \rangle}
\def\o#1{\overline{#1}}

\def\U{u}
\def\V{v}
\def \genE{{\cal E}}
\def \Ea{{E}}
\def \Eb{{\tilde E}}

\def\cL{{\cal L}}

\def\cO{{\cal O}}

\def\cA{{\cal A}}

\def\eps{\epsilon}

\def\eps{\epsilon}
\def\Tr{\operatorname*{Tr}}
\def\tr{\operatorname*{tr}}

\def\angle#1{\langle #1 \rangle}
\def\nn{\nonumber}

\def\cleq{\stackrel{\rm cl}{\sim}}

\def\Tr{\operatorname*{Tr}}
\def\tr{\operatorname*{tr}}

\def\Res_#1{\operatorname*{Res}_{#1}}
\def\sgn{\operatorname*{sgn}}

\def\eps{\epsilon}
\def\ep{\varepsilon}
\def\Fm{{F_{\;\!\!-}}}
\def\Fp{{F_{\;\!\!+}}}
\def\bep{\boldsymbol{\mathbf{\ep}}}

\def\ie{{\it i.e.} }
\def\eg{{\it e.g.} }

\def\eqn#1{eq.~\eqref{#1}}
\def\eqns#1#2{eqs.~\eqref{#1} and~\eqref{#2}}

\def\Eqn#1{Eq.~\eqref{#1}}

\def\fig#1{figure~{\ref{#1}}}

\def\tab#1{table~{\ref{#1}}}

\def\Tab#1{Table~{\ref{#1}}}

\def\sec#1{Section~{\ref{#1}}}
\def\secs#1#2{Sections~{\ref{#1}} and~{\ref{#2}}}
\def\app#1{Appendix~{\ref{#1}}}
\def\rcite#1{ref.~\cite{#1}}
\def\rcites#1{refs.~\cite{#1}}

\def\AHH{$\sqrt{\text{Kerr}}$ }

\def\ranks{ranks }
\def\rank{rank }

\def\vs{\varsigma}
\def\ahalf{\bar{a}}

\newcommand{\ga}{\alpha}
\newcommand{\gb}{\beta}
\newcommand{\gc}{\gamma}

\newcommand{\gad}{{\dot{\alpha}}}
\newcommand{\gbd}{{\dot{\beta}}}

\newcommand{\fud}[2]{{}^{#1}{}_{#2}\,}
\newcommand{\fdu}[2]{{}_{#1}{}^{#2}\,}

\title{From higher-spin gauge interactions to Compton amplitudes for root-Kerr
}

\author[a]{Lucile Cangemi,}
\author[a]{Marco Chiodaroli,}
\author[a,b]{Henrik Johansson,}
\author[c,d]{Alexander Ochirov,}
\author[a,e]{Paolo Pichini,}
\author[f,g]{and Evgeny Skvortsov}

\affiliation[a]{Department of Physics and Astronomy, Uppsala University,
Box 516, 75120 Uppsala, Sweden,}
\affiliation[b]{Nordita, Stockholm University and KTH Royal Institute of Technology, \\
Hannes Alfv\'{e}ns  v\"{a}g 12, 10691 Stockholm, Sweden,}
\affiliation[c]{London Institute for Mathematical Sciences,
21 Albemarle Street, London W1S 4BS, UK,}
\affiliation[d]{School of Physical Science and Technology, ShanghaiTech University, \\
393 Middle Huaxia Road, Shanghai 201210, China,}
\affiliation[e]{Centre for Theoretical Physics, Department of Physics and Astronomy, \\
Queen Mary University of London, Mile End Road, London E1 4NS, UK,}
\affiliation[f]{Service de Physique de l’Univers, Champs et Gravitation, \\
Universit\'{e} de Mons, 20 place du Parc, 7000 Mons, Belgium,}
\affiliation[g]{Lebedev Institute of Physics,
Leninsky avenue 53, 119991 Moscow, Russia}

\emailAdd{lucile.cangemi@physics.uu.se}
\emailAdd{marco.chiodaroli@physics.uu.se}
\emailAdd{henrik.johansson@physics.uu.se}
\emailAdd{ochirov@shanghaitech.edu.cn}
\emailAdd{p.pichini@qmul.ac.uk}
\emailAdd{evgeny.skvortsov@umons.ac.be}

\preprint{UUITP-35/23}

\abstract{
We develop massive higher-spin theory as a framework for describing dynamics of rotating compact objects, such as Kerr black holes. In this paper, we explore gauge interactions up to quartic order and corresponding Compton amplitudes of higher-spin massive objects coupled to electromagnetism and Yang-Mills theory. Their classical counterparts are known as root-Kerr gauge-theory solutions, whose amplitudes are closely related to those of Kerr black holes. We use three distinct approaches: (i) massive higher-spin gauge symmetry to introduce cubic interactions for all spins and the quartic interactions up to spin 3, which is implemented both off shell and via Ward identities; (ii) a chiral higher-spin approach to construct quartic Lagrangians with correct degrees of freedom to all spins; (iii) on-shell functional patterns before and after taking the classical limit to constrain the Compton amplitudes. As final results, we arrive at simple local formulae for the candidate root-Kerr Compton amplitudes both in the quantum regime and classical limit, to all orders in spin. This is a precursor to the gravitational Kerr case, which is presented in a follow-up paper. 
}

\begin{document}
\maketitle
\addtocontents{toc}{\protect\setcounter{tocdepth}{2}}

\section{Introduction}
\label{sec:intro}

Higher-spin theory aims to describe interacting particles with spin quantum numbers exceeding those of the fundamental force carriers --- gauge bosons and gravitons.
At the fundamental level, the existence of such particles seems crucial for theories of quantum gravity, such as string theory (for a review see \eg\rcite{Bekaert:2022poo}). More relevant to this work, theories with massive higher-spin fields in Minkowski background are often understood as effective field theories (EFTs) valid below certain energy scales \cite{Ferrara:1992yc,
Cucchieri:1994tx,
Klishevich:1997pd,
Giannakis:1998wi,
Buchbinder:1999be,
Buchbinder:2000ta,Bekaert:2003uc,
Zinoviev:2001dt,
Zinoviev:2006im,Zinoviev:2008ck,
Francia:2007ee,
Metsaev:2007rn,
Francia:2008ac,
Metsaev:2012uy,
Hassan:2012wr,
Cortese:2013lda,
Bernard:2015uic,
Fukuma:2016rru,
Bonifacio:2018aon,Afkhami-Jeddi:2018apj,Kaplan:2020ldi}.

In recent years, the application of scattering-amplitude methods to gravitational-wave physics (see \eg the review~\cite{Buonanno:2022pgc}) has provided a new arena for higher-spin theory.
Indeed, the long-range physics of compact objects can be captured by spinning point particles.
For example, consider a three-point amplitude involving two particles of spin~$s$ and mass~$m$, and a massless force carrier.
Its dependence on the $2s{+}1$ physical spin states of massive particle~$\bm{1}$ may be encoded in a symmetrized set $\{a_1,\dots,a_{2s}\}$ of ${\rm SU}(2)$ indices, and likewise for particle~$\bm{2}$.
A particularly interesting infinite family of such amplitudes, proposed by Arkani-Hamed, Huang and Huang (AHH) \cite{Arkani-Hamed:2017jhn}, may be written in terms of massive Weyl spinors~\cite{Conde:2016vxs,Arkani-Hamed:2017jhn}
(related to the momenta via
$\ket{p^a}_\alpha [p_a|_{\dot{\beta}}=p_\mu \sigma_{\alpha\dot{\beta}}^\mu$)
simply as\footnote{Pauli matrices appear as
$\sigma^\mu_{\alpha\dot{\beta}} = (1,\sigma^1,\sigma^2,\sigma^3)
= \bar{\sigma}_\mu^{\dot{\alpha}\beta}$.
Lorentz ${\rm SL}(2,\mathbb{C})$ indices and little-group ${\rm SU}(2)$ indices are raised and lowered via left-multiplication by Levi-Civita tensors, normalized as $\epsilon^{12}=1=\epsilon_{21}$.
The Lorentz invariants formed out of Weyl spinors are 
$\braket{1_a 2^b} := \epsilon_{\alpha\beta} \bra{1_a}^\alpha \bra{2^b}^\beta$
and 
$[1_a 2^b] := \epsilon^{\dot{\alpha}\dot{\beta}}
[1_a|_{\dot{\alpha}} [2^b|_{\dot{\beta}}$.
}
\begin{subequations} \begin{align}
\label{AHH3plus}
{\cal A}_\text{AHH}(\bm{1}_{a_1 \dots a_{2s}},\bar{\bm{2}}^{b_1 \dots b_{2s}}\!,3^+) &
= \tfrac{1}{m^{2s}} \braket{1_{(a_1} 2^{(b_1}}\cdots\braket{1_{a_{2s})} 2^{b_{2s})}} {\cal A}_3^{(0)} \,, \\*
\label{AHH3minus}
{\cal A}_\text{AHH}(\bm{1}_{a_1 \dots a_{2s}},\bar{\bm{2}}^{b_1 \dots b_{2s}}\!,3^-) &
= \tfrac{1}{m^{2s}}\,[1_{(a_1} 2^{(b_1}]\cdots[1_{a_{2s})} 2^{b_{2s})}]\,{\cal A}_3^{(0)} \,.
\end{align} \label{AHH}%
\end{subequations}
Here ${\cal A}_3^{(0)}$ corresponds to a minimally coupled massive scalar,
\eg ${\cal A}_3^{(0)} = 2Q\,\varepsilon_3^\pm\!\cdot p_1$
in the electromagnetic (EM) case.
For $s \le 1$, \eqn{AHH} matches all EM interactions in the Standard Model~\cite{Chung:2018kqs}
and, upon switching the gauge group, also the strong interaction~\cite{Ochirov:2018uyq}.

For ${\cal A}_3^{(0)}=-\sqrt{32\pi G_\text{Newton}} (\varepsilon_3^\pm\!\cdot p_1)^2$, \eqn{AHH} gives equally simple amplitudes for massive spin-$s$ particles interacting gravitationally.
Remarkably, they were shown to reproduce the classical behavior of rotating Kerr black holes (BHs)~\cite{Guevara:2018wpp},
which helped solidify the link between quantum scattering amplitudes and classical dynamics of compact objects
\cite{Holstein:2008sw,Holstein:2008sx,Vaidya:2014kza,Guevara:2017csg,Vines:2017hyw,Kosower:2018adc,Guevara:2018wpp,Chung:2018kqs,Bautista:2019tdr,Maybee:2019jus,Guevara:2019fsj,Aoude:2020onz,Aoude:2021oqj}.
In the context of the classical double copy~\cite{Monteiro:2014cda,Arkani-Hamed:2019ymq,Guevara:2020xjx,Bern:2010ue}, this matching to the Kerr metric has a gauge-theory analogue, where the interaction~\eqref{AHH} implies the asymptotic EM field of a rotating charged disk (or ring).  
In view of this link between the Kerr metric and the gauge-theory counterpart, we refer to \eqn{AHH} as the three-point $\sqrt{\text{Kerr}}$ amplitudes~\cite{Arkani-Hamed:2019ymq}.
Note that the $\sqrt{\text{Kerr}}$ solution is expected to also describe the EM field sourced by a Kerr-Newman BH~\cite{Chung:2019yfs}.
Here we will, however, focus on a non-gravitational origin of $\sqrt{\text{Kerr}}$. 

Besides such BH-like objects,
there exist genuine microscopic higher-spin particles that interact electromagnetically, such as atomic nuclei.
For instance, a ground state of ${}^{10}\text{Boron}$ has spin~3, and an observationally stable isotope of ${}^{180}\text{Tantalum}$ has spin~9.
Interactions of a given nucleus state may be described by an EFT, whose validity range depends on its lifetime.
It may be too naive to expect nuclei EFTs to possess any kind of simplicity.
Nonetheless, the exploration of the parameter space of such theories should perhaps start with those EFTs that do exhibit simplicity, such as those related to $\sqrt{\text{Kerr}}$ amplitudes.

Despite the striking all-spin pattern of the three-point AHH amplitudes, the related higher-point amplitudes and higher-order BH interactions are still poorly understood.
An important research direction is the study of Compton amplitudes in the classical limit (see \rcites{
Guevara:2017csg,
Guevara:2018wpp,
Chung:2018kqs,
Aoude:2020onz,
Haddad:2020tvs,
Chiodaroli:2021eug,
Aoude:2022trd,
Bern:2022kto, 
Aoude:2022thd,
Chen:2022clh, 
Cangemi:2022abk, 
Saketh:2022wap, 
Bjerrum-Bohr:2023jau, 
Haddad:2023ylx, 
Brandhuber:2023hhy, 
Alessio:2023kgf, 
Aoude:2023vdk, 
Bern:2023ity, 
Bjerrum-Bohr:2023iey,
Brandhuber:2023hhl}).
These four-point tree amplitudes are building blocks for calculating various observables related to gravitational radiation
(see \eg
\rcites{Maybee:2019jus,
Chung:2019duq,
Bern:2020buy,
Bern:2021xze, 
Kosmopoulos:2021zoq,
Chen:2021kxt,
Alessio:2022kwv,
Bautista:2023szu, 
DeAngelis:2023lvf, 
Aoude:2023dui}). 
Unfortunately, a straightforward extension of \eqn{AHH} to four points~\cite{Arkani-Hamed:2017jhn,Johansson:2019dnu} is affected by local contact-term ambiguities~\cite{Chung:2018kqs,Falkowski:2020aso,Chiodaroli:2021eug,Aoude:2022trd}.
This is often highlighted as a problem of spurious poles in the opposite-helicity Compton amplitudes.
However, the problem is not the presence of the unphysical poles, but that removing them can at best be done up to contact terms that cannot be unambiguously determined~\cite{Chung:2018kqs}.
It is thus important to identify symmetry principles that may guide the identification of the correct contact terms.
In turn, this can potentially lead to new insights into BH physics.

In this paper, we combine several methods in higher-spin field theory in order to construct and constrain the Compton amplitudes that we believe to closely correspond to the gauge interactions of a \AHH object. Our approach starts at the quantum level using off-shell actions either in terms of chiral~\cite{Ochirov:2022nqz} or non-chiral massive higher-spin fields~\cite{Fierz:1939ix,Singh:1974qz,Singh:1974rc}.
The latter are endowed with higher-spin gauge symmetry~\cite{Zinoviev:2001dt}, which we employ as a natural guiding principle for constraining the interactions and amplitude calculations~\cite{Cangemi:2022bew}. 

Quantum field theory of massive higher spins has been a challenging research direction.
This is in large part because the conventionally chosen traceless symmetric tensor field $\Phi_{\mu_1\dots\mu_s}$ contains unphysical longitudinal modes. Even for a free action~\cite{Fierz:1939ix,Singh:1974qz,Singh:1974rc}, ensuring their non-propagation by imposing transversality $\partial^\lambda \Phi_{\lambda\mu_2\ldots \mu_s} = 0$ requires the introduction of auxiliary fields.
The multitude of compensating unphysical degrees of freedom complicates the introduction of consistent interactions.
One significant improvement proposed by Zinoviev~\cite{Zinoviev:2001dt,Zinoviev:2006im,Zinoviev:2008ck,Zinoviev:2009hu,Zinoviev:2010cr,Buchbinder:2012iz}
is to convert all second-class constraints into first-class ones, corresponding to a new \emph{massive higher-spin gauge symmetry} \`a la St\"uckelberg.
In this formalism, any gauge-invariant interaction is consistent.
The price to pay is that the field content describing a single massive spin-$s$ particle is now even larger, including $s{+}1$ double-traceless symmetric tensors $\Phi_{\mu_1\dots\mu_k}$ of rank~$k=0,\ldots,s$.
Complementary to this non-chiral formalism, we will also use a powerful \emph{chiral approach}~\cite{Ochirov:2022nqz} for constructing massive higher-spin theories in $4d$.
It relies on choosing a spinor field $\Phi_{\alpha_1\ldots \alpha_{2s}}$ in the $(2s,0)$ representation of the Lorentz group ${\rm SL}(2,\mathbb{C})$.
This approach sidesteps the issue of unphysical degrees of freedom entirely. Parity, however, is no longer automatic and requires extra care.

Our work combines Zinoviev's massive higher-spin symmetry and the chiral approach with classical-limit considerations and heuristic assumptions, so as to construct the Compton amplitudes, relevant for describing well-separated non-decaying charged massive higher-spin fields.
This paper gives an extended exposition of the results in the letter~\cite{Cangemi:2022bew}, where higher-spin gauge invariance was identified as a key property for elucidating BH dynamics.
We also go well beyond that work and complete the study of Compton amplitudes in the $\sqrt{\text{Kerr}}$ theory.  
We give new perspectives on previous work, such as the current constraint~\cite{Ferrara:1992yc, Porrati:1993in, Cucchieri:1994tx} proposed to single out BH amplitudes~\cite{Chiodaroli:2021eug}, and the multitude of subtleties of the classical limit.
Here we only focus on processes where the spin magnitude of the compact objects is conserved; spin-magnitude-changing processes in the presence of gravity are more subtle \cite{Aoude:2023fdm,
Bern:2023ity}, 
but also an interesting avenue for further research.
Our closely related results, regarding gravitationally interacting massive higher-spin theories and Compton amplitudes for Kerr BHs, will be reported in \rcite{Upcoming2}. 

The paper proceeds as follows.
We warm up in \sec{sec:Spin1} by discussing a $\sqrt{\text{Kerr}}$ model involving a massive charged spin-1 field, which can be derived as a limit of a spontaneously broken Yang-Mills theory.
This is a top-down example for the massive gauge symmetry originating from a fundamental theory with massless gauge symmetry.
Then we flip the perspective and show how to obtain the same result in a bottom-up approach by imposing massive gauge symmetry (off shell), or via the associated Ward identities (on shell).

In \sec{sec:HigherSpinGauge}, we extend these bottom-up techniques to arbitrary spin.
Note that the massive higher-spin symmetry allows for general interactions, so additional constraints are used in concert to restrict to the $\sqrt{\text{Kerr}}$ amplitudes.
These constraints are compatible with imposing an improved high-energy behavior~\cite{Chiodaroli:2021eug,Cangemi:2022bew} via power counting and tree-level unitarity~\cite{Ferrara:1992yc,Porrati:1993in}, similar to the behavior of fundamental theories.
We give an explicit construction of the spin-2 (see also \rcite{Zinoviev:2009hu}) and spin-3 cubic Lagrangians that reproduce the \AHH amplitudes~\eqref{AHH} using off-shell massive gauge symmetry.
We also present compelling evidence that the \AHH three-point amplitudes are unique to any spin, once the massive Ward identities and the additional constraints are employed.
The on-shell part of the cubic action is also worked out explicitly for all spins.
The on-shell method of Ward identities turns out to be more computationally efficient than the Lagrangian approach at the quartic order, and we discuss its implementation for spin-2 and spin-3 theories.

To find contact interactions to arbitrarily high spins, we turn to the chiral approach of \rcite{Ochirov:2022nqz}.
In \sec{sec:chiralform}, we analyze possible parity-preserving interactions that determine the non-minimal cubic $\sqrt{\text{Kerr}}$ terms.
A general-spin chiral Lagrangian is parameterized up to the quartic level, which includes interactions linear in the field strength $F_{\mu\nu}$, as well as generic $F_{\mu\nu}^2$ interactions that are relevant for the opposite-helicity Compton amplitudes.  

In \sec{sec:QuantCompAmps}, using plausible constraints and choices for the contact terms in the $\sqrt{\text{Kerr}}$ theory, we present a manifestly local form for the quantum Compton amplitudes for all spins and gauge groups.
The amplitudes are expressed in terms of complete homogeneous symmetric polynomials, which we observe to be a defining feature of $\sqrt{\text{Kerr}}$ amplitudes.
Contact terms in the abelian sector are needed for consistency of the classical limit, and we build them out of the symmetric polynomials guided by the constraints inherited from massive gauge symmetry and other desirable properties.

The classical limit of the candidate Compton amplitude, studied in \sec{sec:ClassCompAmps}, provides perhaps the strongest physical constraint on the quartic interactions at high spin.
The abelian amplitude develops a divergence in the large-spin limit, unless contact terms are added in a specific way.
We analyze two variations of the classical limit for spin: the $s\to\infty$ limit of quantum spin operators~\cite{Guevara:2019fsj,Cangemi:2022abk,Cangemi:2022bew} and the wavefunction scaling of coherent spin states~\cite{Aoude:2021oqj}.
Given our choice of constraints, we extract from the quantum amplitudes a unique classical color-dressed Compton amplitude to all orders in spin.
Possible modifications of the constraints and the consequences for the classical amplitude are also discussed.

Conclusions can be found in \sec{sec:outro}.
The main text is supplemented with appendices to keep track of the notation in \app{sec:appConventions}; to discuss the contact terms within the chiral approach in \app{app:ChiralQuarticContact}; and to collect some remarks on factorization poles in \app{app:factorisation}.

\section{Spin-one warm-up}
\label{sec:Spin1}

Before delving into the general case of higher-spin fields, let us set the stage using the familiar spin-1 case.
We will discuss three constructive approaches for dealing with massive vector fields coupled to electromagnetism: spontaneous symmetry breaking, massive gauge invariance order by order, and Ward identities. 
The point of this section is to show the importance of a gauge symmetry for constraining the EM interactions of massive fields.

\subsection{St\"uckelberg, Proca and Higgs}
\label{sec:fromHiggs}

A minimal model for our purposes involves EM interactions of a Proca field~$W_\mu$.
In a bottom-up construction, one could start from the free St\"uckelberg theory
\be
{\cal L}_\text{St\"uck.}
= -\frac{1}{2} |W_{\mu\nu}|^2
+ |m W_\mu - \partial_\mu \varphi|^2 , \qquad \quad
W_{\mu\nu} = \partial_\mu W_\nu - \partial_\nu W_\mu \,,
\label{Spin1ActionStuck}
\ee
where $\varphi$ is a complex version of St\"uckelberg's auxiliary field~\cite{Stueckelberg:1938hvi}. We use the mostly-minus metric convention $\eta_{\mu\nu} = \text{diag}(1,-1,-1,-1)$ and work in four spacetime dimensions.
The action is invariant under the symmetry
\be
\delta W_\mu = \partial_\mu \xi \,, \qquad \quad
\delta \varphi = m \xi \,,
\label{StueckelbergSymmetry}
\ee
where $\xi$ is a complex gauge parameter.
It can be used to remove the auxiliary field~$\varphi$  entirely, which gives the Proca Lagrangian 
${\cal L}_\text{Proca}:={\cal L}_\text{St\"uck.}|_{\varphi=0}$.
The equations of motion that follow from ${\cal L}_\text{Proca}$ then generate the standard transversality condition $\partial \cdot W=0$, which is needed to reduce the degrees of freedom down to three (complex) physical polarizations.
This condition can also be obtained as the gauged-fixed version of the field equation for $\varphi$, or as the Noether identities for the above gauge symmetry.  

The Lagrangian~\eqref{Spin1ActionStuck} is also invariant under a global U(1) symmetry, which we can gauge and thus couple it to an EM field $A_\mu$.
However, the standard procedure of adding minimal interactions using covariant derivatives breaks the massive gauge symmetry~\eqref{StueckelbergSymmetry}.
In the next section, we will explain how to introduce EM interactions order by order in the coupling while ensuring compatibility with both the U(1) and massive gauge symmetries.
In the spin-1 case, however, this can be done to all orders at once --- using the knowledge that such interactions should come from a spontaneously broken ${\rm SO}(3)$ gauge theory.

Indeed, consider an unbroken non-abelian theory with gauge field $\vec A_\mu$ and Higgs field~$\vec \phi$,
both in the vector representation of ${\rm SO}(3)$,
\be 
{\cal L}_{{\rm SO}(3)} = -\frac{1}{4} \|\vec F_{\mu\nu}\|^2
+ \frac{1}{2} \|D_\mu \vec \phi\|^2
+ \frac{\mu^2\!}{2} \|\vec{\phi}\|^2 - \frac{\lambda}{4!} \|\vec{\phi}\|^4 \,, \qquad
D_\mu \vec \phi = \partial_\mu \vec \phi - g \vec A_\mu \times \vec\phi \,.
\label{SO3YM}
\ee
Here the non-abelian Lie algebra is encoded in the cross product
$(\vec A_\mu \times \vec\phi)_j = iA_\mu^k t^k_{jl} \phi^l $,
with hermitian generators $t^k_{jl} = -i\epsilon_{jkl}$, and
$\vec F_{\mu\nu}=2\partial_{[\mu} \vec A_{\nu]} - g\vec A_\mu\times\vec A_\nu$.
The minima of the Higgs potential parameterize the sphere of size 
$\|\vec{\phi}\|^2 = v^2 := \frac{6\mu^2\!}{\lambda}$.
Therefore, we may reparameterize the scalars in the standard way,
\be
\vec{\phi} = e^{\frac{i}{v}(\pi_1 t^1 + \pi_2 t^2)}
\left(\begin{smallmatrix} 0 \\ 0 \\ v + h \end{smallmatrix}\right),
\ee
using the Goldstone bosons $\pi_i$ and the massive Higgs boson~$h$.
Since we are not interested in the Higgs boson, we may decouple it in the infinite-mass limit, where $\lambda,\mu \to \infty$
but $v$ stays finite (see \eg \rcite{Schwartz:2014sze}).
The resulting Lagrangian is
\beal
{\cal L}_{\rm U(1)} = & -\frac{1}{4} (F_{\mu\nu})^2\!
- \frac{1}{2} |W_{\mu\nu}|^2\! + |m W_\mu\!- D_\mu \varphi|^2
- iQ \o{W}\!_\mu F^{\mu\nu} W_\nu\!
- Q^2 |\overline W_{[\mu} W_{\nu]}|^2 \\ &
- \frac{1}{4|\varphi|^2}
  \bigg[ \bigg( {\rm sinc} \Big(\!\tfrac{\sqrt{2}Q|\varphi|}{m}\Big)
  [\overline\varphi D^\mu \varphi - \varphi \o{D^\mu \varphi}]
  - m \cos\!\Big(\!\tfrac{\sqrt{2}Q|\varphi|}{m}\Big) 
    [\overline \varphi W^\mu\!- \overline W^\mu\!\varphi]\!\bigg)^{\!2} \\ &
  \qquad \qquad \qquad~\:\qquad
  - \Big( [\overline \varphi D^\mu \varphi
        - \varphi \o{D^\mu \varphi} ]
  - m [\overline \varphi W^\mu\!- \overline W^\mu\!\varphi] \Big)^2
  \bigg] \,,
\label{ActionFromHiggs}
\eeal
where $F_{\mu\nu}=2\partial_{[\mu} A_{\nu]}$ is the field strength of the residual U(1) gauge symmetry (along the VEV direction, $A_\mu = A^3_\mu$).
The remaining two gauge bosons are given mass $m=gv$ and correspond to the complex field
$W_\mu = \tfrac{1}{\sqrt{2}} (A^1_\mu + i A^2_\mu)$,
with field strength $W_{\mu\nu}=2D_{[\mu} W_{\nu]}$.
The Goldstone bosons are equivalent to the complex St\"uckelberg field considered previously,
$\varphi = \tfrac{1}{\sqrt{2}} (\pi_1 + i \pi_2)$.
The covariant derivatives $D_\mu = \partial_\mu - iQ A_\mu$,  with EM charge $Q=g$, now correspond to the ${\rm U}(1)$ symmetry.
The apparent non-locality in $|\varphi|^2$ is canceled by series expanding in $Q$.

Clearly, the Lagrangian~\eqref{ActionFromHiggs} is invariant under the ${\rm U}(1)$ gauge transformations.
However, it also inherits the following massive gauge symmetry:
\beal
\delta W_\mu & = D_\mu \xi = \partial_\mu \xi - i Q A_\mu \xi \,, \\ 
\delta A_\mu & = iQ(\o{W}\!_\mu \xi - \bar\xi W_\mu) \,, \\
\delta \varphi & = m \xi + \frac{m}{2\o{\varphi}}
   \Big( 1 - \tfrac{\sqrt{2}Q|\varphi|/m}{\tan(\sqrt{2}Q|\varphi|/m)}
   \Big) (\bar\xi \varphi - \o{\varphi} \xi)
 = m \xi + \frac{Q^2 \varphi}{3m} (\bar\xi \varphi - \o{\varphi} \xi)
 + {\cal O}(Q^4) \,.
\label{GaugeTransformFromHiggs}
\eeal
Note that the variation $\delta A_\mu$ is not coming from the EM gauge transformation (since it is the massless spin-1 field of the system), which is clear from the fact that the variation is strictly quadratic in the fields.
In contrast, both $\delta W_\mu$ and $\delta \varphi$ contain linear terms, agreeing with the linearized massive gauge symmetry previously considered in \eqn{StueckelbergSymmetry}.

We may use the above symmetry to set $\varphi=0$,
which in the familiar language of weak interactions
corresponds to the unitary gauge, in which the Goldstone bosons
are ``eaten'' by the massive vector field, and only the terms
in the first line of the Lagrangian~\eqref{ActionFromHiggs} survive.
Note that in the context of higher-spin generalizations, however,
not all auxiliary fields can be gauged away in this way
\cite{Fierz:1939ix,Singh:1974qz,Singh:1974rc}.
It is also not clear how to obtain a higher-spin analogue of the Brout-Englert-Higgs mechanism.
What is discussed in the following sections of paper can be viewed as the first terms of a higher-spin generalization of the non-linear sigma model \eqref{ActionFromHiggs}, where the Higgs field has already been decoupled.

It is useful to briefly analyze the theory \eqref{ActionFromHiggs} from the point of view of its high-energy behavior. For small coupling $Q\ll 1$, the tree-level amplitudes do not grow too large given that the characteristic momenta is smaller than the VEV scale, $p \le v= \frac{m}{Q}$. Above this energy scale, tree-level unitarity is violated and instead the amplitudes must be unitarized by loop-level resummation. Because at these energy scales the original Lagrangian \eqref{SO3YM}, with $\mu \to 0$, is the better description, and since $\lambda \to \infty$ the theory must be treated non-perturbatively. Alternatively, when we tune the quartic coupling to the perturbative regime $\lambda< 1$, then tree-level unitarity is restored if we include the Higgs boson, which now can contribute to tree amplitudes at energy scales of the Higgs mass $m_h=\sqrt{2}\mu= \frac{m}{Q} \sqrt{\frac{\lambda}{3}}$. 

Note that the high-energy limit is in the opposite direction of the classical limit that we will consider in \sec{sec:ClassCompAmps}. Indeed, the classical limit is a low-energy limit (large-distance limit), and as such it does not obviously care about tree-level unitarity and completion by the Higgs boson. Nevertheless, we find that the principles that govern the high-energy behavior also appears to be good principles for constraining classical physics.

\subsection{Spin-1 order by order}
\label{sec:spin1offshell}

In the higher-spin case, we do not have access to a mechanism analogous to spontaneous symmetry breaking.
Nevertheless, the concept of massive higher-spin gauge symmetry has proven to be very fruitful, even if it means working out non-linear corrections order by order in the fields.
In this section, we will pretend to be unaware of the Higgs construction for massive spin-1 bosons and will instead build their EM interactions and gauge symmetry anew order by order.
At the same time, we will use this example to introduce the main features of our approach to higher spins used in
\sec{sec:HigherSpinGauge}.

The goal of this paper is to describe {\it well-separated} and {\it non-decaying} massive spinning fields. Therefore, we may omit all self-interactions involving three or more massive fields, physical or auxiliary, as well as interactions with a single massive field (production or decay).
In other words, the matter part of any Lagrangian must be strictly quadratic in the massive fields.
For instance, the contact term $Q^2 |\overline W_{[\mu} W_{\nu]}|^2$ and most of the second and third lines of the spin-1 Lagrangian~\eqref{ActionFromHiggs} are irrelevant for our purposes.
Moreover, since the infinitesimal gauge parameters~$\xi$ shift the auxiliary fields, they should be treated as massive fields, which forces other massive fields to appear at most linearly in combination with~$\xi$.
For example, the non-linear terms in the variation $\delta \varphi$ in \eqn{GaugeTransformFromHiggs} are irrelevant.

The general strategy is to start from a quadratic Lagrangian ${\cal L}_2$ and the linearized massive gauge variation $\delta_0$, both minimally coupled/covariantized with respect to the EM gauge field.
Then we systematically add corrections involving more fields, 
\beal
\label{eq:lagragianapproach}
{\cal L} & = {\cal L}_2 + {\cal L}_3+ {\cal L}_4 + \ldots\,, \\
\delta &= \delta_0+\delta_1+\delta_2 + \ldots\,,
\eeal
where ${\cal L}_n$ and $\delta_{n-1}$ are covariant polynomials of degree~$n$ in the fields\footnote{When the homogeneity degree is mentioned, ``fields'' means both fields and gauge parameters. Since the first step in introducing interactions is to covariantize the derivatives in ${\cal L}_2$, it is convenient to consider these $(D\Phi)^2$-terms still be a part of ${\cal L}_2$, even though it is no longer of degree 2.
} (massive or massless).
The gauge field $A_\mu$ only appears in the form of the field strength $F_{\mu\nu}$ or inside the covariant derivative $D_\mu = \partial_\mu - iQ A_\mu$, thus ensuring U(1) gauge symmetry.
Importantly, we choose to regard $D_\mu$ as a derivative rather than a field when counting the interaction degree.

For spin~1, we start from a minimally coupled St\"uckelberg action~\eqref{Spin1ActionStuck},
\be
{\cal L}_2 = -\frac{1}{4} (F_{\mu\nu})^2
- \frac{1}{2} |W_{\mu\nu}|^2
+ |m W_\mu - D_\mu \varphi|^2\,,
\label{Spin1ActionMin}
\ee
where $W_{\mu\nu}=2D_{[\mu} W_{\nu]}$, and consider the linearized massive gauge transformation~\eqref{StueckelbergSymmetry}, with covariantized derivatives
\be
\label{StueckelbergSymmetry1}
\delta_0 W_\mu = D_\mu \xi \,, \qquad \quad
\delta_0 \varphi = m \xi \,, \qquad \quad
\delta_0 A_\mu = 0 \,.
\ee
However, once the minimal interactions are included, $Q \neq 0$, the massive gauge symmetry~$\delta_0$ is broken by the term $-|W_{\mu\nu}|^2/2$.
To restore it, we add a non-minimal deformation
to both the Lagrangian and the gauge transformations.
This is done by suitable ans\"{a}tze
\be
{\cal L}_3 = -i c_1 Q \overline W^\mu F_{\mu\nu} W^\nu \,, \qquad
\delta_1 W_\mu = 0 =\delta_1 \varphi  \,, \qquad
\delta_1 A_\mu = i c_2 Q [\o{W}\!_\mu \xi -\bar \xi W_\mu] \,,
\label{Spin1Ansatz3pt}
\ee
where $c_i$ are parameters to be determined.
The ans\"{a}tze may appear as fine-tuned, given the few non-vanishing terms.
However, to obtain the non-minimal interactions and transformations we rely on a set of guiding principles, which we will consistently use for higher-spin cases later encountered.
These guiding principles include:
\begin{description}[leftmargin=!,labelwidth=\widthof{\bfseries(PC)}]
\item[\namedlabel{NoSelfInteractions}{(SI)}]
Self-interactions of massive fields are suppressed (as motivated by the classical limit).
\item[\namedlabel{Parity}{(PS)}]
Parity symmetry is imposed on interactions.
Specifically, the Levi-Civita tensor $\epsilon_{\mu\nu\rho\sigma}$ or (anti-)self-dual field strengths do not appear.
\item[\namedlabel{PowerCounting}{(PC)}]
Power counting of derivatives: at most $s + s' - 1$ derivatives in ${\cal L}_3$,
where $s$ and $s'$ are the \ranks of the two massive fields,
\eg \rank 1 for $W_\mu$ and \rank 0 for $\varphi$.
\item[\namedlabel{PowerCountingTransform}{(PC2)}]
Power counting: at most $s + s' - 1$ derivatives in $\delta_1$,
where $s$ and $s'$ are the ranks of the massive fields and their gauge parameters.\footnote{Note that gauge variation thus has at least one less derivative than the corresponding vertex, which is required in order to preserve the structure of the constraints that follow from the gauge symmetry. In fact, it is not necessary to add this constraint if we first fix the on-shell part of ${\cal L}_3$, since this bound comes out of the general properties of the formalism ($k$ derivatives in a vertex plus one derivative in $\delta_0$ minus two derivatives in ${\cal L}_0$ result in $k-1$ derivatives for $\delta_1$, in general), see also below. \label{footPC}}
\item[\namedlabel{MinimalCoupling}{(MC)}]
Minimal-coupling interactions and linearized gauge transformations $\delta_0$ are fixed. This follows from disallowing the kinetic terms ${\cal L}_2$ to be modified.
Combined with \ref{NoSelfInteractions} it implies that the non-minimal interactions behave as ${\cal L}_n \propto (F_{\mu\nu})^{n-2}$, and for the non-minimal gauge transformations,
this implies that $\delta_n \Phi \propto (F_{\mu\nu})^n$, where $\Phi$ is a massive field, \eg $\Phi\in\{W_\mu, \varphi\}$. 
\end{description}
From the power-counting constraint \ref{PowerCounting}, the unique non-minimal interaction is given in~\eqn{Spin1Ansatz3pt}, where we count the derivatives appearing in $F_{\mu \nu}$.
Note that \ref{NoSelfInteractions} implis that the massive field variations, $\delta W_\mu$ and $\delta \varphi$,
should not contain multiple powers of massive fields or gauge parameters,
because they would mix with the massive self-interactions that we have suppressed.
Combined with assumptions~\ref{PowerCountingTransform} and~\ref{MinimalCoupling},
this implies $\delta_1 W_\mu = 0 = \delta_1 \varphi$. 
The variation $\delta_1 A_\mu$ is quadratic in the fields, including the gauge parameters $\xi$, $\bar\xi$ that must appear linearly.
Hence the power-counting constraint~\ref{PowerCountingTransform}, together with the Lorentz index and reality property of the field $A_\mu$, implies the compact ansatz in~\eqn{Spin1Ansatz3pt}.

Finally, the free parameters $c_1$ and $c_2$ are uniquely fixed by requiring massive gauge invariance to linear order in the coupling $Q$,
\be
\label{eq:gaugevar3}
(\delta_0 + \delta_1) ({\cal L}_2 + {\cal L}_3) =  {\cal O}(Q^2)
\qquad \Rightarrow \qquad
c_1 = c_2 = 1 \,.
\ee
This is of course consistent with the all-order results~\eqref{ActionFromHiggs} and \eqref{GaugeTransformFromHiggs}, which in view of condition~\ref{NoSelfInteractions} become
\be
{\cal L}_{n\ge4}=0 \,, \qquad \quad \delta_{n\ge2}=0 \,.
\ee

If we wanted to proceed systematically to ${\cal O}(Q^2)$,
we would need to formulate appropriate ans\"{a}tze for ${\cal L}_4$ and $\delta_2$. For the spin-1 example we can be cavalier about the details, since the correct solution is ${\cal L}_4 = \delta_2 A_\mu = \delta_2 W_\mu = \delta_2 \varphi = 0$, but for higher-spin cases it is unclear what are the appropriate bounds we should impose on  the derivative power counts. We cannot rely on the higher-spin literature for guidance, so we lack clear higher-order analogues of
assumptions~\ref{PowerCounting} and~\ref{PowerCountingTransform}.
Instead, we proceed by gradually increasing
the allowed derivative counts in ${\cal L}_4$ and $\delta_2$ until a solution to the massive gauge transformation is obtained.
However, we require that the pieces of ${\cal L}_4$ that contribute to the Compton four-point amplitude should not exceed the derivative count implied by the ${\cal L}_3$ interactions. Thus the higher-order interactions should not make the high-energy behavior worse than implied by the cubic interactions.

Note that the cubic Lagrangian obeying \eqn{eq:gaugevar3} yields the following three-point amplitude $\cA(\bm{1}^{s=1}\!,\bar{\bm{2}}^{s=1}\!,3)$, for two massive spin-1 fields with momenta $p_1$ and $p_2$ and one photon with momentum $p_3$,
\begin{equation}
\label{eq:spin1threeptcov}
{\cal A}(\bm{1}^{s=1}\!,\bar{\bm{2}}^{s=1}\!,3) =  -2
Q ( \bep_1\!\cdot \bep_2\;\ep_3\!\cdot\!p_1
+ \bep_2\!\cdot \ep_3\;\bep_1\!\cdot\!p_2
+ \ep_3\!\cdot \bep_1\;\bep_2\!\cdot\!p_3 ) \,.
\end{equation}
It is instructive to plug in the on-shell spinor expressions
for the massive polarization vectors, parameterized in \app{sec:appConventions} among other definitions and conventions, allowing us to write the physical amplitudes simply as
\be
{\cal A}(\bm{1}^{s=1}\!,\bar{\bm{2}}^{s=1}\!,3^+)= 2Q (\varepsilon_3^+\cdot p_1)
\frac{\braket{\bm{12}}^2\!}{m^2} \,, \qquad
{\cal A}(\bm{1}^{s=1}\!,\bar{\bm{2}}^{s=1}\!,3^-) = 2Q (\varepsilon_3^-\cdot p_1) 
\frac{[\bm{12}]^2\!}{m^2} \,,
\ee
where $\braket{\bm{12}} := z_{1a} \braket{1^a 2^b} z_{2b}$, $[\bm{12}] := z_{1a} [1^a 2^b] z_{2b}$, and for convenience we used polarizations vectors with saturated SU(2) indices~\cite{Chiodaroli:2021eug},
\be
{\bm \varepsilon}_{i,\mu}:=\varepsilon_{i,\mu}^{a b} z_{ia} z_{ib}
 = \frac{\bra{i^a}\sigma_\mu|i^b]}{\sqrt{2}m}z_{ia} z_{ib}  \,.
\label{polVectors}
\ee
We observe that these spin-1 amplitudes belong
to the \AHH family~\eqref{AHH}.
Notably, the \AHH family does not accommodate~\cite{Chung:2018kqs} the amplitudes from the minimal EM coupling~\eqref{Spin1ActionMin},
which is not entirely surprising: despite the name, the minimal EM coupling is not a consistent interaction, as it violates the massive gauge symmetry or, in other words, it sources unphysical degrees of freedom. As we will discuss in more detail in the rest of this work, the same holds for arbitrary spin.
Namely, the minimal EM coupling for massive particles with spin $s \geq 1$ does not give rise to the \AHH three-point amplitudes and it is found to violate massive gauge invariance. Moreover, we checked up to $s = 4$ that the conditions \ref{NoSelfInteractions}, \ref{Parity}, \ref{PowerCounting}, \ref{PowerCountingTransform} and \ref{MinimalCoupling}, imposed on an off-shell cubic Lagrangian, yield the \AHH three-point amplitudes as unique solution. For $s > 4$ constructing the off-shell massive-gauge-invariant Lagrangians is computationally challenging, so we will instead define a set of on-shell \textit{massive Ward identities}, introduced in the next section.

Note that, although it is a matter of straightforward computation at low spins,
the relationship between the Lagrangians and the on-shell amplitude expressions~\eqref{AHH} tends to be obscure.
A way~\cite{Chalmers:1997ui,Chalmers:2001cy,Ochirov:2022nqz}
to make it more transparent at the expense of explicit parity
is discussed in \sec{sec:spin1chiral}.

\subsection{Massive Ward identities}
\label{sec:spin1ward}

As an alternative to the fully off-shell approach above,
we can phrase massive gauge invariance in terms of (generalized) Ward identities for momentum-space correlation functions that are partially on shell.
To write the sort of identities most convenient for our purposes,
we take the perspective that
\emph{the purpose of Ward identities is to prevent pure-gauge fields to be sourced by physical fields when solving the equations of motion.}
This is of course true for Ward identities in massless gauge theory, which set to zero any scattering amplitude
involving a single longitudinally polarized gauge boson.
In the same way, we may view the massive gauge symmetry~\eqref{StueckelbergSymmetry}
as an unphysical solution $W_\mu = \partial_\mu \xi$, $\varphi = m \xi$ of the free theory~\eqref{Spin1ActionStuck},
which decouples from the remaining fields in any S-matrix element. Note that this unphysical solution is not constrained by the usual mass-shell condition, hence Ward identities typically involve one off-shell leg corresponding to the pure-gauge field.  

For a flexible implementation of such Ward identities, we use
\emph{off-shell vertices} $V(\dots)$,
which we regard as pre-contracted with
the corresponding planewave polarization-tensor \emph{placeholders}
\be \label{eq:placeholder}
\epsilon_i^{\mu_1\dots\mu_{s_i}}
= \epsilon_i^{\mu_1}\!\cdots \epsilon_i^{\mu_{s_i}} 
\qquad \text{for each particle }i.
\ee
For $s>1$, this automatically implies
full symmetrization over the Lorentz indices.
On the mass shell ($p_i^2=m_i^2$, all momenta are understood as incoming),
the placeholders are simply replaced
by the physical polarization vectors:
\beal
&\epsilon_{i,\mu} ~\to~ \varepsilon_{i,\mu}^h(p,q)
\hskip1cm  \text{particle }i\text{ is massless, with helicity }h=\pm , \\
&\epsilon_{i,\mu} ~\to~ \varepsilon_{i,\mu}^{ab}(p)
\hskip1.35cm  \text{particle }i\text{ is massive, with ${\rm SU}(2)$ spin indices }a,b = 1,2.
\eeal
Note that only $\varepsilon_i$ are assumed to obey
the usual properties expected from polarization vectors,
namely $\ep_i \cdot p_i = \ep_i^2 = 0$.
For example, $V(W_1 \o{W}\!_2 A_3)$ is a scalar function of $\epsilon_{i,\mu}$ and momenta $p_i$, $i=1,2,3$,  
containing all off-shell information
about the EM interaction vertex of the spin-1 field,
while the corresponding amplitude is obtained by putting every leg on shell,
which we write as
\be
{\cal A}(\bm{1}^{s=1}\!,\bar{\bm{2}}^{s=1}\!,3)\,:=\,V(W_1 \o{W}\!_2 A_3) \big|_{(1,2,3)} \,.
\ee
In this notation, a spin-1 Ward identity
due to the St\"uckelberg symmetry~\eqref{StueckelbergSymmetry} looks like
\be
\label{WardIdentitySpin1}
m V(\varphi_1 \o{W}\!_2 A_3)
- i p_1^\mu\,\frac{\partial~}{\partial \epsilon_1^\mu} V(W_1 \o{W}\!_2 A_3)
\bigg|_{(2,3)}\!= 0 \,,
\ee
where we used $\frac{\partial}{\partial x^\mu} \leftrightarrow -ip_\mu$
assuming incoming momenta. Throughout this work we use the subscript notation $(i,j,\dots)$ to indicate that the on-shell conditions $p_i^2-m_i^2 = \ep_i\cdot p_i = \ep_i^2 = 0$ are applied to particles $\{i,j,\dots\}$, in this case particles 2 and 3, for all terms displayed.
Note that requiring that the expression vanishes
in the fully off-shell sense would be a different and overly strong claim,
since it is violated even by the Lagrangian~\eqref{ActionFromHiggs}. Indeed, the Ward identity \eqref{WardIdentitySpin1} follows from 
\be
\label{wardnoether}
\left(\delta_0 {\cal L}_3+\delta_1 {\cal L}_2\right)\big|_{\text{on-shell}}=    \delta_0 {\cal L}_3\big|_{\text{on-shell}}={\cal O}(Q^2)\,,
\ee
which is weaker than the full off-shell gauge invariance of the action at this order, as expressed by \eqn{eq:gaugevar3}.
Hereabove, $\delta_1 {\cal L}_2= \delta_1 (\delta S_2 /\delta \Phi)$ is a multiple of the free equations of motion to this order\footnote{It is worth recalling that ${\cal L}_2$ is chosen to contain the minimal interactions introduced via $\partial \to D$. Nevertheless, we can neglect the cubic part of ${\cal L}_2$ in \eqn{wardnoether}, since its variation is multiplied by $\delta_1$, which is bilinear in the fields.
}
in $Q$.
Therefore, the second term vanishes when the corresponding legs are put on-shell. On the other hand, once we found an ${\cal L}_3$ that is gauge invariant on shell, one can take its (non-unique) off-shell extension and read off $\delta_1$, cf. footnote~\ref{footPC}.
Indeed, $\delta_0 {\cal L}_3\big|_{\text{on-shell}}=0$ means that $\delta_0 {\cal L}_3$ is a multiple of the free equations of motion and, hence, can be compensated by a suitable $\delta_1$. 

In order to constructively use \eqn{WardIdentitySpin1},
we first come up with an ansatz for the vertices, schematically, 
\begin{subequations} \begin{align}
V(W_1 \o{W}\!_2 A_3) & \sim m\, \epsilon_1 \, \epsilon_2 \, \epsilon_3\,
\Big(\frac{p}{m}\Big)  \quad \to \quad 6\text{ terms} \,, \\
V(\varphi_1 \o{W}\!_2 A_3) & \sim m\,  \epsilon_2 \, \epsilon_3\,
\big(p^0\big) \qquad\;\,\to \quad 1\text{ term} \,,
\end{align} \label{Spin1Ansatze}%
\end{subequations}
where we specify the power counting in various ingredients
but omit momentum labels, index contractions and free coefficients.
We have already used the derivative-counting assumption~\ref{PowerCounting}
and parity (absence of the Levi-Civita tensor).
Furthermore imposing the reality condition
$V(W_1 \o{W}\!_2 A_3) = -V(W_2 \o{W}\!_1 A_3)$,
as per assumption~\ref{Parity}, determines three free coefficients.
Imposing the Ward identity~\eqref{WardIdentitySpin1}
reduces the remaining four coefficients to one.
This last free parameter is fixed to the coupling $Q$ by identifying the minimal-coupling~\ref{MinimalCoupling} terms that originate from an appropriate kinetic term $|D_{[\mu} W_{\nu]}|^2$.
This gives the final result 
\begin{subequations} \begin{align}
V(W_1 \o{W}\!_2 A_3) & =  -
Q \big[ \eps_1\!\cdot \eps_2\;\eps_3\!\cdot\!(p_1-p_2)
+ \eps_2\!\cdot \eps_3\;\eps_1\!\cdot\!(p_2-p_3)
+ \eps_3\!\cdot \eps_1\;\eps_2\!\cdot\!(p_3-p_1) \big] \,, \\*
V(\varphi_1 \o{W}\!_2 A_3) & = iQ m\,\eps_2\cdot \eps_3 \,.
\end{align} \label{Spin1Solutions}%
\end{subequations}
The kinematic form of the spin-1 vertex
is identical to the 3-gluon Feynman rule,
which is no coincidence in view of the discussed construction
of the Lagrangian~\eqref{ActionFromHiggs} from non-abelian gauge theory,
which results in the same trivalent vertices as above.
The on-shell amplitude $\cA(\bm{1}^{s=1}\!,\bm{2}^{s=1}\!,3)$ can be computed from the vertex $V(W_1 \o{W}\!_2 A_3)$ and it agrees with \eqn{eq:spin1threeptcov}, thus reproducing the \AHH result.

An important advantage of \eqn{WardIdentitySpin1} is that we do not need to construct an ansatz for the non-minimal gauge transformations $\delta_1$, since it only depends on the free-theory transformations $\delta_0|_{Q=0}$, as discussed in \eqn{wardnoether}.
This significantly reduces the number of free parameters and hence allows us to compute amplitudes with higher multiplicity and higher values of the spin $s$.
For the purpose of computing the Compton diagram one does not even have to solve Ward identities for all the vertices: one leg has to be the physical spin-$s$ field, \ie the vertices with two St\"uckelberg fields and one gauge field $A$ can be omitted.
However, such reduced Ward identities are in general less constraining than imposing massive gauge invariance on the full off-shell Lagrangian.
Although \eqn{WardIdentitySpin1} is enough to fix the three-point amplitude uniquely, for $s > 1$ the reduced Ward identities alone are not sufficient, and they will need to be supplemented by additional on-shell constraints.
We will discuss this in detail in the next section.

\section{Higher-spin gauge theory with massive gauge symmetry}
\label{sec:HigherSpinGauge}

Having already outlined the problem of unphysical degrees of freedom,
in this section we discuss Zinoviev's solution to it \cite{Zinoviev:2001dt}
based on massive gauge symmetry,
which intertwines the primary higher-spin field
with its auxiliary-field descendants,
as well as with the gauge field with which it interacts.

\subsection{General Lagrangian, gauge fixing and Feynman propagator}
\label{sec:freelag}

We are now ready to review how to construct general integer-spin theories using massive gauge symmetry. In addition to the physical symmetric spin-$s$ field $\Phi_{\mu_1\dots\mu_s}$, which we will sometimes shorten to $\Phi^s$,
Zinoviev \cite{Zinoviev:2001dt} introduced auxiliary fields $\Phi^{s-1}, \Phi^{s-2}, \dots, \Phi^1, \Phi^0$, where the last one is a scalar field.
These fields are assumed to be complex and double-traceless $\Phi^{\lambda\nu}{}_{\lambda\nu\mu_5\dots\mu_k} = 0$,
such that even the physical, maximal-rank field $\Phi^s$ contains more off-shell degrees of freedom than the field chosen in earlier literature~\cite{Fierz:1939ix,Singh:1974qz,Singh:1974rc}.

The fields $\{\Phi^k\}_{k=0}^s$ are subject to massive gauge transformations 
\be
\label{ZinovievGaugeTransform}
\delta_{0}\Phi_{\mu_1\dots\mu_k}
= \partial_{(\mu_1} \xi_{\mu_2\dots\mu_k)}
+ m \alpha_k \xi_{\mu_1\dots\mu_k}
+ m\beta_k \eta_{(\mu_1\mu_2} \xi_{\mu_3\dots\mu_k)} \,,
\ee
where the complex gauge parameters are traceless,
$\xi^\lambda{}_{\lambda\mu_3\dots\mu_k} = 0$, and the numerical coefficients are
\be
\alpha_k = \sqrt{\frac{(s - k) (s + k + d - 3)}{(k + 1) (2 k + d - 2)}} \,,
\qquad \quad
\beta_k = \frac{ k \alpha_{k - 1}}{2 k + d - 6} \,,
\qquad \quad
d=4 \,.
\label{ZinovievCoefficients}
\ee
Note that $\alpha_s=0$, implying that the physical field $\Phi^s$ cannot be set to zero by a gauge choice, and the relevant parameters are $\{\xi^k\}_{k=0}^{s-1}$.
Of course, for $s=1$ we simply recover St\"uckelberg's massive gauge transformations \eqref{StueckelbergSymmetry}.

In \rcite{Cangemi:2022bew}, we found that the free Lagrangian for a complex massive spin-$s$ field that is invariant under the massive gauge transformations~\eqref{ZinovievGaugeTransform} can be written as
\begin{subequations}
\label{FreeL0}
\be
{\cal L}_\text{free} = {\cal L}_{\rm F} - {\cal L}_\text{gf},
\ee
with
\be
{\cal L}_\text{gf} = -\sum_{k=0}^{s-1} (-1)^k (k+1) \o{G_{\mu(k)}} G^{\mu(k)}\,,
\ee
\end{subequations}
where ${\cal L}_\text{F} = {\cal L}_\text{free} + {\cal L}_\text{gf}$ is the Lagrangian in Feynman gauge, obtained by adding the gauge-fixing term ${\cal L}_\text{gf}$ to the gauge-invariant Lagrangian ${\cal L}_\text{free}$.
The Feynman-gauge Lagrangian ${\cal L}_\text{F}$ is completely diagonal,
\be
\label{QuadraticActionFeynmanGauge}
{\cal L}_\text{F} = -\sum_{k=0}^s (-1)^{k+1} \bigg[ \bar{\Phi}_{\mu(k)} (\Box + m^2) \Phi^{\mu(k)} - \frac{k(k-1)}{4} \bar{\tilde{\Phi}}_{\mu(k-2)}(\Box + m^2)\tilde{\Phi}^{\mu(k-2)} \bigg] \,,
\ee
with $\tilde{\Phi}_{\mu(k-2)} := \Phi^{\lambda}{}_{\lambda \mu_3 \dots \mu_k}$ denoting single traces.
 
The gauge-fixing functions $G^{\mu(k)} = G^{(\mu_1 \dots \mu_k)}$ contain the off-diagonal contributions and are given by 
\be
G^{\mu(k)}\!:=\partial_\lambda \Phi^{\lambda\mu (k)}
- \frac{k}{2} \partial^\mu \tilde{\Phi}^{\mu(k-1)}
+ m\alpha_{k} \Phi^{\mu(k)}
- m \alpha_{k+1}\frac{k{+}1}{2} \tilde{\Phi}^{\mu(k)} 
-  m \alpha_k \frac{k{-}1}{4} \eta^{\mu\mu} \tilde{\Phi}^{\mu(k-2)} \,,
\ee
where we use the convention that repeated all-upper (or all-lower) indices $\eta^{\mu \mu}\Phi^{\mu(k)}$, $\partial^\mu\Phi^{\mu(k)}$ are a compact notation for symmetrization of indistinguishable indices. Note that traces of rank-0,1 fields are defined to be zero, $\tilde{\Phi}^{\mu(-2)}=\tilde{\Phi}^{\mu(-1)} := 0$. 

The gauge-fixing functions have a simple gauge transformation following from \eqn{ZinovievGaugeTransform},  
\be
\delta_0 G^{\mu(k)}=\frac{1}{k+1}(\Box+m^2)\xi^{\mu(k)}\,.
\ee
The appearance of the Klein-Gordon operator $(\Box+m^2)$ implies that gauge-fixing constraints, such as the equivalent of Lorenz gauge $G^{\mu(k)}=0$, are invariant under gauge transformations with parameters $\xi^{\mu(k)}$ that satisfy the mass-shell constraint $p^2=m^2$. Hence, a residual gauge freedom remains present.  

The diagonal form of the gauge-fixed action~\eqref{QuadraticActionFeynmanGauge} is particularly advantageous when extending our analysis to four-point amplitudes, where the propagator for each field is required.
The Feynman-gauge propagator $\Delta^{(s)}$ for a spin-$s$ field is diagonal and has trivial momentum dependence.
It is unique, and we find that in $d=4$ dimensions it is given by the simple generating function~\cite{Cangemi:2022bew}  
\be
\label{full_prop}
\Delta(\epsilon,\bar \epsilon)
 = \sum_{s=0}^{\infty} (\epsilon)^s {\cdot} \Delta^{\!(s)} {\cdot}   
   (\bar\epsilon)^s
 = \frac{i}{p^2{-}m^2{+}i0}\,
   \frac{1 - \frac{1}{4}\epsilon^2 \bar \epsilon^2}
        {1 + \epsilon\cdot\bar\epsilon
           + \frac{1}{4} \epsilon^2 \bar\epsilon^2} \,,
\ee
where, as before, $\epsilon_\mu,\bar \epsilon_\mu$ are auxiliary vectors regarded as independent, and $\epsilon\cdot\bar\epsilon$, $\epsilon^2$, $\bar\epsilon^2$ are standard Lorentz contractions. Remarkably, this propagator already includes the appropriate projectors that enforce the double-tracelessness of the rank-$s$ fields.\footnote{While the Feynman-gauge kinetic term does not a have unique inverse, the additional requirement that the propagator acts as a double-traceless projector gives the unique formula.
}
Note that the Feynman propagator does not discern between physical and auxiliary fields, but treats them all on the same footing, which is also clear from the Feynman-gauge-fixed Lagrangian~\eqref{QuadraticActionFeynmanGauge}.   

In order to introduce interactions in the spin-$s$ theory \eqref{FreeL0} one can first attempt to minimally couple it to electromagnetism.  As before, we covariantize the free Lagrangian, and we obtain the minimally coupled interacting spin-$s$ theory,
\be
\label{eq:L2mincoup}
{\cal L}_2 = {\cal L}_\text{free}\Big|_{ \partial_\mu\to D_\mu = \partial_\mu - iQ A_\mu}\,,
\ee
where the subscript on $\cL_2$ is to emphasize that this comes from the quadratic terms.\footnote{We emphasize that, unlike ${\cal L}_2$, the gauge-fixing term is kept quadratic. Alternatively, one can also covariantize the gauge-fixing term with respect to massless gauge transformations, but this approach will not be pursued here. } 
However, this step alone will not give rise to a consistent theory. The minimal interactions spoil the counting of the degrees of freedom, since we have not ensured that the massive gauge symmetry is preserved.

As was the case in \sec{sec:spin1offshell}, covariantizing ${\cal L}_2$ and $\delta_0$ is not sufficient, and we must add non-minimal interactions\footnote{Note that neither the ``minimal EM'' nor non-minimal terms are consistent interactions when considered independently of each other. Only very specific linear combinations of such terms form consistent interactions. Those can be found be requiring the longitudinal modes to decouple, which is equivalent to imposing the massive higher-spin gauge symmetry.
}, ${\cal L}_3+...$, 
and gauge transformations, $\delta_1+...$.
In \sec{sec:spin1offshell}, we discussed the systematic approach for fixing the theory at the Lagrangian level, as shown in \eqn{eq:lagragianapproach}.
This involves constructing the action and gauge transformations order by order, fully off-shell. For instance, at cubic order, we need to consider two types of contributions to the off-shell cubic vertex,
\be \label{cubicV}
V(\Phi^s_1 \Phi^{s'}_2A_3)= V_{\cL_2}(\Phi^s_1 \Phi^{s'}_2A_3)+ V_{\cL_3}(\Phi^s_1 \Phi^{s'}_2A_3)\,,
\ee
where $V_{\cL_2}(\Phi^s_1 \Phi^{s'}_2A_3)$ is extracted from the minimally-coupled Lagrangian $\cL_2$ in \eqn{eq:L2mincoup}, and $V_{\cL_3}(\Phi^s_1 \Phi^{s'}_2A_3)$ from the non-minimal cubic interactions $\cL_3$ linear in the field strength of the massless boson $F_{\mu\nu}$.
The $\cL_2$ term can be written out explicitly,
\begin{align}
V_{\cL_2}(\Phi^s_1 \bar{\Phi}^{s'}_2A_3) & = 2\delta^{ss'}  (-1)^{s} (i p_2\cdot \epsilon_3) (\epsilon_1\cdot \epsilon_2)^{s-2} \bigg[(\epsilon_1\cdot \epsilon_2)^{2} - \frac{s(s-1)}{4} \epsilon_1^2 \epsilon_2^2 \bigg] \nn \\
& + V_{{\cal L}_\text{gf}}(\Phi^s_1 \bar{\Phi}^{s'}_2A_3)
\end{align}
where the first line comes from ${\cal L}_{\rm F}$ and the second line comes from the $\overline{G_{\mu(k)}}G^{\mu(k)}$ term.
Note that if the gauge-fixing function is covariantized it alter the second line; however, we choose to work with a strictly quadratic gauge-fixing function, which does not alter the interaction terms.

The $V_{\cL_3}$ interactions are constructed as an ansatz, together with the non-minimal gauge transformations $\delta_1$, and the free coefficients are fixed by requiring off-shell massive gauge invariance. We will do this explicitly for spin-2 and spin-3 fields in \sec{sec:massivegaugecubic}, and we will see that it is a significant computational challenge even at low spins.

On the other hand, in \sec{sec:spin1ward} we introduced an approach influenced by on-shell techniques, working with identities at the level of the vertices. This approach only requires the free-field massive gauge transformations, which lets us bypass ansatzing the non-linear gauge transformations and thus provides a significant computational advantage.
Moreover, there is an immediate connection to scattering amplitudes, which is an advantage considering our goal of comparing to the \AHH three-point amplitudes and extending them to four points. We will use this method to compute arbitrary-spin three-point amplitudes in \sec{sec:massivegaugecubic}, and four-point Compton amplitudes for spin-2 and spin-3 particles in \sec{sec:massivegaugequartic}.

Note, however, that solving the Ward identities, which are partially on-shell, and finding a fully off-shell gauge-invariant Lagragian are two equivalent approaches. Indeed, consider the cubic vertex in  \eqn{cubicV}, the Ward identities $\delta_0 V|_{\text{on-shell}}=0$ guarantee that $\delta_0 V$ is a multiple of the free equations of motion and, hence, $\delta_1$ can be read off once the variation is kept off-shell, see \eqn{wardnoether}. Different ways of extending $V_{{\cal L}_3}$ off-shell will lead to different $\delta_1$, which are related via field redefinitions. Therefore, the ambiguity in the off-shell extension does not carry any information about the physics. Let us note, however, that, in practice, we solve the reduced Ward identities where one of the fields is the physical spin-$s$ field and additional constraints are required, which are discussed below.

\subsection{Massive Ward identities again}
The discussion of the Ward identities in \sec{sec:spin1ward} can easily be extended to any spin $s$.
At the cubic order we have the vertices \eqref{cubicV} that encodes interactions between two fields from the massive spin-$s$ multiplets and a gauge field, $V(\Phi^k\!,\o{\Phi}^n\!,A)$, where $k,n=1,...,s$ run over the auxiliary fields $k,n=1,...,s{-}1$ and the physical field $k,n=s$.
In practice, one factorizes fields via the unconstrained polarizations $\epsilon^\mu_i$ introduced in \eqn{eq:placeholder}.
As a result, in momentum space $V(\Phi^k \o{\Phi}^n A)$ becomes a scalar function of $\epsilon_{i,\mu}$ and $p_i$, $i=1,2,3$, which we denote $V(\Phi^k_1 \o{\Phi}^n_2 A_3)$ with the subscript referring to the $i$-index of $\epsilon_{i,\mu}$.\footnote{Note that it is not necessary to impose double-tracelessness of the fields $\Phi_i^{\mu(k)}$, since either the on-shell conditions or the Feynman propagator~\eqref{full_prop} will automatically enforce it.
}
We can and will use the same polarization vectors $\epsilon_i$ to factorize the gauge parameters $\xi^{\mu(k)}$, $k=0,...,s{-}1$ since $\epsilon_{i,\mu}$ is merely a book-keeping device.
The tracelessness of the gauge parameters is ensured by further imposing $\epsilon_i^2=0$.

We then consider the linearized gauge transformation $\delta_0$ in \eqn{ZinovievGaugeTransform}, and exhibit all fields that pick up a change proportional to the gauge parameter $\xi_i^{\mu(k)}$. Recalling that the same polarization vectors can be used for fields and gauge parameters, we arrive at the following implementation of the gauge transformations   
\beal \label{k_specific_transformation}
\delta_0 \Phi^{\mu(k)}_1 &
 = m \alpha_k \xi^{\mu(k)}_1 &&  \to &
   m \alpha_k & V(\Phi^{\mu(k)}_1\!\cdots)
   \Big|_{\epsilon_1^2=0} \,, \\
\delta_0 \Phi^{\mu(k+1)}_1 &
 = \partial^{\mu} \xi^{\mu(k)}_1 &&  \to &
   \frac{1}{k+1} \frac{\partial~}{\partial x_\nu}
   \frac{\partial~}{\partial \epsilon_1^\nu} &
   V(\Phi^{\mu(k+1)}_1\!\cdots)
   \Big|_{\epsilon_1^2=0} \,, \\
\delta_0 \Phi^{\mu(k+2)}_1 &
 = m \beta_{k+2} \eta^{\mu\mu} \xi^{\mu(k)}_1
 && \to & \frac{m \beta_{k+2}}{(k+2)(k+1)}  \frac{\partial~}{\partial \epsilon_1} \cdot
   \frac{\partial~}{\partial \epsilon_1} &
   V(\Phi^{\mu(k+2)}_1\!\cdots)
   \Big|_{\epsilon_1^2=0} \,.
\eeal
Here, we displayed the operators that realize the gauge transformations due to $\xi^{\mu(k)}$ for the $3$-point case. Generalization to $n$-point case is straightforward.

Now we consider the (reduced) massive Ward identity for a process involving three plane waves, and apply the gauge transformation $\delta_0$ that is due to $\xi^{\mu(k)}$ to the first leg. First, using the relations \eqref{k_specific_transformation}, and plugging in $\beta_{k+2}=\frac{1}{2}a_{k+1} (k{+}2)/(k{+}1)$,  we construct an off-shell version of the gauge-transformed three-point vertex,
\be
\label{eq:gaugetransV}
V(\xi_1^k \bar{\Phi}_2^s  A_3) := 
m \alpha_k  V(\Phi_1^k \bar{\Phi}_2^s A_3)
- \frac{i p_1}{k\!+\!1} \cdot \frac{\partial~}{\partial \epsilon_1}
  V(\Phi_1^{k+1} \bar{\Phi}_2^s A_3)
+ \frac{m\alpha_{k+1}}{2(k{+}1)^2} 
  \bigg(\!\frac{\partial~}{\partial \epsilon_1}\!\bigg)^{\!\!2}
  V(\Phi_1^{k+2} \bar{\Phi}_2^s A_3),
\ee
where the notation reminds us that we are effectively scattering the gauge parameter $\xi_1^{\mu(k)}$. Then, the massive Ward identity at three points is given by setting the above vertex to zero after imposing on-shell conditions on legs $2,3$ and tracelessness on leg $1$.

Since we plan to work with the reduced Ward identities, we now state the additional constraints that we will impose on the three point interactions for the spin-$s$ theory:
\begin{description}
\item[\namedlabel{WardIdentity}{(WI)}]
Massive Ward identities due to Zinoviev's gauge symmetry~\eqref{ZinovievGaugeTransform}:
\be
V(\xi_1^k \bar{\Phi}_2^s  A_3) \big|_{(2,3),\epsilon_1^2 = 0} = 0 \,.
\ee
\end{description}
\begin{description}
\item[\namedlabel{CurrentConstraint}{(CC)}]
Current constraint following from high-energy unitarity, as discussed in \rcite{Chiodaroli:2021eug}:
\be
p_1 \cdot \frac{\partial~}{\partial \epsilon_1} V(\Phi_1^s \bar{\Phi}_2^s  A_3) \bigg|_{(2,3),\epsilon_1^2 = 0} = {\cal O}(m) \,.
\label{CurrentConstraintEqn}
\ee
\end{description}
\begin{description}
\item[\namedlabel{NearDiagonal}{(ND)}]
Near-diagonal interactions of two massive fields. The cubic vertices differ at most by one unit:\footnote{In practice, it is applied to $n=s$ since we consider only the reduced Ward identities where the physical spin-$s$ field is present.}
\begin{equation}
V(\Phi_1^{n-k} \bar{\Phi}_2^{n} A_3^h) \big|_{k>h} = 0 \,.
\end{equation}
\end{description}
In addition, the previously introduced constraints \ref{NoSelfInteractions}, \ref{Parity}, \ref{PowerCounting}, \ref{MinimalCoupling} should still be imposed on the spin-$s$ theory. 

Let us discuss the new constraints. As we will see in \sec{sect:3ptallspin}, they are needed, in combination with the reduced massive Ward identities, to fix the three-point \AHH amplitudes uniquely.\footnote{Note that leading-Regge open-string higher-spin amplitudes violate the constraints that we are imposing. Specifically, it is the combination of  \ref{PowerCounting}+\ref{CurrentConstraint} that are too constraining for string amplitudes above $s>2$. In particular, string amplitudes involve non-minimal interactions.
}
However, in the off-shell Lagrangian approach, the \AHH amplitudes follow uniquely from massive gauge invariance without the need to impose additional constraints, as we checked up to $s = 4$.
Therefore, it is possible that \ref{CurrentConstraint} and \ref{NearDiagonal} encode information about massive gauge symmetry, which is somehow missed by the reduced on-shell Ward identities, since they do not probe all three-point vertices but only those with one leg being the physical spin-$s$ field.
To confirm this, we would need to push the off-shell computation to higher values of spin, $s > 4$. 

The \ref{NearDiagonal} constraint is obeyed by the minimally coupled theories that follow from the free Lagrangians~\eqref{FreeL0}, so we propose that it should be extended to the non-minimal interactions.
The current constraint \ref{CurrentConstraint} has been argued~\cite{Ferrara:1992yc, Porrati:1993in, Cucchieri:1994tx} to give rise to scattering amplitudes with an improved high-energy limit.
For instance, in the spin-1 theory discussed in \sec{sec:Spin1}, the massive propagator in unitary gauge (UG) becomes
\be
\Delta_{\mu\nu}^{\rm UG} = -i \frac{\eta_{\mu\nu}-\frac{p_\mu p_\nu}{m^2}}{p^2 - m^2 + i0} \,.
\ee
In the high-energy limit $p/m \to \infty$, it is clear that the second term in the propagator numerator give rise to a divergence.
The implication for a physical process, such as the Compton amplitude
${\cal A}(\bm{1}^{s=1}\!,\bar{\bm{2}}^{s=1}\!,3^-\!,4^+)$, is that such divergences appear in the massive-exchange diagrams and are proportional to the contraction of the exchanged momentum $p_\mu$ and the three-point vertices, namely
$\frac{p}{m} \cdot \frac{\partial~}{\partial \epsilon_{p}} V(\Phi_{p}^s \bar{\Phi}_i^s  A_j) \big|_{(i,j)}$.
The current constraint~\eqref{CurrentConstraintEqn} cures this type of divergences by removing the factor of the mass~$m$ from the denominator.

On the other hand, in Feynman gauge the propagators~\eqref{full_prop} do not contain momentum-dependent terms. In that case, the divergences appear in the vertices of the theory.
For instance, we can modify the $s = 1$ Lagrangian in \sec{sec:spin1offshell} by adding the term
\be
\label{eq:spin1addterm}
i Q \tilde{c}_1 F^{\mu\nu} \left( \overline{W}_\mu - \frac{1}{m} \overline{D_\mu \varphi} \right) \left( W_\nu - \frac{1}{m}  D_\nu \varphi \right) \,,
\ee
where $\tilde{c}_1$ is a free coefficient. The resulting Lagrangian is still gauge-invariant at cubic order, since the structure in the brackets is invariant under the free gauge transformations~\eqref{StueckelbergSymmetry1}. 
However, it violates the current constraint, namely,
\be
p_1\cdot \frac{\partial~}{\partial \epsilon_1} V(W_1 \overline{W}_2  A_3) \Big|_{(2,3)}
= Q \tilde{c}_1 (2 \bep_2\!\cdot\!p_1 \, \ep_3\!\cdot\!p_1 - p_1^2\,\bep_2 \!\cdot\! \ep_3) + {\cal O}(m) \,.
\label{CurrentConstraintViolated}
\ee
In Feynman gauge, the terms in \eqn{eq:spin1addterm} proportional to $1/m$ give rise to high-energy divergences in scattering amplitudes, unless $\tilde{c}_1 = 0$ as required by the current constraint.

Vertex~\eqref{eq:spin1addterm} is also an illustration of the fact that the massive higher-spin gauge symmetry needs to be supplemented with appropriate constraints on the number of derivatives in the interactions.
Indeed, one can take any function of $W_\mu$, $\o{W}_\mu$ and replace them by the expressions in the brackets of \eqn{eq:spin1addterm}, the resulting vertex being gauge-invariant. However, such a cavalier approach induces higher derivatives for the auxiliary fields.
For example, \eqn{eq:spin1addterm} is a three-derivative vertex, while the minimal interaction is a two-derivative one.
The current constraint is a way to improve the derivative counting at least at the cubic order. Black holes should correspond to the best possible scenario, where the number of derivatives is kept the lowest at all orders.

\subsection{Massive gauge symmetry at cubic order}
\label{sec:massivegaugecubic}

\subsubsection{Example: spin-2 $\sqrt{\rm Kerr}$ theory}

In this section, we apply the methods outlined in \sec{sec:spin1offshell} to the case of a massive spin-2 field coupled to a photon. We work in $d = 4$ dimensions and start with the free massive spin-$2$ Lagrangian $\cL_2$ and then add non-minimal terms $\cL_3$. For convenience, and following~\rcite{Zinoviev:2009hu}, we decompose the spin-$s$ Lagrangians as $\cL^{(s)}_{n,k}$, where $n$ is the number of fields (the gauge field in $D_\mu$ is not counted) and $k$ is the number of derivatives (counting both $D_\mu$ and the derivative in $F_{\mu\nu}$). 

The minimally coupled spin-2 Lagrangian is then
\begin{equation}
\cL_2^{(2)} = \cL_{2,2}^{(2)} + \cL_{2,1}^{(2)} + \cL_{2,0}^{(2)}\,,
\end{equation}
where
\begin{align}
\cL_{2,2}^{(2)} & = 
\o{D_\alpha \Phi_{\mu \nu}} D^\alpha \Phi^{\mu \nu}
-2\o{D \cdot \Phi_\mu} D \cdot \Phi^\mu
+\o{D \cdot \Phi_\nu} D^\nu \tilde{\Phi}
+\o{D_\nu \tilde{\Phi}} D \cdot \Phi^\nu
-\o{D_\mu \tilde{\Phi}} D^\mu \tilde{\Phi}
+\o{D_\alpha \varphi} D^\alpha \varphi \nn \\* & \quad
-\o{D_\mu B_\nu} D^\mu B^\nu
+\o{D \cdot B}\,D\cdot B \,, \nn \\
\cL_{2,1}^{(2)} & =
\sqrt{3}m (\o{D \cdot B} \varphi + \bar{\varphi} D \cdot B)
-\sqrt{2} m (\o{D\cdot \Phi_\nu} B^\nu + \o{B}_\nu D\cdot \Phi^\nu)
-\sqrt{2} m (\o{\tilde{\Phi}} D \cdot B + \o{D \cdot B} \tilde{\Phi}) \,, \nn \\
\cL_{2,0}^{(2)} & =
-m^2 (\o{\Phi}_{\mu \nu} \Phi^{\mu \nu} - \o{\tilde{\Phi}} \tilde{\Phi})
-\sqrt{\frac{3}{2}}m^2 (\o{\tilde{\Phi}} \varphi + \bar{\varphi} \tilde{\Phi})
+2m^2 \bar{\varphi} \varphi \,.
\end{align}
Here we recall the shorthand notation $D\cdot \Phi^{\mu_1\dots\mu_k} := D_\rho \Phi^{\rho \mu_1\dots\mu_k}$ and $\tilde{\Phi}^{\mu_1\dots\mu_k} := \eta_{\rho\sigma} \Phi^{\rho\sigma\mu_1\dots\mu_k}$, and for clarity we renamed the St\"{u}ckelberg fields to $B_\mu$ and $\varphi$. 

The minimally coupled massive gauge transformations $\delta_0$ are given by
\beal
\label{eq:delta0spin2}
\delta_0 \Phi_{\mu \nu} &= 
D_{(\mu}\xi_{\nu)} + \frac{m}{\sqrt{2}}\eta_{\mu \nu} \zeta \,, \\
\delta_0 B_\mu &=
D_\mu \zeta + \frac{m}{\sqrt{2}} \xi_\mu \,, \\
\delta_0 \varphi &=
\sqrt{3}m \zeta \,.
\eeal
As before, for $Q \neq 0$ this gauge symmetry is broken and we need to add non-minimal deformations to restore it.
Assuming \ref{NoSelfInteractions}, \ref{MinimalCoupling}, \ref{PowerCounting} and \ref{PowerCountingTransform} the allowed ansatz for the Lagrangian contains 37 free parameters, whereas the ansatz for the gauge transformations contains 42 free parameters.
Imposing massive gauge invariance leaves 12 free parameters in the Lagrangian and 18 in the gauge transformations.
The leftover free parameters arise due to field-redefinition redundancies and higher-point effects, since none of them contributes to the on-shell three-point amplitude ${\cal A}(\Phi_1 \bar{\Phi}_2 A_3^-)$, which is uniquely predicted and it matches the known \AHH answer,
\begin{align}
\label{eq:ampl3spin2}
{\cal A}(\bm{1}^{s=2}\!,\bar{\bm{2}}^{s=2}\!,3^+) & = 
\frac{4Q}{m^2} {\bep_1}\!\cdot\!{p_2} {\bep_2}\!\cdot\!{p_1} (
{\bep_1}\!\cdot\!{\ep_3}\, {\bep_2}\!\cdot\!{p_3}
+{\bep_2}\!\cdot\!{\ep_3}\, \bep_1\!\cdot\!p_2
+ {\bep_1}\!\cdot\!{\bep_2}\, {\ep_3}\!\cdot\!{p_1} ) \\*
&\quad + Q {\bep_1}\!\cdot\!{\bep_2} ( 4{\bep_1}\!\cdot\!{\ep_3}\, {\bep_2}\!\cdot\!{p_3}
+4 {\bep_2}\!\cdot\!{\ep_3}\, {\bep_1}\!\cdot\!{p_2}
+2 \bep_1\!\cdot\!\bep_2\, {\ep_3}\!\cdot\!{p_1} )
= 2Q\,\ep_3^+\!\cdot\!p_1 \frac{\braket{\bm{12}}^4\!}{m^4} . \nn
\end{align}
One explicit solution for the full cubic Lagrangian $\cL_3$ is
\begin{equation}
\cL_3^{(2)} = \cL_{3,3}^{(2)} + \cL_{3,2}^{(2)} + \cL_{3,1}^{(2)} \,,
\end{equation}
where
\begin{subequations} \begin{align}
\cL_{3,3}^{(2)} & =
\frac{2 i Q}{m^2} F^{\mu \nu} \Big(
\o{D_\mu \Phi_{\alpha \beta}} D^\alpha \Phi\fud{\beta}{\nu}
-\o{D_\alpha \Phi_{\beta \nu}} D_\mu \Phi^{\alpha \beta}
+ \o{D_\alpha \Phi_{\beta \mu}} D^\alpha \Phi\fud{\beta}{ \nu}
-2\o{D_\alpha \Phi_{\beta \mu}} D^\beta \Phi\fud{\alpha}{ \nu} \nn \\* & \quad
-\o{D_\mu \Phi_{\alpha \beta}} D_\nu \Phi^{\alpha \beta}
+\o{D_\mu \Phi_{\nu \alpha}} D \cdot \Phi^\alpha
-\o{D \cdot \Phi_\alpha} D_\mu \Phi\fdu{\nu}{ \alpha}
+\o{D \cdot \Phi_\mu} D \cdot \Phi_\nu \nn \\* & \quad
-\o{D \cdot \Phi_\mu} D_\nu \tilde{\Phi}
+\o{D_\nu \tilde{\Phi}} D \cdot \Phi_\mu
-\o{D_\mu \Phi_{\nu \alpha}} D^\alpha \tilde{\Phi}
+\o{D_\alpha \tilde{\Phi}} D_\mu \Phi\fdu{\nu}{ \alpha}
+\o{D_\mu \Phi} D_\nu \tilde{\Phi}
\Big) \,, \\
\cL_{3,2}^{(2)} & =
\frac{i \sqrt{2} Q}{m} F^{\mu \nu}
\left(
\o{D \cdot \Phi_\mu} B_\nu
-\bar{B}_\nu D \cdot \Phi_\mu
-\o{D_\mu \tilde{\Phi}} B_\nu
+ \bar{B}_\nu D_\mu \tilde{\Phi}
\right) \\
\cL_{3,1}^{(2)} & =
i Q F^{\mu \nu} \left(2 \bar{\Phi}_{\mu \alpha} \Phi\fdu{\nu}{ \alpha}
+ \bar{B}_\mu B_\nu
\right)\,,
\end{align} \end{subequations}
where we recall that $\tilde \Phi := \Phi^\mu_\mu$ and $F_{\mu \nu} := 2 \partial_{[\mu} A_{\nu]}$.

The non-minimal gauge transformations $\delta_1$ are
\begin{equation}
\begin{split}
\delta_1 \Phi_{\mu \nu} = &\, 
{-}\frac{i Q}{m^2} \Big\{ (F\fdu{\mu}{ \alpha} D_{[\alpha} \xi_{\nu]} + F\fdu{\nu}{ \alpha} D_{[\alpha} \xi_{\mu]})
+\frac{1}{2} \eta_{\mu \nu} F^{\alpha \beta} D_{[\alpha} \xi_{\beta]}
\Big\},\\
\delta_1 A_\mu = &\,
{-}\frac{i Q}{m^2} \Big\{
2\o{D_\alpha \Phi_{\beta \mu}} D^{[\alpha} \xi^{\beta]}
-2\o{D_{[\alpha} \xi_{\beta]}} D^\alpha \Phi\fud{\beta}{ \mu}
+\sqrt{2}m (\o{B}^\alpha D_{[\mu} \xi_{\alpha]} - \o{D_{[\mu} \xi_{\alpha]}} B^\alpha) \\ & \qquad \quad
+m^2 (\bar{\Phi}_{\mu \alpha} \xi^\alpha - \bar{\xi}^\alpha \Phi_{\mu \alpha})
-m^2 (\bar{B}_\mu \zeta - \bar{\zeta} B_\mu)
\Big\}\,.
\end{split}
\end{equation}

The next step is to study the spin-2 theory in the language of massive Ward identities, as outlined in \sec{sec:HigherSpinGauge}.
From the free-theory part of \eqn{eq:delta0spin2}, we compute the gauge-transformed vertices
\begin{subequations} \begin{align}\!\!
V(\xi_1 \bar{\Phi}_2 A_3) & = -\frac{i p_1}{2} \!\cdot\!\frac{\partial}{\partial \epsilon_1} V(\Phi_1,\bar{\Phi}_2,A_3) + \frac{m}{\sqrt{2}} V(B_1,\bar{\Phi}_2,A_3) \,, \\\!\!
V(\zeta_1 \bar{\Phi}_2 A_3) & = \frac{m}{2\sqrt{2}} \frac{\partial}{\partial \epsilon_1}\!\cdot \!\frac{\partial}{\partial \epsilon_1} V(\Phi_1 \bar{\Phi}_2 A_3) - i p_1 \!\cdot \frac{\partial}{\partial \epsilon_1}\! V(B_1 \bar{\Phi}_2 A_3) + \sqrt{3} m V(\varphi_1 \bar{\Phi}_2 A_3) \,,
\end{align} \end{subequations}
and impose the massive Ward identities
\be
V(\xi_1 \o{\Phi}_2 A_3) \big|_{(2,3)} = V(\zeta_1 \o{\Phi}_2 A_3) \big|_{(2,3)} = 0
\ee
together with the assumptions~\ref{NoSelfInteractions}, \ref{MinimalCoupling}, \ref{PowerCounting} and \ref{PowerCountingTransform}.
The solution yields a three-point amplitude ${\cal A}(\bm{1}^{s=2}\!,\bar{\bm{2}}^{s=2}\!,3)$ with a single free parameter $c$:
\begin{align}
\label{spin2ansatzgeneral3pt}
{\cal A}(\bm{1}^{s=2}\!,\bar{\bm{2}}^{s=2}\!,3) = &\,Q \bigg(
   \frac{c}{m^2} \bep_1\!\cdot\!p_2\,\bep_2\!\cdot\!p_1
 + \frac{4\!+\!c}{2} \bep_1\!\cdot\!\bep_2\!\bigg)\!
   \big(
   \bep_1\!\cdot\!\bep_2 \, \ep_3\!\cdot\!p_1 + \bep_2\!\cdot\!\ep_3 \,
   \bep_1\!\cdot\!p_2 + \ep_3\!\cdot\!\bep_1 \, \bep_2\!\cdot\!p_3
   \big) \nn \\ &
 - \frac{Q \alpha}{2} (\bep_1\!\cdot\!\bep_2)^2 \, \ep_3\!\cdot\!p_1 \,.
\end{align}
If we additionally impose the constraints \ref{CurrentConstraint} and \ref{NearDiagonal}, we find $c = 4$, matching \eqn{eq:ampl3spin2}.

In summary, the combination of massive Ward identities, the current constraint and the \ref{NearDiagonal} assumption encodes the same information on the cubic theory as the off-shell Lagrangian gauge invariance, but the latter is considerably simpler to implement and easier to push to higher orders.

\subsubsection{Example: spin-3 $\sqrt{\rm Kerr}$ theory}

\def\spa12{\langle 12 \rangle}
\def\spb12{[12]}
\def\SI{\ref{NoSelfInteractions}}
\def\MC{\ref{MinimalCoupling}}
\def\PC{\ref{PowerCounting}}
\def\PCT{\ref{PowerCountingTransform}}
\def\WI{\ref{WardIdentity}}
\def\CC{\ref{CurrentConstraint}}
\def\ND{\ref{NearDiagonal}}

We can repeat the above analysis for a massive spin-$3$ field coupled to electromagnetism. We begin with the minimally coupled Lagrangian
\begin{equation}
\cL_2^{(3)} = \cL_{2,2}^{(3)} + \cL_{2,1}^{(3)} + \cL_{2,0}^{(3)}\,,
\end{equation}
where
\begin{subequations} \begin{align}
\cL_{2,2}^{(3)} = &
- \o{D_\mu \Phi_{\nu\rho\sigma}} D^\mu \Phi^{\nu\rho\sigma}
+3 \o{D\!\cdot\!\Phi_{\mu\nu}} D\!\cdot\!\Phi^{\mu\nu}
+\frac{3}{2} \o{D\!\cdot\!\tilde{\Phi}} D \!\cdot\!\tilde{\Phi}
+3 \o{D_\mu \tilde{\Phi}_\nu} D^\mu \tilde{\Phi}^\nu \\* &
-3 \o{D_\mu \tilde{\Phi}_\nu} D\!\cdot\!\Phi^{\mu\nu}
-3 \o{D\!\cdot\!\Phi^{\mu\nu}} D_\mu \tilde{\Phi}_\nu
+ \o{D_\mu H_{\nu\rho}} D^\mu H^{\nu\rho}
-2 \o{D\!\cdot\!H_\mu} D\!\cdot\!H^\mu \nn \\* &
+ \o{D_\mu \tilde{H}} D\!\cdot\!H^\mu
+ \o{D\!\cdot\!H^\mu} D_\mu \tilde{H}
- \o{D_\mu \tilde{H}} D^\mu \tilde{H}
- \o{D_\mu B_\nu} D^\mu B^\nu
+ \o{D\!\cdot\!B} D\!\cdot\!B \nn
+ \o{D_\mu \varphi} D^\mu \varphi \,, \nn \\
\cL_{2,1}^{(3)} = &
\,2\sqrt{3} m \o{\tilde{\Phi}}_\mu D\!\cdot\!H^\mu
+2\sqrt{3} m \o{D\!\cdot\!H^\mu} \tilde{\Phi}_\mu
+\sqrt{3} m \o{D\!\cdot\!\Phi_{\mu\nu}} H^{\mu\nu}
+\sqrt{3} m \o{H^{\mu\nu}} D\!\cdot\!\Phi_{\mu\nu} \nn \\* &
+\frac{\sqrt{3}}{2} m \o{D\!\cdot\!\tilde{\Phi}} \tilde{H}
+\frac{\sqrt{3}}{2} m \o{\tilde{H}} D\!\cdot\!\tilde{\Phi}
-\sqrt{5} m \o{\tilde{H}} D\!\cdot\!B
-\sqrt{5} m \o{D\!\cdot\!B} \tilde{H}
-\sqrt{5} m \o{D\!\cdot\!H_\mu} B^\mu \nn \\* &
-\sqrt{5} m \bar{B}^\mu D\!\cdot\!H_\mu
+\sqrt{6} m \o{D\!\cdot\!B} \varphi
+\sqrt{6} m \bar{\varphi} D\!\cdot\!B \,, \\
\cL_{2,0}^{(3)} = &
\,m^2 \bar{\Phi}_{\mu\nu\rho} \Phi^{\mu\nu\rho} 
- 3m^2 \o{\tilde{\Phi}}_\mu \tilde{\Phi}^\mu
+\frac{3}{2} m^2 \o{\tilde{H}} \tilde{H}
-\frac{3}{2} m^2 \bar{B}_\mu B^\mu
+5 m^2 \bar{\varphi} \varphi
+\frac{\sqrt{15}}{2} m^2 \o{\tilde{\Phi}}_\mu B^\mu \nn \\* &
+\frac{\sqrt{15}}{2} m^2 \bar{B}^\mu \tilde{\Phi}_\mu
-\sqrt{\frac{15}{2}} m^2 \o{\tilde{H}} \varphi
-\sqrt{\frac{15}{2}} m^2 \bar{\varphi} \tilde{H} \,.
\end{align} \end{subequations}
As before, we use the shorthands $D\cdot \Phi^{\mu_1\dots\mu_k} := D_\rho \Phi^{\rho \mu_1\dots\mu_k}$ and $\tilde{\Phi}^{\mu_1\dots\mu_k} := \eta_{\rho\sigma} \Phi^{\rho\sigma\mu_1\dots\mu_k}$.
The minimally coupled gauge transformations are
\beal
\label{eq:delta0spin3}
\delta \Phi_{\lambda\mu\nu} & = \partial_{(\lambda} \xi_{\mu\nu)} + \frac{\sqrt{3}}{4} m \eta_{(\lambda\mu} \xi_{\nu)} \,, \\ 
\delta H_{\mu\nu} & = \partial_{(\mu} \xi_{\nu)} + \frac{m}{\sqrt{3}} \xi_{\mu\nu}
+  \frac{\sqrt{5}}{2}  m \eta_{\mu\nu} \xi \,, \\ 
\delta B_\mu & = \partial_\mu \xi + \frac{\sqrt{5}}{2}m \xi_\mu \,, \\ 
\delta \varphi & = \sqrt{6} m \xi \,,
\eeal
in terms of the gauge parameters $\xi_{\mu\nu}, \xi_\mu$ and $\xi$, where $\xi^\mu_\mu = 0$.

In order to restore the gauge symmetry for $Q \neq 0$, we construct an ansatz for the non-minimal cubic Lagrangian $\cL_{3}^{(3)}$ and the non-minimal gauge transformations $\delta_1$.
Assuming \ref{NoSelfInteractions}, \ref{MinimalCoupling}, \ref{PowerCounting} and \ref{PowerCountingTransform}, the allowed ansatz contains 432 free parameters in the Lagrangian and 436 in the gauge transformations.
Imposing massive gauge invariance leaves 141 free parameters in the Lagrangian and 260 in the gauge transformations.
As before, the leftover free parameters are due to field redefinitions and higher-point effects, since they do not contribute to the on-shell three-point amplitude given by
\begin{align}
\label{eq:ampl3spin3}
{\cal A}(\bm{1}^{s=3}\!,\bar{\bm{2}}^{s=3}\!,3^+) = 
&-\frac{8 Q}{m^4} (\bep_1\!\cdot\!p_2)^2 (\bep_2\!\cdot\!p_1)^2 (
\bep_1\!\cdot\!{\ep_3}\, {\bep_2}\!\cdot\!{p_3}
+{\bep_2}\!\cdot\!{\ep_3}\, \bep_1\!\cdot\!p_2
+ {\bep_1}\!\cdot\!{\bep_2}\, {\ep_3}\!\cdot\!{p_1} ) \nn\\*
&-\frac{4 Q}{m^2} {\bep_1}\!\cdot\!{\bep_2}\, {\bep_1}\!\cdot\!{p_2}\, {\bep_2}\!\cdot\!{p_1} \left( 
4 {\bep_1}\!\cdot\!{\ep_3}\, {\bep_2}\!\cdot\!{p_3}
+4 {\bep_2}\!\cdot\!{\ep_3}\, {\bep_1}\!\cdot\!{p_2}
+3 {\bep_1}\!\cdot\!{\bep_2}\, {\ep_3}\!\cdot\!{p_1} 
\right) \nn\\*
&-2 Q({\bep_1}\!\cdot\!{\bep_2})^2\left(
3 {\bep_1}\!\cdot\!{\ep_3}\, {\bep_2}\!\cdot\!{p_3}
+3 {\bep_2}\!\cdot\!{\ep_3}\, {\bep_1}\!\cdot\!{p_2}
+ {\bep_1}\!\cdot\!{\bep_2}\, {\ep_3}\!\cdot\!{p_1}
\right) \nn\\
=& \,\, 2 Q\,\ep_3^+\!\cdot\!p_1 \frac{\braket{\bm{12}}^6\!}{m^6} \,.
\end{align}
This matches the known \AHH amplitude~\eqref{AHH}. 

The explicit solutions for the non-minimal Lagrangian and gauge transformations are rather cumbersome, so we omit them in this work. Instead, we can reproduce the above results in a simpler fashion using massive Ward identities.
From \eqn{eq:delta0spin3} we read off the gauge-transformed vertices
\begin{align}
V(\xi_1^{\mu\nu} \bar{\Phi}_2 A_3) & = -\frac{i p_1}{3} \!\cdot\!\frac{\partial}{\partial \epsilon_1} V(\Phi_1,\bar{\Phi}_2,A_3) + \frac{m}{\sqrt{3}} V(H_1,\bar{\Phi}_2,A_3) \,,
\\
V(\xi_1^\mu \bar{\Phi}_2 A_3) & = \frac{m}{8\sqrt{3}} \frac{\partial}{\partial \epsilon_1}\!\cdot \!\frac{\partial}{\partial \epsilon_1} V(\Phi_1 \bar{\Phi}_2 A_3) - \frac{i p_1}{2} \!\cdot \frac{\partial}{\partial \epsilon_1}\! V(H_1 \bar{\Phi}_2 A_3) + \frac{\sqrt{5} m}{2} V(B_1 \bar{\Phi}_2 A_3) \,, \nn
\\
V(\xi_1 \bar{\Phi}_2 A_3) & = \frac{\sqrt{5}m}{4} \frac{\partial}{\partial \epsilon_1}\!\cdot \!\frac{\partial}{\partial \epsilon_1} V(H_1 \bar{\Phi}_2 A_3) - i p_1 \!\cdot \frac{\partial}{\partial \epsilon_1}\! V(B_1 \bar{\Phi}_2 A_3) + \sqrt{6} m V(\varphi_1 \bar{\Phi}_2 A_3) \,, \nn
\end{align}
and impose the (reduced) massive Ward identities
\begin{equation}
V(\xi_1^{\mu\nu} \bar{\Phi}_2 A_3) \big|_{(2,3)} = V(\xi_1^\mu \bar{\Phi}_2 A_3) \big|_{(2,3)} = V(\xi_1 \bar{\Phi}_2 A_3) \big|_{(2,3)} = 0 \,,
\end{equation}
together with the assumptions \ref{NoSelfInteractions}, \ref{MinimalCoupling}, \ref{PowerCounting}, \ref{PowerCountingTransform}. This leaves two free parameters in the on-shell amplitude. Additionally imposing \CC\ and \ND\ fixes the leftover freedom and reproduces the amplitude~\eqref{eq:ampl3spin3}.

\subsubsection{Uniqueness of spin-$s$ $\sqrt{\rm Kerr}$}
\label{sect:3ptallspin}

In the arbitrary-spin case it is hard to obtain expressions for the non-minimal off-shell Lagrangian and gauge transformations.
The main issue is that such objects are not uniquely defined, because they depend on field redefinitions.
One option is to study fixed values of spin and choose field redefinitions that manifest some nice properties of the Lagrangians, in the hope that such special choices can be extrapolated to arbitrary spin.
In this work, we instead focus on reduced Ward identities, which produced the same on-shell results as the Lagrangian approach in the cases discussed above, but are much simpler to implement for arbitrary spin.
We write down ans{\"a}tze for the vertices $V(\Phi_1^k \Phi_2^s A_3^h)$, use them to compute the gauge variations $V(\xi_1^k \Phi_2^s A_3^h)$ as discussed in \sec{sec:freelag}, and impose combinations of the conditions \MC, \PC, \WI, \CC, \ND.

Imposing \MC+\PC+\WI\ in spin-$s$ $\sqrt{\rm Kerr}$  theory and extracting the on-shell three-point amplitude, we find
\begin{equation} \label{Unfixed_Root_Kerr}
{\cal A}(\bm{1}^s\!,\bar{\bm{2}}^s\!,3^+)  = 
{\cal A}_3^{(0)} \frac{ \langle \bm{12}\rangle^{2s}}{m^{2s}}
 \left\{ 1 +  \sum_{k=1}^{s-1} c_k  \left( \frac{[\bm{12}]^k}{\langle \bm{12}\rangle^k}
  - 1\right) \right\} ,
\end{equation}
where $c_k$ are free parameters and $\cA_3^{(0)} = 2Q (\ep_3\cdot p_1)$ is the scalar amplitude.
Imposing \ND\ does not modify \eqn{Unfixed_Root_Kerr}, whereas \CC\ requires $\sum_k c_k=0$. However, \CC+\ND\ fix $c_k=0$ and yield \eqn{eq:ampl3spin3}. 

The calculations outlined above for the \AHH theories were performed explicitly up to $s\le10$.
However, we do not expect any new feature beyond this spin and believe the above results to be valid for any spin.
Note that the off-shell analysis done in the previous sections lands us unambiguously on the \AHH amplitude, and it only required the conditions \ref{NoSelfInteractions}, \ref{Parity}, \ref{PowerCounting}, \ref{PowerCountingTransform} and \ref{MinimalCoupling}.
We believe that a complete set of Ward identities is equivalent to the off-shell constraints, and hence it will only require the conditions \MC+\PC+\WI\ to derive the \AHH amplitudes. By contrast, the reduced Ward identities (with one leg always having spin~$s$) leave us with free parameters, which are, nevertheless, fixed with the help of the constraints \CC\ and \ND.
Therefore, provided we follow the off-shell approach or impose all Ward identities, the latter constraints are redundant.

\subsubsection{On-shell cubic $\sqrt{\rm Kerr}$  Lagrangians}
\label{sec:onshellLag}
As discussed, it is difficult to obtain arbitrary-spin expressions for off-shell Lagrangians, due to their redundant nature. However, it is much easier to write down the part of the arbitrary-spin Lagrangian that contributes to the on-shell three-point amplitude. In this subsection, we derive expressions for arbitrary-spin on-shell Lagrangians that reproduce the \AHH amplitudes~\eqref{AHH}, in terms of a convenient choice of variables.

It is convenient to re-express the spin-$s$ field as a tensor product of $s$ spin-1 fields in the (generalized) double-copy sense:
\be \label{Eq:multi-copy}
\Phi^s = \Phi^{\mu_1 \mu_2 \ldots \mu_s}
= W^{(\mu_1} \otimes W^{\mu_2} \otimes \cdots \otimes W^{\mu_s)} \,,
\ee
where the tensor product is understood to hold in momentum space, or equivalently as a convolution in position space. We expand the $\sqrt{\rm Kerr}$  gauge theory Lagrangian in terms of operators contributing at different orders in the coupling $g$
\be
{\cal L}_{\sqrt{\rm Kerr}}^{(s)}
= -\frac{1}{4}(F_{\mu\nu})^2+{\cal L}_{2}^{(s)}
+ ig {\cal L}_{3}^{(s)} + g^2 {\cal L}_{4}^{(s)} + \ldots \, .
\ee
Up to cubic order we may assume, without loss of generality, that the non-trivial interactions are abelian and hence $g=Q$. Moreover, we will only analyze the Lagrangian terms that contribute on shell at three points. 

The minimal Lagrangian, obtained from the covariantization of \eqn{FreeL0}, is
\be
{\cal L}_{2}^{(s)}
= (-1)^s\Big(\overline{D_{\mu_0}  \Phi_{\mu_1 \ldots \mu_s}}
  D^{\mu_0} \Phi^{\mu_1 \ldots \mu_s}
- m^2  \bar \Phi_{\mu_1 \ldots \mu_s} \Phi^{\mu_1 \ldots \mu_s} + \ldots  \Big)\,,
\ee
where the ellipsis correspond to terms that vanish on shell. Recall that the covariant derivative is 
$D_{\mu_0} \Phi_{\mu_1 \ldots \mu_s} =  \partial_{\mu_0} \Phi_{\mu_1 \ldots \mu_s}- i Q A_{\mu_0} \Phi_{\mu_1 \ldots \mu_s}$ thus the minimal-coupling cubic interaction terms coming from ${\cal L}_{2}^{(s)}$ are
\be
\label{eq:A3minimal}
 (-1)^s i Q \partial_{\mu_0}  \Phi_{\mu_1 \ldots \mu_s}  A^{\mu_0}  \bar \Phi^{\mu_1 \ldots \mu_s} \!- (\Phi \leftrightarrow \bar \Phi) ~\to~  2 Q  (\varepsilon_3 \cdot p_1)   (-{\bm \varepsilon}_1 \cdot {\bm \varepsilon}_2)^{s}\,.
\ee 
In terms of spinor-helicity variables, the minimal-coupling gauge-theory amplitude for a positive-helicity photon is then
\be
{\cal A}_\text{min}(\bm{1}^s\!,\bar{\bm{2}}^s\!,3^+)
= 2 Q (\varepsilon_3^+\!\cdot p_1) \frac{\braket{\bm{12}}^s [\bm{12}]^s}{m^{2s}} \,.
\ee
and the negative-helicity case is simply obtained by $\varepsilon_3^+\to \varepsilon_3^-$. There is no additional information in the negative-helicity amplitude since we will only use parity-invariant operators, hence we can focus on the positive-helicity case. 

It is now an easy task to compute the difference between the above amplitude and the \AHH amplitude~\eqref{AHH} and reverse-engineer the cubic operators that correspond to the mismatch. We decorate the on-shell part of the cubic Lagrangians with a tilde to emphasize that they need to be supplemented by off-shell-completion terms to restore Zinoviev's massive gauge invariance,
\be
{\cal L}_3^{(s)} = \tilde{\cal L}_3^{(s)} + \text{(gauge-invariance completion terms)}\,.
\ee
At spin $s=1$, one can identify two independent candidate operators that contribute on shell,
\beal
{\cal O}_1 & = -\overline W_\mu F^{\mu\nu}  W_\nu ~~~~~~~~~\to~~~~~\,
  \frac{2Q}{m^2} (\varepsilon_3^+\!\cdot p_1) \braket{\bm{12}} (\braket{\bm{12}}-[\bm{12}])\,, \\
{\cal O}_2 & = -\frac{1}{m^2} \overline W_{\rho\mu} F^{\mu\nu} W_\nu{}^\rho ~~\to~~
   -\frac{2Q}{m^2} (\varepsilon_3^+\!\cdot p_1) [\bm{12}] (\braket{\bm{12}}-[\bm{12}]) \,,
\eeal
where $F_{\mu\nu}= 2\partial_{[\mu}A_{\nu]} $ and $W_{\mu\nu}=2\partial_{[\mu}W_{\nu]}$ are (linearized) field strengths for the massive and massless fields (here $\Phi^\mu=W^\mu$). Using the spinor-helicity representations on the right-hand side, it is clear that $\tilde{\cal L}_3^{(1)}= {\cal O}_1$ reproduces the $\sqrt{\rm Kerr}$ amplitude.

For $s>1$ we can construct the $\tilde{\cal L}_3^{(s)}$ operators recursively, by tensoring the ${\cal O}_1$ and ${\cal O}_2$ operators with spin-1 building blocks that are insensitive to the photon helicity. For three-point on-shell kinematics there are only two such independent building blocks,
\beal
\U & = -\frac{1}{m^2} \overline W_{\mu\nu} W^{\mu\nu}  \to
\frac{1}{m^2} \big(\braket{\bm{12}}^2 + [\bm{12}]^2\big)\,, \\
\V & =  -\overline W_\mu W^\mu  \to
   \frac{\braket{\bm{12}} [\bm{12}]}{m^2} \,.
\eeal

Using \eqn{Eq:multi-copy} to decompose $\Phi_{\mu_1 \ldots \mu_s}$ into spin-1 fields, we can now write down $\tilde{\cal L}_{3}^{(s)}$ to any spin.
Through spin 3, the non-minimal interaction terms are
\beal
\tilde{\cal L}_3^{(0)} & = 0 \,, \qquad~~\:\!\quad
\tilde{\cal L}_3^{(2)} = (\U+\V) {\cal O}_1 + \V {\cal O}_2 \,, \\
\tilde{\cal L}_3^{(1)} & = {\cal O}_1 \,, \qquad \quad
\tilde{\cal L}_3^{(3)} = (\U+\V) \big( \U {\cal O}_1 + \V {\cal O}_2 \big) \,.
\eeal
When combined with the the minimally-coupled interactions~\eqref{eq:A3minimal}, these terms reproduce the \AHH amplitudes up to spin 3.

For spin 4 and higher, we find that the non-minimal \AHH interaction terms satisfy the following recurrence formula,
\begin{align} \label{L3recurrence}
\tilde{\cal L}^{(s)}_3 = (\U^2  - 2 \V^2) \tilde{\cal L}^{(s-2)}_3  - 
 \V^4 \tilde{\cal L}^{(s-4)}_3 + (2\V + \U) \V^{s - 2} \big({\cal O}_1+ {\cal O}_2\big) \,.
\end{align}
For example, for spin $s=4,5,6$ we have the non-minimal interactions
\begin{subequations} \begin{align}
\tilde{\cal L}_3^{(4)} & = u (u^2 + u v - v^2)  {\cal O}_1  + u v (u + v)  {\cal O}_2 \,, \\
\tilde{\cal L}_3^{(5)} & =(u^2 - v^2)(u^2 + u v - v^2) {\cal O}_1 + u v (u^2 + u v - v^2) {\cal O}_2 \,, \\
\tilde{\cal L}_3^{(6)} & = 
(u^2 - v^2)(u^3 + u^2 v - 2 u v^2 - v^3) {\cal O}_1 +v (u^2 - v^2) (u^2 + u v - v^2) {\cal O}_2 \,. 
\end{align} \end{subequations}
As an alternative to the recursive formula, we can also explore a generating-function description of the cubic on-shell Lagrangian $\tilde{\cal L}_3$.
Indeed, it is straightforward to obtain the generating function by summing \eqn{L3recurrence} over all spins, giving an equation for the sum with solution
\be
\sum_{s=0}^\infty \tilde{\cal L}^{(s)}_3 =  \frac{{\cal O}_1 + \V {\cal O}_2}{(1-v)(1-\U+\V^2)} \,
\ee
For simplicity, we have suppressed the auxiliary generating variable, since this leads to no ambiguity.

In terms of standard Lagrangian notation, using the fields $\Phi^{\mu_1 \cdots \mu_s}$ and covariantizing the derivatives, the non-minimal cubic interactions up to spin-3 take the form
\begin{subequations} 
\label{eq:L3onshellexample}
\begin{align}
\tilde{\cal L}^{(1)}_3=& -\bar \Phi_\mu F^{\mu\nu}  \Phi_\nu\,,  \\
\tilde{\cal L}^{(2)}_3=&\;\bar \Phi_{\mu \nu} F^\nu{}_{\rho}  \Phi^{\rho \mu }+\frac{4}{m^2}\overline {D_{[\mu} \Phi_{\nu] \rho}} F^{\rho \sigma}  D^{[\mu} \Phi^{\nu]}_{\ \ \sigma}+\frac{4}{m^2}\overline {D_{[\mu} \Phi_{\nu] \rho}} F^\nu{}_\sigma  D^{[\sigma} \Phi^{\mu] \rho}\,, \\
\tilde{\cal L}^{(3)}_3= & 
-\frac{4}{m^2}\overline {D_{[\mu} \Phi_{\nu] \rho \lambda}} F^{\rho}{}_\sigma  D^{[\mu} \Phi^{\nu]\sigma \lambda}
-\frac{4}{m^2}\overline {D_{[\mu} \Phi_{\nu] \rho \lambda}} F^\nu{}_\sigma  D^{[\sigma} \Phi^{\mu] \rho \lambda} \\*
& \null - \frac{2^4}{m^4}\overline {D^{[\nu} D_{[\mu} \Phi_{\rho]}{}^{\sigma] \lambda}} F_{\lambda}{}^\kappa  {D_{[\nu} D^{[\mu} \Phi^{\rho]}{}_{\sigma] \kappa}} 
- \frac{2^4}{m^4}\overline {D^{[\sigma} D_{[\mu} \Phi_{\rho]}{}^{\nu] \lambda}} F_\nu{}^\kappa  {D_{[\kappa} D^{[\mu} \Phi^{\rho]}{}_{\sigma] \lambda}} \,. \nn
\end{align}
\end{subequations}
Note that the above Lagrangians have derivatives appearing only as curls of the spin-$s$ field, which is the general structure of interactions of massless fields, see \eg \rcite{Grigoriev:2020lzu}.

As mentioned above, \eqn{eq:L3onshellexample} contains the operators that contribute to the on-shell three-point amplitude. In order to have the full cubic Lagrangian ${\cal L}^{(s)}_3$, we would need to add additional terms that vanish on shell, yet they are needed to preserve massive gauge invariance. For the highest derivative terms in ${\cal L}^{(s)}_3$ we can find the completions, since they correspond to having gauge invariance in the high-energy limit. The following leading-derivative terms are compatible with the above on-shell analysis and high-energy gauge invariance:  
\begin{align}
{\cal L}^{(s)}_3\!= (-1)^s\frac{2^{2s-2}}{m^{2s-2}}& \bigg(\,\overline{\Phi^{[\mu_1;\nu_1][\mu_2;\nu_2] \cdots [\mu_{s-1};\nu_{s-1}]\rho}} F_\rho{}^\sigma \Phi_{[\mu_1;\nu_1][\mu_2;\nu_2] \cdots [\mu_{s-1};\nu_{s-1}]\sigma} \nn \\
&+\overline{\Phi^{[\mu_1;\sigma][\mu_2;\nu_2] \cdots [\mu_{s-1};\nu_{s-1}]\rho}} F_{\mu_1}{}^{\nu_1} \Phi_{[\sigma;\nu_1][\mu_2;\nu_2]  \cdots [\mu_{s-1};\nu_{s-1}]\rho}\\
&+\frac{1}{2}\overline{D_\rho\Phi^{[\mu_1;\nu_1][\mu_2;\nu_2]  \cdots [\mu_{s-1};\nu_{s-1}]\rho}} F_{\mu_1}{}^{\sigma} \Phi_{[\mu_2;\nu_2]  \cdots [\mu_{s-1};\nu_{s-1}]\sigma \nu_1}\nn\\
&+\frac{1}{2}\overline{\Phi^{[\mu_2;\nu_2] \cdots [\mu_{s-1};\nu_{s-1}]\rho\nu_1}\!} F_{\rho}{}^{\mu_1} D^\sigma\Phi_{[\mu_1;\nu_1] [\mu_2;\nu_2] \cdots [\mu_{s-1};\nu_{s-1}]\sigma} \bigg)\!
+ {\cal O}\Big(\frac{1}{m^{2s-4}}\!\Big) , \nn
\end{align}
where the $[\mu;\nu]$ notation is shorthand for curls, \eg with one curl we have $\Phi^{\mu_1\mu_2 \ldots [\mu_i;\nu_i]  \ldots \mu_s}:= D^{[\nu_i}\Phi^{\mu_1\mu_2 \ldots \mu_i]\ldots \mu_s}$, and similarly for taking curls repeatedly. The St\"{u}ckelberg fields all enter ${\cal L}^{(s)}_3$ as lower derivative terms, thus they are hidden in the ${\cal O}({m^{-2s+4}})$ corrections. 

\subsection{Massive gauge symmetry at quartic order}
\label{sec:massivegaugequartic}

Finding consistent quartic interactions for massive higher-spin fields is a notoriously hard problem. In the St{\"u}ckelberg formalism discussed in this work, it requires solving gauge invariance to quadratic order in the coupling $Q$,
\be
\label{eq:gaugevar4}
(\delta_0+\delta_1+\delta_2)(\cL_2+\cL_3+\cL_4) = \cO(Q^3) \,.
\ee
At linear order, the above equation reduces to \eqn{eq:gaugevar3} and the implementation of gauge invariance has already been discussed. The new information comes at $\cO(Q^2)$, where the non-minimal quartic interactions in $\cL_4$ starts contributing. At this order, we need to compute the variation of the quartic vertices under the free-theory gauge transformations, namely $\delta_0 \cL_4$ and also $\delta_0(\cL_2+\cL_3)$  from expanding the covariant derivatives. We also need to vary the free Lagrangian $\cL_2|_{Q=0}$ under $\cO(Q^2)$ corrections to the gauge transformations, $\delta_2$, which are quadratic in the fields and linear in the gauge parameters. Both $\cL_4$ and $\delta_2$ require writing down an ansatz and hence introducing new free parameters.

The number of derivatives required in the ansatz, and hence the number of free parameters, increase rapidly with spin. To see this, we consider the contribution of the three-point vertices to the four-point Compton amplitude, given by
\begin{equation}
\label{eq:dercountingcompton}
   \bigg\{\Delta_P\Big(\frac{\partial}{\partial\epsilon_{P}}, \frac{\partial}{\partial\epsilon_{{-}P}}\Big) \sum_{l=0}^s V(\Phi^s_1 \bar{\Phi}^l_P A_4) V(\Phi^l_{{-}P} \bar{\Phi}^s_2 A_3) 
~ + ~ (3 \leftrightarrow 4) \bigg\} \bigg|_{\epsilon_P,\epsilon_{{-}P} \to0} \,,
\end{equation}
where the Feynman propagator-generating function $\Delta_P(\epsilon, \bar \epsilon)$ \eqref{full_prop} has become an operator through the substitution of its arguments by derivatives, which act on the polarization vectors of the internal line.
The lowest-derivative solution for the three-point vertices contains $2s{-}1$ derivatives, as discussed in the previous sections.
Since \eqn{eq:dercountingcompton} contains two such vertices and a propagator, and the latter subtracts two derivatives, we expect at least $4s{-}4$ derivatives in the Compton amplitude and in the four-point contact terms in $\cL_4$.
This will produce up to $4s{-}3$ derivatives in the term $\delta_0 \cL_4$.
We expect the same power of derivatives in the term $\delta_2 \cL_2$, and since $\cL_2$ has two derivatives, $\delta_2$ should have $4s{-}5$. 

Moreover, we need to consider the variation of the cubic vertices under the first non-linear corrections to the gauge transformations, for instance in the term $\delta_1 \cL_3$.
Both $\delta_1$ and $\cL_3$ are partially fixed by the higher-spin generalization of \eqn{eq:gaugevar3}, but in general there are leftover free parameters.
Provided we fix the cubic amplitude, all the leftover parameters are associated with field redefinitions and with redefinitions of gauge parameters, \ie they represent no physical information and can be set at will, for instance, to zero.
Nevertheless, \eqn{eq:gaugevar4} at $\cO(Q^2)$ is a system of coupled equations containing many tensor structures, and it is very hard to solve in general. 

Alternatively, we can resort to the same Ward identities that provided simplification in the cubic case. This requires computing a four-point Compton amplitude using the three-point solutions to the cubic Ward identities and the Feynman-gauge propagators defined in \eqn{full_prop}. 
More precisely, we need to compute the abelian amputated correlation function\footnote{We refer to \eqn{eq:fourpointoffshellexchange} as an \textit{amputated} correlation function, since it can be obtained from an ordinary correlation function by stripping the propagators off the external legs.
}
\begin{align}
\label{eq:fourpointoffshellexchange}
\langle \Phi^k_1 \bar \Phi^s_2 A_3 A_4 \rangle_{\rm U(1)}\!:=\,& \bigg\{
\Delta_P\Big(\frac{\partial}{\partial\epsilon_{P}}, \frac{\partial}{\partial\epsilon_{{-}P}}\Big) \sum_{l=0}^s V(\Phi^k_1 \bar{\Phi}^l_P A_4) V(\Phi^l_{{-}P} \bar{\Phi}^s_2 A_3) 
+  (3 \leftrightarrow 4) \bigg\}\Big|_{\epsilon_P,\epsilon_{{-}P} \to 0} \nn \\*
& \null + V_{{\cal L}_{2}}(\Phi^k_1 \bar \Phi^s_2 A_3 A_4)+ V_{{\cal L}_{3}}(\Phi^k_1 \bar \Phi^s_2 A_3 A_4) + V_{{\cal L}_{4}}(\Phi^k_1 \bar \Phi^s_2 A_3 A_4)  \,.
\end{align}
The terms $V_{{\cal L}_{2}}$ and $V_{{\cal L}_{3}}$ pick up the quartic terms from ${\cal L}_2\sim (D\Phi)^2$ and ${\cal L}_3$ respectively, which are due to $A_\mu$ inside the covariant derivatives $D_\mu$. In addition, we need the four-point contact terms of form
\be
\label{eq:fourpointoffshellcontact}
V_{{\cal L}_4}(\Phi^k_1 \bar{\Phi}^s_2 A_3 A_4) = {\rm Ansatz}^{(k,s)}(p_i,\epsilon_i) \,,
\ee
where the ansatz for the contact terms must involve two field strengths $\cL_4 \propto F_{\mu\nu} F_{\rho\sigma}$.

Let us also define a gauge-transformed amputated correlator for each gauge parameter~$\xi^k$, in the same fashion as in \eqn{eq:gaugetransV}, 
\be
\label{eq:gaugetrans4Points}
\langle \xi_1^k \bar{\Phi}_2^s  A_3A_4 \rangle\!:= 
m \alpha_k  \langle \Phi_1^k \bar{\Phi}_2^s A_3A_4\rangle 
- \frac{i p_1}{k\!+\!1} \cdot \frac{\partial~}{\partial \epsilon_1}
  \langle \Phi_1^{k+1} \bar{\Phi}_2^s A_3A_4\rangle
+ \frac{m}{2} \beta_{k+2}
  \bigg(\!\frac{\partial~}{\partial \epsilon_1}\!\bigg)^{\!2}\!
  \langle \Phi_1^{k+2} \bar{\Phi}_2^s A_3A_4\rangle .
\ee
The four-point Ward identities then become
\be
\label{eq:4ptward}
\langle \xi^k_1 \bar{\Phi}^s_2 A_3 A_4 \rangle \big|_{(2,3,4), \epsilon_1^2=0} ~ = ~ 0\,,
\ee
where legs 2, 3 and 4 are on-shell states, meaning that they obey $p_i^2-m_i^2 = \ep_i^2 = \ep_i\cdot p_i = 0$, and the leg 1 is traceless $\epsilon_1^2=0$. Likewise, the spin-$s$ amplitude can be obtained from the amputated abelian correlator, where all legs are taken on shell:
\be
A_{\rm U(1)}(\bm{1}^s\!,\bar{\bm{2}}^s\!,3,4)
:= -\frac{1}{2}\langle \Phi^s \bar \Phi^s A_3 A_4 \rangle_{\rm U(1)} \Big|_{(1,2,3,4),\;Q \to 1} \,,
\ee
where have removed the coupling constant and tweaked the overall normalization to be compatible with later formulae of color-stripped Compton amplitudes.
In general, $A(...)$ denote amplitudes without couplings and color factors, as opposed to full amplitudes ${\cal A}(...)$.

When using Ward identities, only the free-theory gauge transformations $\delta_0\big|_{Q=0}$ contribute, so we do not have to worry about non-linear corrections $\delta_1$ and $\delta_2$. As discussed, the number of terms in the ansatz for $\delta_2$ increase rapidly with increasing spins, so this provides a major simplification. In general, \eqn{eq:4ptward} is also a quadratic equations, since the three-point free parameters in $V(\Phi_i^{s'} \bar\Phi_j^{s} A_k)$ get squared in \eqn{eq:fourpointoffshellexchange}. In practice, however, once we fix the on-shell three-point amplitude as described in previous sections, the free parameters left in $V(\Phi_i^{s'} \bar\Phi_j^{s} A_k)$ are four-point contact terms, since they vanish in the factorization limit.
Therefore, we can choose to reabsorb them in the ansatz for $V(\Phi^k_1 \bar{\Phi}^s_2 A_3 A_4)$ and thus linearize the system of equations~\eqref{eq:4ptward}.

We can also study the non-abelian case. We work at the level of color-ordered amplitudes, hence the cubic vertices are the same as the ones discussed in previous sections, once the abelian charge $Q$ is replaced by the non-abelian gauge coupling $Q\rightarrow g/\sqrt{2}$.
Then we compute the non-abelian color-ordered amputated correlation function
\beal
\label{eq:fourpointoffshellexchangeNA}
\langle \Phi^k_1 \bar{\Phi}^s_2 A_3 A_4 \rangle_\text{ord} :=\,&\bigg\{
   \Delta_P\bigg(\frac{\partial}{\partial\epsilon_{P}}, \frac{\partial}{\partial\epsilon_{{-}P}}\bigg) \sum_{l=0}^s V(\Phi^k_1 \bar{\Phi}^l_P A_4) V(\Phi^l_{{-}P} \bar{\Phi}^s_2 A_3) \bigg\} \bigg|_{\epsilon_P,\epsilon_{{-}P} \to0} \\
&+ \frac{1}{s_{12}} \frac{\partial}{\partial\epsilon_q} {\cdot} \frac{\partial}{\partial\epsilon_{-q}}  V(\Phi^k_1 \bar{\Phi}^s_2 A_q) V_g(A_{-q} A_3 A_4) \\
& \null + \widetilde V_{{\cal L}_{2}}(\Phi^k_1 \bar \Phi^s_2 A_3 A_4 ) + \widetilde V_{{\cal L}_{3}}(\Phi^k_1 \bar \Phi^s_2 A_3 A_4 ) + \widetilde V_{{\cal L}_{4}}(\Phi^k_1 \bar \Phi^s_2 A_3 A_4 ) \,.
\eeal
Note that in this case we only need one massive-exchange diagram, given by the first line, since the other one in \eqn{eq:fourpointoffshellexchange} contributes to a different color ordering. Instead, we need to add the gluon-exchange diagram, given by the second line, where $V_g(A_i A_j A_k)$ is the color-ordered three-gluon vertex and the derivative operator $(\partial/\partial{\epsilon_q})\cdot( \partial/\partial{\epsilon_{-q}})$ contracts the Lorentz indices corresponding to $A_q$ and $A_{-q}$ polarizations.

As before, $\widetilde{V}_{{\cal L}_{2}}$ and $\widetilde{V}_{{\cal L}_{2}}$ are the quartic terms coming from $\cL_2$ and $\cL_3$.
Then we need the four-point contact terms involving two field strengths,
\be
\label{}
\widetilde{V}_{{\cal L}_4}(\Phi^k_1 \bar{\Phi}^s_2 A_3 A_4) = {\rm Ansatz}^{(k,s)}(p_i,\epsilon_i) \,.
\ee
Note that in the abelian case~\eqref{eq:fourpointoffshellcontact} the vertex $V_{{\cal L}_4}(\Phi^k_1 \bar{\Phi}^s_2 A_3 A_4)$ must be symmetric under $1\leftrightarrow 2$ exchange and $3\leftrightarrow 4$ exchange separately. In the non-abelian case, instead, the vertex is only symmetric under the combination of both exchanges, therefore the ansatz will contain additional free parameters.

In the rest of this section, we will apply the Ward identities outlined above to the cases of spin-2 and spin-3 particles and will derive explicit abelian and non-abelian Compton amplitudes.
The latter are color-decomposed as follows:
\begin{align}
&{\cal A}(\bm{1}^s\!,\bar{\bm{2}}^s\!,3,4)
:= 2g^2 \Big[ T^{c_3} T^{c_4} A(\bm{1}^s\!,\bar{\bm{2}}^s\!,3,4)
            + T^{c_4} T^{c_3} A(\bm{1}^s\!,\bar{\bm{2}}^s\!,4,3) \Big]
 = \langle \Phi^s \bar \Phi^s A_3 A_4 \rangle\big|_{(1,2,3,4)} \,, \nn \\
&\langle \Phi^s \bar \Phi^s A_3 A_4 \rangle:=  2T^{c_3} T^{c_4} \langle \Phi^s \bar \Phi^s A_3 A_4 \rangle_{\rm ord}+2 T^{c_4} T^{c_3} \langle \Phi^s \bar \Phi^s A_4 A_3 \rangle_{\rm ord} \,.
\label{ColorDressing}
\end{align}
Namely, when extracting the ordered amplitudes from the four-point color-dressed amplitude, we choose to strip $2g^2$ along with the color factor. We can also relate the amputated correlators of the abelian and ordered type, using $g^2=2Q^2$,
\be
 \langle \Phi^s \bar \Phi^s A_3 A_4 \rangle_{\rm U(1)}= \langle \Phi^s \bar \Phi^s A_3 A_4 \rangle_{\rm ord} +\langle \Phi^s \bar \Phi^s A_4 A_3 \rangle_{\rm ord}\,,
\ee
which is an echo of the usual Kleiss-Kuijf relation for color-ordered partial amplitudes.

\subsubsection{Quartic spin-2 interactions}
\label{sec:massivegaugequarticSpin2}
We can consider the opposite-helicity amplitude $A(\bm{1}^{s=2}\!,\bar{\bm{2}}^{s=2}\!,3^-\!,4^+)$ between two massive spin-2 fields and two massless spin-1 bosons.
We first construct an ansatz for the off-shell quartic vertices with the following schematic derivative counts:
\beal
\label{eq:spin2comptonder}
V(\Phi^2_1 \bar{\Phi}^2_2 A_3 A_4) & \sim \partial^4 \,, \\
V(\Phi^1_1 \bar{\Phi}^2_2 A_3 A_4) & \sim \partial^5 \,, \\
V(\Phi^0_1 \bar{\Phi}^2_2 A_3 A_4) & = 0 \,,
\eeal
where the last vertex is set to zero from the start, as it is consistent with our guiding principles. 

In the abelian case, imposing Ward identities yields the amplitude~\cite{Cangemi:2022bew}
\begin{align}
A_{\rm U(1)}(\bm{1}^{s=2}\!,\bar{\bm{2}}^{s=2}\!,3^-\!,4^+)
 = \frac{\bra{3}1|4]^2 (U+V)^4}{m^8 t_{13} t_{14}} 
 - \frac{\braket{\bm{1}3} \bra{3}1|4] [4\bm{2}]}{m^8 t_{13}}
   4 U (U^2+V^2) & \\
+\,\frac{\braket{\bm{1}3} \braket{3\bm{2}} [\bm{1}4] [4\bm{2}]}{m^8}
   (3 U^2+V^2) & + C^{(2)}_{-+} \,, \nn
\end{align}
where we use the variables $U = \frac{1}{2}(\bra{\bm1}4|\bm2]-\bra{\bm2}4|\bm1])-m[\bm{12}]$, $V = \frac{1}{2}(\bra{\bm1}4|\bm2]+\bra{\bm2}4|\bm1])$ introduced in \rcite{Cangemi:2022bew}, and the contact term is
\begin{align}
\label{eq:s2abContacts}
C^{(2)}_{-+}
 = \frac{c_1}{m^6} & (\braket{\bm{12}}+[\bm{12}])^2
   \braket{\bm{1}3}\braket{3\bm{2}} [\bm{1}4] [4\bm{2}]
 + \frac{c_2}{m^6} (\braket{\bm{12}} - [\bm{12}])^2
   \braket{\bm{1}3}\braket{3\bm{2}}
   [\bm{1}4] [4\bm{2}] \nn \\
 + \frac{c_3}{m^8} &
   \Big( \langle 3|1|4] \braket{\bm{12}}
   \big(\langle 3|1|4] [\bm{12}] \braket{\bm{12}}^2 +(\braket{\bm{12}} + [\bm{12}]) (\langle \bm{1}|3|\bm{2}] \braket{\bm{2}3} [\bm{1}4] +  \langle \bm{2}|3|\bm{1}] \braket{\bm{1}3} [4\bm{2}])\big) \nn \\ &~\;
 + s_{12} \braket{\bm{12}}[\bm{12}]
   \braket{\bm{1}3}\braket{3\bm{2}} [\bm{1}4] [4\bm{2}] \Big) \,.
\end{align}
The three coefficients $c_n$ are free parameters not fixed by the Ward identities, nor by off-shell gauge invariance.  In the non-abelian case, were we use the same power counting for the derivative in the vertices $\tilde V(\Phi^k \bar \Phi^2 A_3 A_4)$ as above, and after imposing the Ward identities we obtain the color-ordered amplitude
\beal
 & A(\bm{1}^{s=2}\!,\bar{\bm{2}}^{s=2}\!,3^-\!,4^+)
 = \frac{\bra{3}1|4]^2 (U+V)^4}{m^8 s_{12} t_{14}}
 - \frac{\braket{\bm{1}3} \bra{3}1|4] [4\bm{2}]}{m^8 s_{12}}
   4U (U^2+V^2) \\ & \qquad
 + \frac{\braket{\bm{1}3} \braket{3\bm{2}} [\bm{1}4] [4\bm{2}]}{m^{4s} s_{12}} 
   \big[ t_{13} (3U^2+V^2+W_-^2-W_+^2) - 4m^2V(U+W_+)\big]
 + \tilde{C}^{(2)}_{-+} \,,
\eeal
where we use the shorthand $W_{\pm} = \frac{m}{2}(\braket{\bm{12}}\pm[\bm{12}])$. The contact term now contains one additional free parameter,
\be \label{eq:nonabquarticContact}
\tilde{C}^{(2)}_{-+}  = C^{(2)}_{-+}
 + \frac{c_4}{m^6} \braket{\bm{12}}[\bm{12}]
   \big( \braket{\bm{2}3}^2 [\bm{1}4]^2
       - \braket{\bm{1}3}^2 [4\bm{2}]^2 \big) \,,
\ee
which we cannot fix in the absence of further constraints. 

We can also consider the consequences for other helicity sectors, since our procedure gives amplitudes expressed in terms of covariant polarization vectors. Specializing to the positive same-helicity amplitude gives
\be
A^{\rm abelian} (\bm{1}^{s=2}\!,\bar{\bm{2}}^{s=2}\!,3^+\!,4^+)
= \frac{\braket{\bm{12}}^4 [34]^2}{m^2 t_{13} t_{14}}
+ C^{(2)}_{++} \,,
\ee
where the contact term contains an additional five free parameters,
\beal
C^{(2)}_{++}   = 
&\,\frac{[34]^2}{m^6}(c_5 \braket{\bm{12}}^3[\bm{12}] + c_6 \braket{\bm{12}}^2[\bm{12}]^2 + c_7 \braket{\bm{12}}[\bm{12}]^3) \\*
&+ \frac{\braket{\bm{12}}}{m^6}(c_8 [\bm{12}]+ c_9 \braket{\bm{12}})([\bm13]^2[\bm24]^2+[\bm14]^2[\bm23]^2) \,.
\eeal
Finally, the non-abelian color-ordered amplitude, in the positive same-helicity sector, is given by
\begin{equation}
A (\bm{1}^{s=2}\!,\bar{\bm{2}}^{s=2}\!,3^+\!,4^+)
= \frac{\braket{\bm{12}}^4 [34]^2}{m^2 s_{12} t_{14}}
+ \tilde{C}^{(2)}_{++} \,,
\end{equation}
and there are no new free parameters compared to the abelian case, \ie $\tilde{C}^{(2)}_{++}  = C^{(2)}_{++}$.

In later sections of this paper additional structures and patterns will be used, such that we can fix these free parameters. In particular, we will choose to use $c_{n>2}=0$ and $c_1=c_2=-1$.

\subsubsection{Quartic spin-3 interactions}

We can also consider the amplitude $A (\bm{1}^{s=3}\!,\bar{\bm{2}}^{s=3}\!,3,4)$ between two massive spin-3 fields and two photons or gluons. The Lagrangian can be written in terms of a rank-3 field $\Phi^3_{\mu\nu\rho}$ and three lower-rank St{\"u}ckelberg fields $\{ \Phi^2_{\mu\nu}, \Phi^1_\mu, \Phi^0 \}$.
In the abelian case we consider an ansatz for the off-shell quartic vertices with the following derivative count,
\beal
\label{eq:spin3comptonder}
V(\Phi^3_1 \bar{\Phi}^3_2 A_3 A_4) & \sim \partial^8 \,,\\
V(\Phi^2_1 \bar{\Phi}^3_2 A_3 A_4) & \sim \partial^9 \,,\\
V(\Phi^1_1 \bar{\Phi}^3_2 A_3 A_4) & = 0 \,,\\
V(\Phi^0_1 \bar{\Phi}^3_2 A_3 A_4) & = 0 \,.
\eeal
We will not show the details of this calculation, as it is much more involved compared the lower-spin cases that we already elaborated on.  

Solving the massive Ward identities yields 53 free parameters in the on-shell amplitude $A (\bm{1}^{s=3}\!,\bar{\bm{2}}^{s=3}\!,3,4)$, with 21 free parameters that only appear in the opposite-helicity amplitude $A (\bm{1}^{s=3}\!,\bar{\bm{2}}^{s=3}\!,3^-\!,4^+)$ and 32 free parameters  that only appear in the same-helicity amplitude $A (\bm{1}^{s=3}\!,\bar{\bm{2}}^{s=3}\!,3^+\!,4^+)$. 
We omit the explicit expressions since they are rather cumbersome, and we will study Compton amplitudes for spins $s\geq 3$ in more detail in the next section.

\section{Chiral higher-spin gauge theory}
\label{sec:chiralform}

In the previous section, we have seen how to simplify the study of massive-gauge-invariant Lagrangians through the use of on-shell Ward identities. We have applied these identities at three points to re-derive the \AHH amplitudes from first principles, and we have extended them to four points to compute new results for the Compton amplitudes. Ward identities require constructing ans{\"a}tze for all vertices $V(\Phi^k_1\bar\Phi_2^s\dots)$ and fields $\Phi^k_1$ with $0\leq k \leq s$. This quickly becomes computationally intense, and we were able to find explicit results at four points only up to spin $s = 3$.

In this section, we make use of a complementary new approach to massive higher-spin theory~\cite{Ochirov:2022nqz},
which in four dimensions allows us to forego auxiliary fields altogether.
The approach amounts to trading the SO(1,3) symmetric tensors $\Phi_{\mu_1\dots\mu_s}$ for ${\rm SL}(2,\mathbb{C})$ chiral symmetric tensors $\Phi_{\alpha_1\dots\alpha_{2s}}$.
Recall that SO(1,3) symmetric traceless tensors correspond to non-chiral tensors  
$\Phi_{\alpha_1\dots\alpha_s\dot{\beta}_1\dots\dot{\beta}_s} :=
\Phi_{\mu_1\dots\mu_s} \sigma_{\alpha_1\dot{\beta}_1}^{\mu_1}\!\cdots
\sigma_{\alpha_s\dot{\beta}_s}^{\mu_s} $ in ${\rm SL}(2,\mathbb{C})$ notation,
hence switching to $\Phi_{\alpha_1\alpha_2\dots\alpha_{2s}}$ as the fundamental field is a radical reformation, that nevertheless encodes the same on-shell degrees of freedom.
In principle, any rank-$2s$ ${\rm SL}(2,\mathbb{C})$ tensor field is capable of describing the degrees of freedom of a spin-$s$ particle.
However, in general, the off-shell field contains a lot of redundant components upon reduction to the Wigner little group ${\rm SU}(2)$, and eliminating them requires additional constraints.
Preserving such constraints when interactions are turned on has been the main difficulty that we were dealing with in the previous sections.
The advantage of the chiral formulation is that $\Phi_{\alpha_1\dots\alpha_{2s}}$ contains exactly $2s+1$ components, which is the number of physical degrees of freedom of the massive spin-$s$ field.
Therefore, no further constraints are needed. The price to pay is that the parity properties of the interactions become highly non-trivial, and it is best to treat the theory perturbatively as an expansion around the chiral sector. Even the familiar spin-1 $W$-boson Lagrangian~\eqref{ActionFromHiggs} develops spurious poles in the mass and has an infinite geometric series expansion in the fields, as we will see below.

\subsection{Spin-1 in chiral formulation}
\label{sec:spin1chiral}

As an illuminating example of the chiral framework,
let us again consider the spin-1 case, for which the transition from the non-chiral formulation is well established~\cite{Chalmers:2001cy}.
For simplicity, we use a real uncharged Proca field, to be generalized shortly.
Consider the standard Proca Lagrangian and add a topological term (proportional to the derivative
$\partial_\mu [\epsilon^{\mu\nu\rho\sigma} W_\nu \partial_\rho W_\sigma]$),
\be
 -\frac{1}{4} (W_{\mu\nu})^2+ \frac{m^2\!}{2} (W_\mu)^2
  - \frac{i}{8} \epsilon_{\mu\nu\rho\sigma} W^{\mu\nu} W^{\rho\sigma}
  = -\frac{1}{4}  W^{\alpha\beta} W_{\alpha \beta}
  + \frac{m^2\!}{2} (W_\mu)^2 \,.
\label{Spin1ActionProca}
\ee
The mass-independent kinetic term now only depends on the anti-self-dual field strength in the ${\rm SL}(2,\mathbb{C})$ notation
$W_\alpha{}^\beta :=(\sigma^{\mu\nu})_\alpha{}^\beta \partial_\mu W_\nu$,
where $\sigma^{\mu\nu} = \sigma^{[\mu} \bar{\sigma}^{\nu]}$.
We then ``integrate in'' a new chiral symmetric spinor field
$\Phi_{\alpha\beta}$ by effectively adding the quadratic form
\be
\frac{1}{4} \big(\sqrt{2}m \Phi^{\alpha\beta} - W^{\alpha\beta}\big)
            \big(\sqrt{2}m \Phi_{\alpha\beta} - W_{\alpha\beta}\big)
\ee
to the action~\eqref{Spin1ActionProca}.
This corresponds to multiplying the path integral by an irrelevant Gaussian integral.
The new Lagrangian can be integrated by parts and rewritten as
\be
\frac{m^2\!}{2} \Phi^{\alpha\beta} \Phi_{\alpha\beta}
+ \frac{m^2\!}{2} (W_\mu)^2
- \frac{m}{\sqrt{2}} W_\mu \tr(\sigma^{\mu\nu} \partial_\nu \Phi) \,,
\ee
where the trace is over the ${\rm SL}(2,\mathbb{C})$ indices.
We can now integrate out the original vector field $W_\mu$, which gives the new kinetic term
$-\frac{1}{4} \tr(\sigma^{\mu\nu} \partial_\nu \Phi)
  \tr(\sigma_{\mu\rho} \partial^\rho \Phi)
=-\frac{1}{2} \partial^\mu \Phi^{\alpha\beta}\partial_\mu \Phi_{\alpha\beta}$.
We thus obtain a chiral spin-1 Lagrangian\footnote{The overall sign of the Lagrangian~\eqref{Spin1ActionProca} comes from that of \eqn{Spin1ActionProca},
which includes the same overall $(-1)^{\lfloor s \rfloor}$ oscillation that can be observed \eg in the non-chiral action~\eqref{QuadraticActionFeynmanGauge} and which comes from the mostly-minus metric convention.
In the chiral formulation, we choose to drop it starting from \eqn{Spin1ActionChiral}.
}
\be
- \frac{1}{2} \partial^\mu \Phi^{\alpha\beta}\partial_\mu \Phi_{\alpha\beta}
+ \frac{m^2\!}{2} \Phi^{\alpha\beta} \Phi_{\alpha\beta} \,=:\,
{-}\frac{1}{2} \braket{\partial^\mu \Phi|\partial_\mu \Phi}
+ \frac{m^2\!}{2} \braket{\Phi|\Phi} \,.
\label{SpinorConventions1}
\ee
where in the last step we converted to bra--ket notation, $|\Phi\rangle := \Phi_{\alpha \beta}$,  instead of writing out the spinor indices.
This is convenient when the number of spinorial indices becomes high.

The chiral gauge-interacting Lagrangian for massive spin-1
was first introduced in \rcite{Chalmers:2001cy}
for the purpose of describing electroweak vector bosons.
Adapting it to a general non-abelian gauge group, with $|\Phi \rangle$ transforming in a matter representation, we find the chiral version of the action~\eqref{ActionFromHiggs} to be given in the bra--ket notation by
\begin{subequations} \begin{align}
\label{Spin1ActionChiralFull}
{\cal L}^{(1)} &
 = \bra{\Phi} \bigg\{
   |\overset{\leftarrow}{D}|\overset{\rightarrow}{D}| \otimes
      \frac{1}{1-\tfrac{ig}{m^2}|\Fm|}
   \bigg\} \ket{\Phi} - m^2 \braket{\Phi|\Phi} + {\cal O}(\Phi^4) \\ &
\label{Spin1ActionChiral4pt}
 = \braket{D_\mu \Phi|D^\mu \Phi} - m^2 \braket{\Phi|\Phi}
 + ig \bra{\Phi}\Fm\ket{\Phi}
 + \frac{ig}{m^2} \bra{\Phi} \Big\{
   |\overset{\leftarrow}{D}|\overset{\rightarrow}{D}| \otimes |\Fm|
   \Big\} \ket{\Phi} \\ & \qquad \qquad \qquad \qquad \qquad~\;\quad
 - \frac{g^2}{m^4} \bra{\Phi} \Big\{
   |\overset{\leftarrow}{D}|\overset{\rightarrow}{D}| \otimes |\Fm|\Fm|
   \Big\} \ket{\Phi}
 + {\cal O}(F^3) + {\cal O}(\Phi^4) . \nn
\end{align} \label{Spin1ActionChiral}%
\end{subequations}
The arrows over the covariant derivatives
\be
\begin{aligned}
\overset{\leftarrow}{D}_{\alpha \dot\beta} &
 = \overset{\leftarrow}{\partial}_{\!\alpha \dot\beta}
 + ig A_{\alpha \dot\beta} \,, \\
\overset{\rightarrow}{D}{}^{\dot\alpha \beta} &
 = \partial^{\dot\alpha \beta} - ig A^{\dot\alpha \beta} ,
\end{aligned} \qquad
A_\mu = A_\mu^c T^c , \qquad T^{c\dagger} = T^c , \qquad
\tr T^c T^{c'}= \frac{1}{2} \delta^{cc'} ,
\label{CovDerivative}
\ee
indicate on which matter field they act, and this notation implicitly assumes that the covariant derivatives do not act on the field strength factors $|F_-|$ of the massless gauge boson. 
The minus label emphasizes that this is the anti-self-dual part of the field strength, $|F_-|:= F_{-\,\alpha}{}^\beta := \frac{1}{2}
(\sigma^\mu\bar{\sigma}^\nu)_\alpha{}^\beta F_{\mu\nu}$, where the (anti-)self-dual parts satisfy $\frac{1}{2} \epsilon_{\mu\nu\rho\sigma} F_\pm^{\rho\sigma} = \pm i F^\pm_{\mu\nu}$. 
These labels are also correlated with the gluon helicities that are allowed to appear in the tree-level amplitude. Hence, at leading order interaction terms with $k$ factors of $|F^-|$ only contribute to amplitudes that have $k$ negative-helicity gluons.  

The gauge-group indices are subsumed by the indexless bra--ket notation, and we can assume without loss of generality that kets $|\Phi\rangle$ belong to the fundamental representation and bras $\langle\Phi|$ to the anti-fundamental representation of the gauge group.
(The Lagrangian is also valid for general complex or real representations.)
Although the indexless notation is convenient for dealing with ${\rm SL}(2,\mathbb{C})$ indices, it admittedly obscures the multiplication order of the non-abelian gauge-group generators. To be more precise, let us write the generic term in the expansion of \eqn{Spin1ActionChiralFull} such that the non-abelian order is displayed, at the expense of having explicit spinorial indices:
\be\!\!
\bra{\Phi} \Big\{ |\overset{\leftarrow}{D}|\overset{\rightarrow}{D}| \otimes
   \underbrace{|\Fm|\cdots|\Fm|}_{n} \Big\} \ket{\Phi}
:= (D_{\alpha \dot\varepsilon} \Phi^{\alpha\beta})^\star
   (\Fm\;\!\!\!\!^{c_1})_\beta{}^{\zeta_2} \cdots (\Fm\;\!\!\!\!^{c_n})_{\zeta_n}\!{}^\delta
   T^{c_1}\!\cdots T^{c_n}
   (D^{\dot\varepsilon\gamma} \Phi_{\gamma\delta}) \,,\!
\label{SpinorConventions2}
\ee
Here and below, the star is used to flip $g \to -g$ and conjugate the gauge-group indices
--- but without conjugating the spacetime indices
(which would otherwise imply an unwanted chirality switch).
For instance,
\be
(D_{\gamma \dot\delta} \Phi^{\alpha\beta})^\star_{\bar \jmath}
 = \big(\partial_{\gamma \dot\delta} \Phi^{\alpha\beta}\!
 - ig A_{\gamma \dot\delta}^c T^c \Phi^{\alpha\beta}\big)_{\bar \jmath}^\star
:= \partial_{\gamma \dot\delta} (\Phi^\star)^{\alpha\beta}_{\bar \jmath}\!
 + ig (\Phi^\star)^{\alpha\beta}_{\bar\imath} T^c_{i\bar\jmath} A_{\gamma \dot\delta}^c \,.
\ee

Note that the action~\eqref{Spin1ActionChiral} obeys the following reality condition:
\be
\int\!d^4x\,{\cal L}^\star = \int\!d^4x\,{\cal L} ,
\label{QuasiReality}
\ee
which we will always demand of a chiral Lagrangian, and this will imply that the theory can be equivalently described in a conventional (\ie non-chiral) formulation using a real-valued Lagrangian.\footnote{Strictly speaking, the equivalence between the chiral and the usual tensorial approaches has been shown only for the lower spins and minimal interactions \cite{Chalmers:1997ui,Chalmers:2001cy}. However, given that we do not see any obstruction in getting any consistent cubic and quartic amplitudes within the chiral approach, it seems plausible that it covers all possible interactions at all orders.
}
In particular, this condition allows for a consistent projection of the theory to the case of real-valued generators,\footnote{In our current conventions, which are tailored to hermitian generators, the real representations of ${\rm SU}(N_\text{c})$ or ${\rm SO}(N_\text{c})$ actually have imaginary-valued generators.
}
such as the adjoint representation of ${\rm SU}(N_\text{c})$
or the fundamental representation of ${\rm SO}(N_\text{c})$, considered in \rcite{Ochirov:2022nqz}.

\subsection{Cubic $\sqrt{\text{Kerr}}$ Lagrangian for arbitrary spin}
\label{sec:ChiralCubicLag}
After having introduced the chiral framework in the familiar spin-1 context, it is now straightforward to generalize it to the spin-$s$ case. In particular, we combine the kinetic terms and minimal cubic interactions of \rcite{Ochirov:2022nqz}, and then add appropriate cubic non-minimal interactions to restore parity invariance of three-point amplitudes. The resulting cubic Lagrangians are simple, and we give the new result without further ado. 

The family of chiral spin-$s$ Lagrangians that reproduce the three-point $\sqrt{\text{Kerr}}$ gauge-theory amplitudes~\eqref{AHH} can be written compactly as
\be
\label{ActionGaugeAHH}
{\cal L}^{(s)} = \braket{D_\mu \Phi|D^\mu \Phi} - m^2 \braket{\Phi|\Phi}
 + \sum_{k=0}^{2s-1} \frac{ig}{m^{2k}} \bra{\Phi}
   \Big\{ |\overset{\leftarrow}{D}|\overset{\rightarrow}{D}|^{\odot k}\!
          \otimes |\Fm| \Big\} \ket{\Phi} + {\cal O}(F^2) \,.
\ee
We have used the $\odot$ symbol to denote symmetrized tensor product~\cite{Guevara:2017csg}. Here we additionally define it to symmetrize the internal spinor indices.
In the higher-spin multi-index notation, in which all indices denoted by the same letter are understood as symmetrized among themselves, we may write more explicitly
\begin{align}
\label{SpinorConventions3}
\braket{\Phi|\Phi} &\:\!\!:= \big(\Phi^{\alpha(2s)}\big)^{\!\star}
   \Phi_{\alpha(2s)} \,, \\*
\braket{D_\mu \Phi|D^\mu \Phi} &\:\!\!:=  \big(D_\mu\Phi^{\alpha(2s)}\big)^{\!\star}
   D^\mu\Phi_{\alpha(2s)} \,, \nn \\*
\bra{\Phi} \big\{|\overset{\leftarrow}{D}|\overset{\rightarrow}{D}|^{\odot k}\!\otimes\!|\Fm|\big\} \ket{\Phi} &\:\!\!:= \big(D_{\alpha(k)\dot{\gamma}(k)}
   \Phi^{\alpha(k+1)\gamma(2s-k-1)}\big)^{\!\star}
   (\Fm\;\!\!\!\!^c)_{\alpha}{}^{\beta} T^c D^{\dot{\gamma}(k)\beta(k)}
   \Phi_{\beta(k+1)\gamma(2s-k-1)} , \nn
\end{align}
where
$D_{\alpha(k)\dot{\beta}(k)}\!:= D_{(\alpha_1 (\dot{\beta}_1}\!\cdots D_{\alpha_k) \dot{\beta}_k)}$ and
$D^{\dot{\alpha}(k)\beta(k)}\!:= D^{(\dot{\alpha}_1 (\beta_1} \cdots D^{\dot{\alpha}_k) \beta_k)}$.

As explained in \rcite{Ochirov:2022nqz}, the on-shell wavefunctions
for the massive particles, described by the chiral fields, are simply
$\ket{\Phi} \to \ket{{\bm p}}^{2s}\!/m^{s}$.
To guide the reader, it may be helpful to evaluate some of the Lagrangian terms by plugging in on-shell wavefunctions, and obtain explicit matrix elements.
For example, consider on-shell three-point kinematics applied to the first two terms of \eqn{ActionGaugeAHH},
\be
\label{AlmostAHH}
\braket{D_\mu \Phi|D^\mu \Phi}
   \Big|_{(2,3^{\pm}\!,1)}\!
 = ig \big( \braket{A_\mu \Phi|\partial^\mu \Phi}
          - \braket{\partial_\mu \Phi|A^\mu \Phi} \big)
   \Big|_{(2,3^{\pm}\!,1)}\!
 = 2g T^{c_3} (p_1\cdot\,\varepsilon_3^\pm)
   \frac{\braket{\bm{21}}^{2s}\!}{m^{2s}} \,.
\ee
Here we generalize the subscript notation $(i_1,\dots,i_n)$ introduced in \eqn{WardIdentitySpin1}, to indicate that we extract the $n$-particle component of the corresponding Lagrangian operators, and we impose the usual on-shell conditions. \Eqn{AlmostAHH} shows that the chiral ``minimal-coupling'' terms reproduce the \AHH amplitudes~\eqref{AHH} for a positive-helicity~\cite{Ochirov:2022nqz}, but not for a negative-helicity gluon. Since the two \AHH amplitudes are related by a parity transformation, the minimal-coupling chiral Lagrangian explicitly breaks parity.
However, let us show now that the non-minimal terms restore it.  

Let us evaluate the three-point matrix element of the non-minimal coupling terms.
They are non-vanishing only for the minus-helicity gluon and evaluate to
\be
\frac{1}{m^{2k}}
\bra{\Phi} \big\{|\overset{\leftarrow}{D}|\overset{\rightarrow}{D}|^{\odot k}\!\otimes\!|\Fm|\big\} \ket{\Phi} \Big|_{(2,3^-\!,1)}\!
= -i\sqrt{2} T^{c_3}
   \braket{\bm{2}3} \braket{3\bm{1}} [\bm{21}]^k
   \frac{\braket{\bm{21}}^{2s-k-1}}{m^{2s}} \,.
\ee
We can now sum the kinematic factors over all powers $k$:
\be
\sum_{k=0}^{2s-1} \braket{\bm{2}3} \braket{3\bm{1}} [\bm{21}]^k
   \braket{\bm{21}}^{2s-k-1}
 = \braket{\bm{2}3} \braket{3\bm{1}}
   \bigg(\frac{[\bm{21}]^{2s}\!- \braket{\bm{21}}^{2s}}
              {[\bm{21}] - \braket{\bm{21}}} \bigg)
 = \sqrt{2} (p_1\!\cdot\varepsilon_3^-)
   \big([\bm{21}]^{2s}\!- \braket{\bm{21}}^{2s}\big) ,
\ee
where we used the identity
$\braket{\bm{2}3} \braket{3\bm{1}} = \sqrt{2}(p_1\!\cdot\varepsilon_3^-)
 \big( [\bm{21}] - \braket{\bm{21}} \big)$ to make locality manifest. 
When the non-minimal matrix element is combined appropriately with the negative-helicity amplitude~\eqref{AlmostAHH}, the $\braket{\bm{21}}^{2s}$ factor is canceled out, and instead replaced by $[\bm{21}]^{2s}$.
We get the chiral-Lagrangian amplitudes as
\be
{\cal A}(\bm{1}^s\!,\bar{\bm{2}}^s\!,3^+)
 = 2 g T^{c_3} (p_1\!\cdot\varepsilon_3^+)
   \frac{\braket{\bm{21}}^{2s}\!}{m^{2s}} \,,
\qquad \quad
{\cal A}(\bm{1}^s\!,\bar{\bm{2}}^s\!,3^-)
 = 2 g T^{c_3} (p_1\!\cdot\varepsilon_3^-) \frac{[\bm{21}]^{2s}\!}{m^{2s}} \,.
\label{AmpGaugeAHH3}
\ee
In summary, we have shown that the action~\eqref{ActionGaugeAHH} restores parity at the three-point level and yields the known \AHH amplitudes~\cite{Arkani-Hamed:2019ymq}.

Although we refer to \eqn{ActionGaugeAHH} as the cubic Lagrangian, it also automatically incorporates higher-point vertices.
For instance, the non-minimal cubic terms can be shown to give the following contact-term contributions to the opposite-helicity Compton amplitude:
\begin{align} \label{ContactMatrixElement}\!\!
\frac{ig}{m^{2k}} \bra{\Phi}
   \big\{|\overset{\leftarrow}{D}|\overset{\rightarrow}{D}|^{\odot k}\!
         \otimes\!|\Fm|\big\} \ket{\Phi} \Big|_{(2,3^-\!,4^+\!,1)}\!
 = \frac{g^2}{2m^{2s}} \braket{\bm{21}}^{2s-k-1} [\bm{21}]^k
   \braket{\bm{2}|[\varepsilon^-_3, \varepsilon^+_4]|\bm{1}}
   [T^{c_3},& T^{c_4}] \nn \\
+\,\frac{\sqrt{2}g^2}{m^{2s+k}}
   \braket{\bm{21}}^{2s-k-1} \braket{\bm{2} 3} \braket{3 \bm{1}}
   \bigg\{ \frac{ [\bm{2}|\varepsilon_4^+\ket{\bm{1}} }
                { [\bm{2}|4\ket{\bm{1}}}
           \big( [\bm{2}|1\!+\!4\ket{\bm{1}}^k {-} m^k [\bm{21}]^k \big)
           T^{c_3} & T^{c_4} \\*
         - \frac{ \bra{\bm{2}}\varepsilon_4^+|\bm{1}] }
                { \bra{\bm{2}}4|\bm{1}] }
           \big( \bra{\bm{2}}1\!+\!3|\bm{1}]^k {-} m^k [\bm{21}]^k \big)
           T^{c_4} & T^{c_3}\!
   \bigg\} \,. \nn
\end{align}
The first line comes from the non-linearity of the self-dual field strength, and the other two lines are more complicated as they come from the gauge-field dependence of the powers of covariant derivatives.
Note that the expression is local, as the spurious denominators can be canceled against the numerators after expanding out the powers in $k$.
Furthermore, note that this contact term does not respect massless gauge invariance by itself, as it needs to be combined with the corresponding cubic exchange diagrams in a gauge-invariant amplitude.
The gauge-invariant contact terms will be discussed in the next subsection. 

\Eqn{ContactMatrixElement} illustrates a general property that the (anti-)self-duality of the explicit field strength in the Lagrangian gives rise to a helicity-selection rule, but only at the lowest multiplicity for which it starts to contribute to an amplitude, whereas at higher multiplicities it includes both helicities.
Namely, the operator on the left-hand side contains only a negative-helicity gluon at three points, but both helicities at four points and higher.
However, the leading-order helicity separation of the terms
involving $|F_\pm|$ is very helpful, since the strategy is to add such terms only when they are needed.
For instance, to restore parity in amplitudes with $k$ negative-helicity gluons one should consider operators involving $k$ field strengths $|F_-|$.

\subsection{Parity of Compton amplitudes}
\label{sec:Parity}

Before we proceed to discussing quartic interactions in more detail, let us recall that, by construction~\cite{Ochirov:2022nqz}, the first two ``minimal-coupling'' terms of the Lagrangian~\eqref{ActionGaugeAHH} imply
\begin{subequations} \label{AmpGaugeAHH4idh}
\be
{\cal A}(\bm{1}^s\!,\bar{\bm{2}}^s\!,3^+\!,4^+) = -2g^2
\frac{\braket{\bm{21}}^{2s} [34]^2}{m^{2s-2} s_{12}}
\bigg[ \frac{T^{c_3} T^{c_4}\!}{t_{14}}
     + \frac{T^{c_4} T^{c_3}\!}{t_{13}} \bigg] \,,
\label{AmpGaugeAHH4pp}
\ee
where the Mandelstam variables are
$s_{12} = (p_1 + p_2)^2$ and $t_{ij} = 2p_i\cdot p_j$.
This same-helicity amplitude is known to follow from its factorization limits~\cite{Arkani-Hamed:2017jhn,Johansson:2019dnu,Lazopoulos:2021mna} onto the three-point amplitudes~\eqref{AmpGaugeAHH3}, as well as to have a well-behaved classical limit~\cite{Aoude:2020onz}, given below in \eqn{AmpGaugeAHH4ppClassical}.
From minimality considerations, it is expected to correspond to a $\sqrt{\text{Kerr}}$ object,
so we do not wish to modify it in any way.
In particular, we have checked that the tower of non-minimal cubic terms in \eqn{ActionGaugeAHH} does not contribute to this amplitude,
which is not trivial in the non-abelian case, in which $|F_-|$ contains terms quadratic in the potential~$A_\mu$.

Although we are discussing a chiral framework, we maintain our parity assumption~\ref{Parity}, as stated during the non-chiral construction of gauge interactions in the previous sections.
Parity demands that the negative-helicity counterpart of \eqn{AmpGaugeAHH4pp} be
\be
{\cal A}(\bm{1}^s\!,\bar{\bm{2}}^s\!,3^-\!,4^-) \stackrel{!}{=} -2g^2
\frac{[\bm{21}]^{2s} \braket{34}^2}{m^{2s-2} s_{12}}
\bigg[ \frac{T^{c_3} T^{c_4}\!}{t_{14}}
     + \frac{T^{c_4} T^{c_3}\!}{t_{13}} \bigg] \,.
\label{AmpGaugeAHH4mm}
\ee
\end{subequations}
The cubic Lagrangian~\eqref{ActionGaugeAHH} alone is consistent with this amplitude for $s=0,1/2$,
but already for $s=1$ it must be corrected by a quartic contact term involving $|F_-|F_-|$, as seen in \eqn{Spin1ActionChiral}.
Since the desired same-helicity amplitude~\eqref{AmpGaugeAHH4mm} for $\sqrt{\text{Kerr}}$ is known from non-Lagrangian considerations, we leave specifying higher-spin $F_-^2$ terms for future work.

From now on we focus on the opposite-helicity sector.
We have explicitly computed the corresponding amplitude from the cubic Lagrangian~\eqref{ActionGaugeAHH} through $s=5$.
Thus we found the following closed-form expression for the color-ordered amplitude for any $s \geq 1$:
\def\N{{\bm N}}
\def\W{{\bm R}}
\beal
\label{OriginalAHHCompton}
A(\bm{1}^s\!,\bar{\bm{2}}^s\!,3^-\!,4^+)
 = \frac{\N^2 \W^{2s-2}\!}{s_{12} t_{14}}
 + \frac{t_{14}}{s_{12}} \braket{\bm{1} 3}^2 [4 \bm{2}]^2\!
   \sum_{j=0}^{2s-4} \frac{\W^{2s-4-j}\!}{m^{j+6}} 
   \sum_{i=0}^j \braket{\bm{21}}^{i+1} & [\bm{21}]^{j-i+1} \\
-\,\frac{\braket{\bm{1} 3} [4 \bm{2}]}{s_{12}}
   \bigg[ \N\!\sum_{j=1}^{2s-2} \!\frac{{\W}^{2s-2-j}\!}{m^{j+2}} 
   \big(\braket{\bm{21}}^j\!+[\bm{21}]^j\big) 
 - \frac{\bra{3}1|4]}{m^{2s+1}} 
   \sum_{i=0}^{2s-3} \braket{\bm{21}}^{i+1} & [\bm{21}]^{2s-i-2}
   \bigg]\,,
\eeal
where
$\N= \braket{\bm{1} 3} [4 \bm{2}] + [\bm{1} 4] \braket{3 \bm{2}}$ and
$\W = \bra{\bm{1}}1\!+\!4|\bm{2}]/m^2= (U+V)/m^2$.
The result for the second color ordering may be obtained from the above by relabeling:
\be
A(\bm{1}^s\!,\bar{\bm{2}}^s\!,4^+\!,3^-)
 = (-1)^{2s} A(\bm{2}^s\!,\bar{\bm{1}}^s\!,3^-\!,4^+)
:= (-1)^{2s} A(\bm{1}^s\!,\bar{\bm{2}}^s\!,3^-\!,4^+)
   \big|_{\bm{1} \leftrightarrow \bm{2}} \,.
\label{Relabeling}
\ee
Recall that a color-stripped amplitude is sensitive to the ordering but not to the color representation of the matter particles, so we may move the bar freely between them.

Interestingly, these results automatically satisfy
the amplitude-level parity constraint
\be
\big[{\cal A}(\bm{1}_{\bar\imath_1}^{\{a\}}\!, \bar{\bm{2}}_{i_2}^{\{b\}}, 3^{h_3}_{c_3}\!, 4^{h_4}_{c_4}\!,\dots)\big]^*
 = 
   {\cal A}(\bar{\bm{1}}_{{\{a\}}i_1}, \bm{2}_{{\{b\}}\bar\imath_2}, 3^{-h_3}_{c_3}\!, 4^{-h_4}_{c_4}\!,\dots) \,.
\label{ParityConstraintFull}
\ee
This roughly means that flipping helicities amounts to swapping left- and right-handed Weyl spinors.
For example, the parity equivalence of \eqns{AmpGaugeAHH4pp}{AmpGaugeAHH4mm} is implied by the spinor conjugation properties
$\braket{1^a 2^b}^*\!= [1_a 2_b]$ and $[34]^*\!= \braket{34}$.
Here and below, we assume that energies satisfy $p_1^0,p_3^0>0>p_2^0,p_4^0$, which is the interesting kinematics for the Compton scattering process.\footnote{For generic real kinematics, there is a subtle, convention-dependent overall sign in \eqn{ParityConstraintFull} and below, which depends on particles' energies due to spinor conjugation properties
\be
\ket{k}^*\!= \sgn(k^0) [k| \,, \qquad \quad
\ket{p^a}^*\!= \sgn(p^0) [p_a|
\qquad \Leftrightarrow \qquad
[p^a|^*\!= -\sgn(p^0) \ket{p_a} \,. \nn
\ee
}
In view of the relabeling property~\eqref{Relabeling}, the color-stripped version of \eqn{ParityConstraintFull} may be rewritten as a constraint on a single ordered amplitude:
\be
\big[A(\bm{1}^s\!,\bar{\bm{2}}^s\!,3^-\!,4^+)\big]^* \Big|\:\!\!\!
   \phantom|_{\substack{\bar z_{1 a} \to z_{1 a}\\
                        \bar z_{2 b} \to z_{2 b}}}
 = (-1)^{2s} 
   A(\bm{1}^s\!,\bar{\bm{2}}^s\!,3^-\!,4^+) \Big|\:\!\!\!
   \phantom|_{\substack{\bm{1} \leftrightarrow \bm{2}\\
                              3 \leftrightarrow 4}} \,,
\label{ParityConstraintOrdered}
\ee
which can be directly verified.
Here we have reinstated the little-group contraction with the SU(2) spinors $z_{i a}$, and in this purely bookkeeping context they are treated as real so as to implement the index conjugation of \eqn{ParityConstraintFull}.

At four points, we can rearrange the conjugation and two-fold relabeling of the parity constraint~\eqref{ParityConstraintOrdered} into two $\mathbb{Z}_2$ operations:
$(\bm{1} \leftrightarrow \bm{2})$ and $(3 \leftrightarrow 4)|_\text{c.c.}$.
Any kinematic block may be decomposed into the even and odd pieces with respect to these operations, and the parity property will follow, as shown in \tab{tab:Parity}.
In the following, we will therefore organize our contact terms so as to make these $\mathbb{Z}_2$ properties explicit.
\begin{table}[t]
\centering
\begin{tabular}{|cc||c||c|}
\cline{1-4}
\cline{1-2}~~~kinematic\!\!&\!\!\!\!\!\!\!\!\!
$\mathbb{Z}_2\times\mathbb{Z}_2$~& ~~resulting~~ & consistent \\ 
\cline{1-2}
\multicolumn{1}{|c|}{$1 \leftrightarrow 2$} & $(3 \leftrightarrow 4)|_\text{c.c.}$ & parity & color factor \\ \hline \hline
\multicolumn{1}{|c|}{$+$} & $+$ & even & $\{T^{c_3},T^{c_4}\}$ \\ \cline{1-4}
\multicolumn{1}{|c|}{$-$} & $-$ & even & $[T^{c_3},T^{c_4}]$ \\ \cline{1-4}
\multicolumn{1}{|c|}{$+$} & $-$ & odd & $\{T^{c_3},T^{c_4}\}$ \\ \cline{1-4}
\multicolumn{1}{|c|}{$-$} & $+$ & odd & $[T^{c_3},T^{c_4}]$ \\ \cline{1-4}
\end{tabular}
\caption{The kinematic building blocks of the Compton amplitude can be assigned $\mathbb{Z}_2\times\mathbb{Z}_2$ charges. The consistent color factor follows from the first charge, whereas the (spacetime) parity of the amplitude is the product of the two charges. The first column assumes bosonic spin-$s$ particles, the fermionic case is obtained by reversing the signs of this column (without affecting parity).}
\label{tab:Parity}
\end{table}

Note that the choice of color factors in \tab{tab:Parity} follows from the eigenvalue of the permutation $(\bm{1} \leftrightarrow \bm{2})$.
This is because, as shown in \eqn{Relabeling}, the $(\bm{1} \leftrightarrow \bm{2})$ exchange relates the two color-ordered amplitudes $A(\bm{1}^s\!,\bar{\bm{2}}^s\!,3\,,4)$ and $A(\bm{1}^s\!,\bar{\bm{2}}^s\!,4\,,3)$, which multiply the color factors $T^{c_3}T^{c_4}$ and $T^{c_4}T^{c_3}$, respectively, and hence it determines the relative sign between these color structures.
This argument is easiest seen once the matter particle is taken to be self-conjugate and, for instance, charged with respect to ${\rm SO}(N_\text{c})$ gauge group.

\subsection{Quartic interactions for $\sqrt{\text{Kerr}}$}
\label{sec:ChiralQuarticOps}

Let us discuss the quartic contact interactions that should be added to the cubic chiral Lagrangian~\eqref{ActionGaugeAHH}.
A fully general approach to the contact-term freedom is described in \app{app:ChiralQuarticContact}.
Here we focus on operators involving two field strengths of opposite helicity, and we choose them such that at four points they give rise to kinematic structures with well-defined properties under the two $\mathbb{Z}_2$ transformations described in \tab{tab:Parity}.

\def\FF{\mathfrak F}
\def\DD{\mathfrak D}
We consider the four-point contact interactions of the following schematic form,\footnote{Note that the displayed operators $\bra{\Phi} \DD_{jkl} \odot \FF_i \,\ket{\Phi}$ form a slightly overcomplete basis, see \app{app:ChiralQuarticContact} for further details.}
\be \label{eq:genL4Op}
{\cal L}_4 = \frac{g^2}{m^4} \sum_{i,j,k,l} c_{ijkl}
   \bra{\Phi} \DD_{jkl} \odot \FF_i \,\ket{\Phi}
 + (\text{higher-derivative Mandelstam terms}) \,,
\ee
where the operators $\FF_i$ contain combinations the self-dual and anti-self-dual field strengths $|F_\pm|$, $\DD_{jkl}$ are constructed only out of covariant derivatives, and $c_{ijkl}$ are the free Wilson coefficients.
${\rm SL}(2,\mathbb{C})$ covariance requires the self-dual field strength $F_+$ to be contracted with derivatives whenever it appears inside $\FF_i$, which we choose to act only on the matter fields, namely 
$|\overset{\leftarrow}{D}|F_+|\overset{\rightarrow}{D}|$.
Similarly, $\DD_{jkl}$ have an even number of derivatives.
We furthermore split $\FF_i$ into the two types of operators that either depend on the commutator or anticommutator of the gauge-group generators:
\be
\FF_i: = \{ T^c, T^{c'} \} \FF^{c c'}_i \qquad \text{or} \qquad
\widetilde \FF_i :=  [ T^c, T^{c'}] \FF^{c c'}_i \,.
\ee
Let us consider the following six distinct structures quadratic in the field strength:
\beal
\label{ContactTermsF}
\FF_1^{cc'} & = \frac{1}{2} \Tr\!\big(
   |F_-^c|\overset{\leftarrow}{D}|F_+^{c'}|\overset{\rightarrow}{D}|
   \big) \,, 
\\
\FF_2^{cc'} & = \frac{1}{4}
   \big( |F_-^c|\overset{\leftarrow}{D}|F_+^{c'}|\overset{\rightarrow}{D}| 
       + |\overset{\leftarrow}{D}|F_+^{c'}|\overset{\rightarrow}{D}|F_-^c|
   \big)_{\alpha_1}{}^{\beta_1} \,, 
\\
\FF_3^{cc'} & = \frac{1}{4}
   \big( |F_-^c|\overset{\leftarrow}{D}|F_+^{c'}|\overset{\rightarrow}{D}| 
       - |\overset{\leftarrow}{D}|F_+^{c'}|\overset{\rightarrow}{D}|F_-^c|
   \big)_{\alpha_1}{}^{\beta_1} \,, 
\\
\FF_4^{cc'} & = \frac{1}{4} \Big\{
   (F_-^c)_{\alpha_1\alpha_2}
   (\overset{\rightarrow}{D}|F_+^{c'}|\overset{\rightarrow}{D})^{\beta_1\beta_2}
 - (\overset{\leftarrow}{D}|F_+^{c'}|\overset{\leftarrow}{D})_{\alpha_1\alpha_2}
   (F_-^c)^{\beta_1\beta_2}
   \Big\} \,, 
\\
\FF_5^{cc'} & = \frac{1}{4} \Big\{
   (F_-^c)_{\alpha_1\alpha_2}
   (\overset{\rightarrow}{D}|F_+^{c'}|\overset{\rightarrow}{D})^{\beta_1\beta_2}
 + (\overset{\leftarrow}{D}|F_+^{c'}|\overset{\leftarrow}{D})_{\alpha_1\alpha_2}
   (F_-^c)^{\beta_1\beta_2}
   \Big\} \,, 
\\
\FF_6^{cc'} & = \frac{1}{4} |F_-^c|_{\alpha_1}{}^{\beta_1}
    |\overset{\leftarrow}{D}|F_+^{c'}|\overset{\rightarrow}{D}|_{\alpha_2}{}^{\beta_2} \,. 
\eeal
Here $\FF_1^{cc'}$ is a scalar, as all of its spinorial indices are traced over.
Operators $\FF_2^{cc'}$ and $\FF_3^{cc'}$ are spin-$1/2$ operators with indices $\alpha_1$ and $\beta_1$ contracting into $\bra{\Phi}$ and $\ket{\Phi}$, respectively. 
Operators $\FF_4^{cc'}$, $\FF_5^{cc'}$ and $\FF_6^{cc'}$ are spin-$1$ operators,
as they have two pairs of indices contracting into the bra and ket states.
All these $\FF_i$ operators have derivative counting~$4$,
and at four points they correspond to the following spinor structures
(after setting $T^c=1/\sqrt{2}$):
\beal
\label{ContactTermsF4pt}
\bra{\Phi} \FF_1 \ket{\Phi}\big|_{(2,3^-\!,4^+\!,1)} &
 = \bra{3}1|4]^2 \frac{\braket{\bm{21}}^{2s}}{m^{2s}} \,,
   \quad & (+,+) \,,
\\
\bra{\Phi} \FF_2 \ket{\Phi}\big|_{(2,3^-\!,4^+\!,1)} & = \frac{1}{2} \bra{3}1|4]
   \big( \braket{\bm{2}3} [4\bm{1}] + [\bm{2}4] \braket{3\bm{1}} \big)
   \frac{\braket{\bm{21}}^{2s-1}}{m^{2s-1}} \,,
   \quad & (-,+) \,,
\\
\bra{\Phi} \FF_3 \ket{\Phi}\big|_{(2,3^-\!,4^+\!,1)} & = \frac{1}{2} \bra{3}1|4]
   \big( \braket{\bm{2}3} [4\bm{1}] - [\bm{2}4] \braket{3\bm{1}} \big)
   \frac{\braket{\bm{21}}^{2s-1}}{m^{2s-1}} \,,
   \quad & (+,-) \,,
\\
\bra{\Phi} \FF_4 \ket{\Phi}\big|_{(2,3^-\!,4^+\!,1)} & = \frac{1}{2}
   \big(\braket{\bm{2}3}^2 [4\bm{1}]^2 - [\bm{2}4]^2 \braket{3\bm{1}}^2\big)
   \frac{\braket{\bm{21}}^{2s-2}}{m^{2s-2}} \,,
   \quad & (-,-) \,,
\\
\bra{\Phi} \FF_5 \ket{\Phi}\big|_{(2,3^-\!,4^+\!,1)} & = \frac{1}{2}
   \big(\braket{\bm{2}3}^2 [4\bm{1}]^2 + [\bm{2}4]^2 \braket{3\bm{1}}^2\big)
   \frac{\braket{\bm{21}}^{2s-2}}{m^{2s-2}} \,,
   \quad & (+,+) \,,
\\
\bra{\Phi} \FF_6 \ket{\Phi}\big|_{(2,3^-\!,4^+\!,1)} &
 = \frac{1}{2} \braket{\bm{2}3} \braket{3\bm{1}} [\bm{2}4] [4\bm{1}]
   \frac{\braket{\bm{21}}^{2s-2}}{m^{2s-2}}
   \quad & (+,+) \,.
\eeal
Here the $(\pm,\pm)$ charges indicate the properties under the $\mathbb{Z}_2\times\mathbb{Z}_2$ exchange of particles at the lowest relevant spin.
The first charge determines the color factor, as shown in \tab{tab:Parity}.
The second charge then indicates if, at the lowest relevant spin, the consistently dressed contact term obeys the parity constraint~\eqref{ParityConstraintFull}. 
Assuming that contact terms come with $s$-independent Wilson coefficients,
we observe that neither of the terms~\eqref{ContactTermsF} is ruled out.

In addition, we should allow for additional derivatives in the form of the $\DD_{jkl}$ operators,
\be
\DD_{jkl} := \frac{1}{m^{2(j+k+l)}}
|\overset{\leftarrow}{D}|\overset{\rightarrow}{D}|^{\odot j} \odot
|\overset{\leftarrow}{D}|\overset{\rightarrow}{D}_+|^{\odot k} \odot
|\overset{\leftarrow}{D}_+|\overset{\rightarrow}{D}|^{\odot l} \,.
\ee
Apart from the matter derivatives $|\overset{\leftarrow}{D}|\overset{\rightarrow}{D}|$, that are already familiar from the cubic Lagrangian~\eqref{ActionGaugeAHH}, the building blocks of $\DD_{jkl}$ can be written out with spinorial indices
\be
|\overset{\leftarrow}{D}|\overset{\rightarrow}{D}_+|
:= \big( |\overset{\leftarrow}{D}|\overset{\rightarrow}{D}|
       + |\overset{\leftarrow}{D}|D_+| \big)_{\alpha_1}\!{}^{\beta_1} \,, 
\qquad \quad
|\overset{\leftarrow}{D}_+|\overset{\rightarrow}{D}|
:= \big( |\overset{\leftarrow}{D}|\overset{\rightarrow}{D}|
       + |D_+|\overset{\rightarrow}{D}| \big)_{\alpha_1}\!{}^{\beta_1} \,,
\ee
where $|D_+|$ acts only on the field $|F_+|$ and its derivatives via the adjoint-representation action.
For example, if we make all indices explicit, we can write
\be
D^+_{\alpha \dot\alpha} \odot F^{+c}_{\beta \dot\beta}
:= \partial_{(\alpha(\dot\alpha} F^{+c}_{\beta)\dot\beta)}
 + g f^{cde}  A^d_{(\alpha(\dot\alpha} F^{+e}_{\beta)\dot\beta)} \,.
\ee
Higher derivatives $|D_+|^{\odot k} \odot |F_+|$ are defined recursively in an analogous way.
Most importantly, we define $|D_+|$ not to act on any of the other fields, $|F_-|$, $\bra{\Phi}$ and $\ket{\Phi}$, nor on the matter derivatives $|\overset{\leftarrow}{D}|$ and $|\overset{\rightarrow}{D}|$.

Let us consider an explicit operator and show how it is used.
Taking $\FF_6$ with abelian generators $T^c=1/\sqrt{2}$ and assuming $2s\ge j+k +l+2$, the four-point matrix element is
\be
\label{ContactTerm6}
\bra{\Phi} \DD_{jkl} \odot \FF_6 \ket{\Phi}|_{(2,3^-\!,4^+\!,1)}
 = \frac{\braket{\bm{1}3}\braket{3\bm{2}}[\bm{1}4][4\bm{2}]}{2m^{2s+k+l-2}}
   [\bm{21}]^{j} [\bm{2}|1{+}4\ket{\bm{1}}^k
   \big({-}\bra{\bm{2}}2{+}4|\bm{1}]\big)^l \braket{\bm{21}}^{2s-j-k-l-2} .
\ee

Now that we have discussed the generic structure of operators that contribute to the opposite-helicity Compton amplitude, we move on to studying a specific class of such operators that are connected to the \AHH amplitudes in the next section.

\section{Quantum Compton amplitudes for $\sqrt{\text{Kerr}}$}
\label{sec:QuantCompAmps}

In this section, we will present our final formulae for the opposite-helicity spin-$s$ Compton amplitudes in the $\sqrt{\text{Kerr}}$ theory, which combine the results obtained in \secs{sec:Spin1}{sec:HigherSpinGauge} using the principle of massive gauge symmetry~\cite{Zinoviev:2001dt}, and those obtained in \sec{sec:chiralform} using the chiral formalism~\cite{Ochirov:2022nqz}.
In addition, we constrain our Compton amplitudes using ans\"{a}tze based on observed patterns and classical-limit analysis. The latter is considered in~\sec{sec:ClassCompAmps}, whereas in this section we focus on the quantum Compton amplitudes.  

In order to set the stage, we quote a chiral Lagrangian that is fully compatible with the final spin-$s$ Compton amplitudes, given later in this section. That is, the $\sqrt{\text{Kerr}}$ theory that we choose to work with in this section is 
\begin{align}
\label{FinalRootKerrLagrangian}
{\cal L} & = \braket{D_\mu \Phi|D^\mu \Phi} - m^2 \braket{\Phi|\Phi}
 + \sum_{k=0}^{2s-1} \frac{ig}{m^{2k}}
   \bra{\Phi} \Big\{
   |\overset{\leftarrow}{D}|\overset{\rightarrow}{D}|^{\odot k}\!\odot |\Fm|
   \Big\} \ket{\Phi} + {\cal O}(|\Fm|^2) \\* &
 -\!\!\sum_{k \le l=0}^{2s-4}\!\sum_{j=0}^{2s-3-l}\!\!\!\!
   \frac{g^2}{m^{2(j+l)+6}} \bra{\Phi} \Big\{\!
   \big(|\overset{\leftarrow}{D}|\overset{\rightarrow}{D}|{+} m^2 \big)\!\odot\!
   |\overset{\leftarrow}{D}|\overset{\rightarrow}{D}|^{\odot j}\!\odot\!
   |\overset{\leftarrow}{D}|\overset{\rightarrow}{D}_+|^{\odot k}\!\odot\!
   |\overset{\leftarrow}{D}_+|\overset{\rightarrow}{D}|^{\odot(l-k)}\!\odot\!
   \FF_6\!
   \Big\} \ket{\Phi} \,. \nn
\end{align}
The second line involving $\FF_6 = \frac{1}{4} \{ T^c, T^{c'} \} |F_-^c|\!\odot |\overset{\leftarrow}{D}|F_+^{c'}|\overset{\rightarrow}{D}|$ corresponds to a chosen non-minimal ${\cal L}_{4}$ completion in the opposite-helicity sector, which we will motivate in the following.
The fully negative-helicity sector needs further operators ${\cal O}(|\Fm|^2)$ to restore parity, however, we will not need them in this paper.

While we have shown in \sec{sec:HigherSpinGauge} that such principles as massive gauge symmetry, parity and power counting, among others, can be used to fix the cubic interactions uniquely, the contact-term freedom in the opposite-helicity Compton amplitudes, or in the ${\cal L}_{4}$ interactions, is substantially more difficult to pin down from first principles.  Nevertheless, in the discussion below we will make a heuristic choice and commit to explicit contact terms that are simple and give consistent classical properties, discussed in detail in \sec{sec:ClassCompAmps}. Other choices of contacts terms are possible, which we will briefly discuss later.

\subsection{Abelian Compton amplitude}
\label{sec:abAmp}

We now consider the spin-$s$ Compton amplitudes for the abelian $\sqrt{\text{Kerr}}$ theory.
They can be obtained by plugging in U(1) generators $T^c=1/\sqrt{2}$, and for convenience also setting $g^2=2Q^2=-1$, in the chiral Lagrangian~\eqref{FinalRootKerrLagrangian}.

After some non-trivial work, the family of spin-$s$ abelian Compton amplitudes in the opposite-helicity sector can be written in a compact and manifestly local form:
\begin{align}
\label{abGTCompton}
A_{\rm U(1)}(\bm{1}^s\!,\bar{\bm{2}}^s\!,3^-\!,4^+) &
 = \frac{\bra{3}1|4]^2 (U + V)^{2s}}{m^{4s} t_{13} t_{14}}
 - \frac{\braket{\bm{1}3}\bra{3}1|4][4\bm{2}]}{m^{4s} t_{13}} P_2^{(2s)}
 + \frac{\braket{\bm{1}3} \braket{3\bm{2}} [\bm{1}4] [4\bm{2}]}{m^{4s}}
   P_2^{(2s-1)} \nn \\* & \quad\,
 - \frac{\braket{\bm{1}3} \braket{3\bm{2}} [\bm{1}4] [4\bm{2}]}{m^{4s-2}}
   \braket{\bm{12}} [\bm{12}] P_4^{(2s-1)}
 + C^{(s)} \,.
\end{align}
The quadratic and cubic part of the chiral Lagrangian~\eqref{FinalRootKerrLagrangian},
give rise to the first four terms above, of which the first three terms were obtained already in \rcite{Cangemi:2022bew}.
The additional contact term~$C^{(s)}$ accounts for non-minimal quartic interactions, and below we will find an expression that explains the choice made in \eqn{FinalRootKerrLagrangian}.

\paragraph{Notation.}
Let us explain the notation used in \eqn{abGTCompton}. $P_n^{(k)}$ are local polynomials that are central to our presentation of the amplitude. They are functions of the spin-dependent variables $U,V,\braket{\bm{12}}, [\bm{12}]$, where we recall that
\be
U = \frac{1}{2 }\big(\bra{\bm{1}}4|\bm{2}] - \bra{\bm{2}}4|\bm{1}]\big)
  - m[\bm{12}] \,, \qquad \quad
V = \frac{1}{2}\big(\bra{\bm{1}}4|\bm{2}] + \bra{\bm{2}}4|\bm{1}]\big) \,.
\ee
A priori, there is little reason to expect that the polynomials that multiply the helicity-dependent factors in \eqn{abGTCompton} have a simple structure, given the intricate form of the amplitude in~\eqn{OriginalAHHCompton}.
However, from explicit calculations at fixed $s$, we find that the $P_n^{(k)}$ polynomials are exceedingly simple when expressed in appropriate variables.  

For integers $n>0$ and $k\ge0$, we define\footnote{For $k<0$ we define $P_n^{(k<0)}=0$. Note that $P_n^{(k)}$ appears in the mathematics literature under the name ``complete homogeneous symmetric polynomials'', although we do not make use of any established property.
}
the symmetric homogeneous polynomials
\be
\label{PolyDef}
P_n^{(k)}
:= \sum_{i=1}^n \frac{\vs_i^k}{\prod_{j \neq i}^n (\vs_i-\vs_j)}
 = \frac{\vs_1^k}{(\vs_1 - \vs_2)(\vs_1 - \vs_3) \cdots (\vs_1 - \vs_n)}
 + \text{perms} \,,
\ee
where ``$+\,\text{perms}$'' is shorthand for inequivalent terms obtained by permuting the $\vs_i$ variables.
The appearance of denominators is innocuous, since one can easily check that the residue of a potential pole $1/(\vs_i - \vs_j)$ is always zero, hence they are polynomials of degree $k-n+1$.

For the Compton amplitude~\eqref{abGTCompton}, we only need $P_{n\le 4}^{(k)}$ that depend on up to four variables, and we globally identify them as
\beal
\label{VarSigmaDef}
\vs_1 & := U + V = \bra{\bm{1}}4|\bm{2}] - m[\bm{12}] \,, \qquad~\;\,\quad
\vs_3 := -m \braket{\bm{12}} \,, \\*
\vs_2 & := U - V = - \bra{\bm{2}}4|\bm{1}] - m [\bm{12}] \,, \qquad \quad
\vs_4 := -m [\bm{12}] \,.
\eeal
To be clear, let us write some of the relevant polynomials in various forms:
\beal
\label{PolyExpl}
P_1^{(2s)} & = \vs_1^{2s} = (U+V)^{2s} \,, \\ 
P_2^{(2s)} & = \frac{\vs_1^{2s}}{\vs_1-\vs_2}+\frac{\vs_2^{2s}}{\vs_2-\vs_1}
 = \frac{(U+V)^{2s}- (U-V)^{2s}}{2 V}
 =\!\sum_{i+j = 2s-1}\!\vs_1^i \vs_2^j \,, \\
P_4^{(2s-1)} & =\!\sum_{i+j+k+l=2s-4}\!\vs_1^i \vs_2^j \vs_3^k \vs_4^l \,.
\eeal
In particular, note that the $(U+V)^{2s}$ factor in \eqn{abGTCompton} is simply the one-variable polynomial~$P_1^{(2s)}$, so every term in the Compton amplitude is multiplied by some polynomial~$P_n^{(k)}$.
Indeed, this observation, which was already made in \rcite{Cangemi:2022bew} for the case $n=2$, is crucial for constraining the contact term~$C^{(s)}$.
The ubiquitous appearance of these polynomials motivates us to conjecture that:
\begin{quote}
\emph{the complete spin-$s$ Compton amplitude for $\sqrt{\text{Kerr}}$ theory should be a finite superposition of only the symmetric homogeneous polynomials $P_n^{(k)}$}.
\end{quote}

\paragraph{Constraints.}
Therefore, to determine $C^{(s)}$ we assume that it is a linear combination of these polynomials:
\be \label{eq:C(s)ansatz}
C^{(s)}= C^{(s)}[P^{(k)}_{n}] \,.
\ee
Furthermore, we impose the following heuristic constraints on the complete amplitude:
\begin{itemize}
\item well-behaved classical limit  $s \to \infty$;
\item amplitudes with $s<2$ not modified: $C^{(s<2)}=0$\,;
\item compatible with massive higher-spin gauge invariance;
\item $s$-independent numerical coefficients;
\item parity invariance imposed, see \eqn{ParityConstraintOrdered};
\item all contact terms have spinor-helicity structure  $\sim \braket{\bm{1}3} \braket{3\bm{2}} [\bm{1}4] [4\bm{2}]$;
\item classical spin quadrupole fixed by $s=1$ amplitude;
\item no dissipation effects, nor contributions from non-perturbative considerations.
\end{itemize}
We find that the simplest contact term $C^{(s)}$ that satisfies all of the above constraints is
\be
\label{eq:abContact}
C^{(s)}
 = -\frac{\braket{\bm{1}3} \braket{3\bm{2}} [\bm{1}4] [4\bm{2}]}{2 m^{4s-1}}
   (\braket{\bm{12}} + [\bm{12}]) \Big(P_4^{(2s)}-P_2^{(2s-2)} \Big) \,.
\ee
The overall normalization is fixed by the classical limit, which will be further discussed in \sec{sec:ClassCompAmps}.
The relative normalization of the two $P^{(k)}_{n}$ is fixed by the requirement that for $s=3/2$ they cancel each other out.
The appearance of the overall $\braket{\bm{12}}+ [\bm{12}]$ factor comes from compatibility with the massive higher-spin gauge invariance, specifically, the $s = 2$ case in \eqn{eq:abContact} matches the $c_1$ contact term in \eqn{eq:s2abContacts}. 
The absence of dissipation effects is equivalent to considering hermitian (or CPT-invariant) interaction terms, corresponding to imposing exchange symmetry of the massive states, $1 \leftrightarrow 2$, on the contact term $C^{(s)}$. While this constraint appears compulsory for a QFT, it can be relaxed for the purpose of describing irreversible processes, such as classical dissipation of the Kerr Compton amplitude~\cite{Bautista:2022wjf}.
Similarly, here we will not consider terms analogous to polygamma contributions that appeared in concert with non-perturbative contributions in the Kerr case of \rcite{Bautista:2022wjf}.

Note that while the third and fourth terms of the amplitude \eqref{abGTCompton} are also contact terms, they originate from the cubic part of the action, as discussed around \eqn{ContactMatrixElement}.
Furthermore, they are related to pole terms of the non-abelian $\sqrt{\text{Kerr}}$ amplitude via the Kleiss-Kuijf relation, as will be discussed in the next subsection.

The contact term $C^{(s)}$ has the same helicity structure, $\braket{\bm{1}3} \braket{3\bm{2}} [\bm{1}4] [4\bm{2}]$, as the third and fourth terms in the amplitude~\eqref{abGTCompton}, which is no accident, as the classical limit is not well-behaved without such a contribution.
Specifically, the fourth term in~\eqn{abGTCompton} generates a divergence in the limit $s \to \infty$ that needs to be canceled\footnote{Alternative contact terms can be found that satisfy all the desired properties, with identical classical limit as $C^{(s)}$. A simple alternative such choice is
\be
-\frac{\braket{\bm{1}3} \braket{3\bm{2}} [\bm{1}4] [4\bm{2}]}{m^{4s-2}} \braket{\bm{12}} [\bm{12}] \Big(2P_4^{(2s-1)}
 + \frac{m}{2}\big(\braket{\bm{12}} + [\bm{12}]\big)P_4^{(2s-2)} \Big) \,. \nn
\ee
}
by $C^{(s)}$, as we will discuss in \sec{sec:OppositeHelicity}. 
We also checked that in the  $s \to \infty$ limit our choice of contact term give rise to a spin quadrupole moment $\sim S^2$ that matches the corresponding spin-1 result, which is the preferred choice in the literature~\cite{Aoude:2022trd}. However, in principle this condition can be relaxed independently of the other constraints, as discussed in 
\sec{sec:spinuniversality}.

It is possible to add further contact terms that have other helicity structures, but given our current limited analysis, we do not see the need for such terms. However, if dissipation effects were to be taken seriously, for example from analysis of some putative classical $\sqrt{\text{Kerr}}$ system, then additional helicity structures would likely be needed. 

\begin{table}
\centering
\begin{tabular}{|c||c|c|c|c|c|c|c|c|} 
 \hline
$s$ & 0 & \!$\,^1\!/\!_2$\! & $1$ & $\,^3\!/\!_2$ & $2$ & $\,^5\!/\!_2$ \\
 \hline
 \hline
$P^{(2s)}_2$ & 0 & $1$ & \!$2U$\!  & \!$3 U^2 {+} V^2$\! & \!$4 U (U^2 {+} V^2)$\! & 
\!$5 U^4 {+} 10 U^2 V^2 {+} V^4$\! \\ [0.5ex]   \hline
\!$P^{(2s-1)}_2$\! & 0  & $0$ & $1$ & $2U$  & $3 U^2 {+} V^2$ & $4 U (U^2 {+} V^2)$  \\ [0.5ex] 
 \hline
\!$P^{(2s-1)}_4$\! & 0 & $0$ & $0$ & $0$ & $1$ & $
2(U {+} W_{+})$ \\ [0.5ex] 
 \hline
$P^{(2s)}_4-P^{(2s-2)}_2$\! & 0 & $0$ & $0$ & $0$ & $-2 W_{+}$ & $W_{-}^2 - 4 U W_{+} + 3 W_{+}^2$  \\ [0.5ex] 
 \hline
\end{tabular}
\caption{Low-spin examples of how the polynomials $P_n^{(2s)}$ contribute to the abelian spin-$s$ Compton amplitude. Last two rows contribute starting at $s=2$, compatible with the analysis of \rcite{Cangemi:2022bew} Here $W_\pm= \frac{m}{2} (\braket{\bm{12}} \pm [\bm{12}])$.}
\label{tab:Polynomials}
\end{table}

\paragraph{Checking contact term.}
We can now check that the ${\cal L}_4$ contact term in the chiral Lagrangian \eqref{FinalRootKerrLagrangian} has the correct matrix elements to reproduce the $C^{(s)}$ contact term.
Adapting \eqn{ContactTerm6}, we find that the matrix element is given by
\begin{align}
\sum_{j,k,l} \frac{1}{m^{2(j+l)+6}} & \bra{\Phi} \Big\{\!
   \big(|\overset{\leftarrow}{D}|\overset{\rightarrow}{D}|{+} m^2 \big)\!\odot\!
   |\overset{\leftarrow}{D}|\overset{\rightarrow}{D}|^{\odot j}\!\odot\!
   |\overset{\leftarrow}{D}|\overset{\rightarrow}{D}_+|^{\odot k}\!\odot\!
   |\overset{\leftarrow}{D}_+|\overset{\rightarrow}{D}|^{\odot(l-k)}\!\odot\!
   \FF_6\!
   \Big\} \ket{\Phi} \Big|_{(2,3^-\!,4^+\!,1)} \nn \\*
=\,& \frac{\braket{\bm{1}3} \braket{3\bm{2}} [\bm{1}4] [4\bm{2}]}{2m^{4s}}
   \sum_{k \le l=0}^{2s-4}\!\sum_{j=0}^{2s-3-l}\!(\vs_3+\vs_4)
   \vs_1^k \vs_2^{l-k} \vs_3^{2s-j-l-3} \vs_4^j \,,
\label{ContactTermComputation}
\end{align}
where we used the shorthand notation~\eqref{VarSigmaDef} for the spin-dependent variables~$\vs_i$.
The sum can be further rewritten in terms of the promised combination of polynomials:
\be
\label{ContactTermSum}
(\vs_3+\vs_4) \sum_{k \le l=0}^{2s-4}\!\sum_{j=0}^{2s-3-l}\!
   \vs_1^k \vs_2^{l-k} \vs_3^{2s-j-l-3} \vs_4^j
 = -m(\braket{\bm{12}} + [\bm{12}]) \Big(P_4^{(2s)}-P_2^{(2s-2)} \Big) \,.
\ee
This matrix-element contribution should be further multiplied by $g^2 \{ T^{c_3}, T^{c_4} \}$,
which in the abelian amplitude we replace by $2Q^2=-1$, thus matching \eqn{eq:abContact}.

\begin{figure}[t]
\centering
\includegraphics[width = 0.8\textwidth,trim=0 0 10pt 0, clip=true]{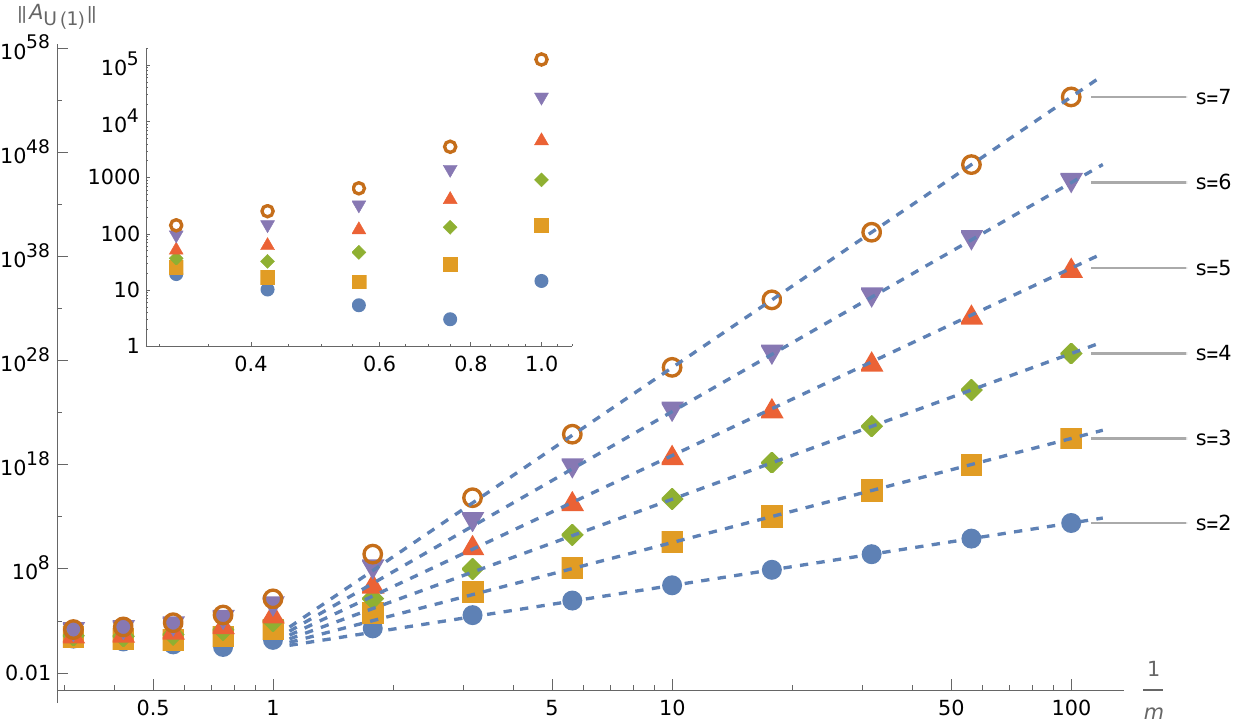}
\vspace{-3pt}
\caption{In the massless limit, the abelian opposite-helicity Compton amplitude~\eqref{abGTCompton} grows as $1/m^{4s-2}$.
Its same-helicity counterparts~\eqref{AmpGaugeAHH4idh} have a tamer, explicitly $1/m^{2s-2}$, dependence and are thus not shown.
The data were obtained for the kinematics
$t_{13}=2$, $t_{14}=-1$, $s_{12} = -1$, but the plots look similar for other generic kinematic points, as well as for the non-abelian amplitude~\eqref{eq:nabGaugeAmp}.
The dashed straight lines are exact power-law functions $\propto 1/m^{4s-2}$. The inset shows the zoomed region of ${\cal O}(1)$ masses, where the mass dependence is no longer described by a power law.
}
\label{fig:mpMassScaling}
\end{figure}

Note that the maximum derivative count of the chosen contact-interaction~${\cal L}_4$ is $\frac{\partial^{4s}}{m^{4s}}$,
as also seen on the left-hand side of \eqn{ContactTermComputation}.
However, all terms in the sum~\eqref{ContactTermSum} have at least two powers of $\vs_3$ or $\vs_4$ (each proportional to $m$), which reduce the maximum derivative count down to $\frac{\partial^{4s-2}}{m^{4s-2}}$.
This agrees with our expectation from the pole contributions to the opposite-helicity Compton amplitude.
Indeed, the minus-helicity cubic interaction in the Lagrangian~\eqref{FinalRootKerrLagrangian} contains $4s{-}1$ derivatives,
its plus-helicity counterpart has the minimal one derivative,
and the massive propagator reduces the number of derivatives by two.
The resulting derivative count corresponds to the leading divergence at high energies, which can be probed numerically by taking $m \to 0$ with fixed momentum invariants.
The amplitude's behavior in such a limit is illustrated in \fig{fig:mpMassScaling}.

As a summary of the abelian calculation, we note that the chiral Lagrangian \eqref{FinalRootKerrLagrangian} gives abelian amplitudes that reproduce the established results for $s \le 3/2$ in the literature.
For spins  $0 \le s \le 1$, the amplitudes agree with those constructed from factorization properties in \rcites{Arkani-Hamed:2017jhn,Johansson:2019dnu}, and for $s=3/2$ the amplitude \eqref{abGTCompton} reproduces the one obtained in \rcite{Chiodaroli:2021eug} using the current constraint~\eqref{CurrentConstraintEqn}.
At higher spins, $s \geq 2$, the first three terms of \eqn{abGTCompton} were obtained in \rcite{Cangemi:2022bew}, but it was suggested that new terms would be needed to restore massive gauge symmetry.
We have now seen that the cubic chiral Lagrangian~\eqref{ActionGaugeAHH} automatically gives the fourth term, while the fifth ``true contact'' term $C^{(s)}$ comes from the chosen quartic interaction, both of which start to contribute at $s=2$, as expected.
\Tab{tab:Polynomials} shows how the relevant polynomials $P_n^{(2s)}$ contribute to the low-spin abelian amplitudes.

\subsection{Non-abelian Compton amplitude}
\label{sec:nonabAmp}

Here we consider the non-abelian spin-$s$ Compton amplitude generated by the chiral Lagrangian \eqref{FinalRootKerrLagrangian}.
Let us focus on the color-stripped amplitude that multiplies the color factor~$2g^2 T^{c_3} T^{c_4}$.
We have found a compact and manifestly local form for it:
\begin{align}
\label{eq:nabGaugeAmp}
A(\bm{1}^s\!,\bar{\bm{2}}^s\!,3^-\!,4^+) &
 = \frac{\bra{3}1|4]^2 (U + V)^{2s}}{m^{4s} s_{12} t_{14}}
 - \frac{\braket{\bm{1}3} \bra{3}1|4] [4\bm{2}] }{m^{4s} s_{12}} P_2^{(2s)}
 + \frac{\braket{\bm{1}3} \braket{3\bm{2}} [\bm{1}4] [4\bm{2}]}{m^{4s} s_{12}} t_{13} P_2^{(2s-1)} \nn \\* & \quad\,
 - \frac{\braket{\bm{1}3} \braket{3\bm{2}} [\bm{1}4] [4\bm{2}]}{m^{4s-2} s_{12}}
   \big(t_{13} \braket{\bm{12}} [\bm{12}] P_4^{(2s-1)}\!+ 2 V P_4^{(2s)}\big)
 - \frac{1}{2} C^{(s)} \,,
\end{align}
where the polynomials $P_n^{(k)}$ and the contact term $C^{(s)}$ are identical to the ones given for the abelian amplitude, see \eqns{PolyExpl}{eq:abContact}, respectively.

If $C^{(s)}$ is omitted, the amplitude~\eqref{eq:nabGaugeAmp} matches numerically the presentation~\eqref{OriginalAHHCompton} of the same amplitude.
However, the contact term is necessary for consistency between the abelian and non-abelian results, as expressed by a Kleiss-Kuijf relation~\cite{Kleiss:1988ne}:
\be
A_{\rm U(1)}(\bm{1}^s\!,\bar{\bm{2}}^s\!,3^-\!,4^+)
 ={-}A(\bm{1}^s\!,\bar{\bm{2}}^s\!,3^-\!,4^+)
   - A(\bm{1}^s\!,\bar{\bm{2}}^s\!,4^+\!,3^-) \,.
\ee
The two orderings on the right-hand side are related by the relabeling property~\eqref{Relabeling}, which follows from the assumed consistency of the amplitude in the adjoint representation of ${\rm SU}(N_\text{c})$.
Since the contact term~\eqref{eq:abContact} is symmetric under the same relabeling $1 \leftrightarrow 2$, it contributes to each of the two orderings equally, with a factor of $-1/2$.
One could consider including further contact terms that are odd under this $\mathbb{Z}_2$ symmetry, in which case they would not contribute to the abelian amplitude, only the non-abelian one.
Currently, we have no good reason to include such terms, but neither can we exclude their existence.
In particular, the consistency of the classical limit for the non-abelian amplitude does not require additional contact terms.
In fact, as will be discussed in \sec{sec:contacttermbreaking}, contact terms that are odd under this $\mathbb{Z}_2$ symmetry will not contribute classically, unless some of the assumptions used for $C^{(s)}$ in \sec{sec:abAmp} are relaxed. 

Note that the above non-abelian ordered amplitude~\eqref{eq:nabGaugeAmp} matches the known $\sqrt{\text{Kerr}}$ results for $s\leq 1$~\cite{Arkani-Hamed:2017jhn,Johansson:2019dnu}, obtained by the on-shell sewing of \AHH three-point amplitudes.
It also matches the non-abelian $s=3/2$ amplitude that can be computed from the cubic spin-3/2 Lagrangian given in \rcite{Chiodaroli:2021eug}, and this is compatible with the absence of a non-minimal four-point contact term $C^{(s\leq 3/2)}=0$.
For $s=2$, the amplitude is consistent with the analysis of massive gauge invariance done in \rcite{Cangemi:2022bew} and in \sec{sec:massivegaugequarticSpin2}.

It might be helpful to display the full amplitude assembled from \eqns{abGTCompton}{eq:nabGaugeAmp}:
\begin{align}
\label{eq:ColorDressedCompton}\!\!
{\cal A}(\bm{1}^s\!,\bar{\bm{2}}^s\!,3^-\!\!,4^+) = 2g^2\!\Bigg\{\!
   \bigg[ \frac{T^{c_3} T^{c_4}\!}{s_{12} t_{14}}
       {+}\frac{T^{c_4} T^{c_3}\!}{s_{12} t_{13}} \bigg]\!
   \bigg[ \frac{\bra{3}1|4]^2\!}{m^{4s}} (U\!+\!V)^{2s}
{-}\frac{\braket{\bm{1}3} \bra{3}1|4] [4\bm{2}]}{m^{4s}}
   t_{14} & P_2^{(2s)} \\*
+\,\frac{\braket{\bm{1}3} \braket{3\bm{2}} [\bm{1}4] [4\bm{2}]}{m^{4s}}
   t_{13} t_{14} \big( P_2^{(2s-1)}\!
                     - m^2 \braket{\bm{12}} [\bm{12}] & P_4^{(2s -1)} \big)
   \bigg] \nn \\*
-\,\frac{\braket{\bm{1}3} \braket{3\bm{2}} [\bm{1}4] [4\bm{2}]}{m^{4s-2}} 
   \bigg( \frac{[T^{c_3}\!,T^{c_4}]}{s_{12}} 2 V P_4^{(2s)}\!
        - \frac{\{T^{c_3}\!,T^{c_4}\}}{4m}
          \big( \braket{\bm{12}} {+} [\bm{12}] \big)
          \big( P_4^{(2s)}\!- & P_2^{(2s-2)} \big)\!
   \bigg)\!\Bigg\} \,. \nn
\end{align}
This is the complete opposite-helicity spin-$s$ Compton amplitude that can be computed from the chiral Lagrangian~\eqref{FinalRootKerrLagrangian}.
In particular, the $\{T^{c_3}\!,T^{c_4}\}$ contribution in the last line corresponds to the contact term $C^{(s)}$ that follows from the chosen ${\cal L}_4$ interaction.

As can be deduced from the factorization of color and kinematic structures in \eqn{eq:ColorDressedCompton}, the first two lines of the amplitude obey the BCJ partial-amplitude relations~\cite{Bern:2008qj,Johansson:2015oia},
\be
\left[
\begin{aligned}
A(\bm{1}^s\!,\bar{\bm{2}}^s\!,4^+\!,3^-)\!&
 = \tfrac{t_{14}}{t_{13}}A(\bm{1}^s\!,\bar{\bm{2}}^s\!,3^-\!,4^+) \\
A_{\rm U(1)}(\bm{1}^s\!,\bar{\bm{2}}^s\!,3^-\!\!,4^+) &
 = \tfrac{s_{12}}{t_{13}}A(\bm{1}^s\!,\bar{\bm{2}}^s\!,3^-\!\!,4^+)
\end{aligned}
\right]_\text{lines 1-2 of \eqn{eq:ColorDressedCompton}} ,
\ee
whereas the third line, containing the commutator and anticommutator contributions, does not fully adhere to these relations.
The third line contributes for $s \ge 3/2$ and, curiously, still exhibits an enhanced behavior in the classical limit that partially satisfies\footnote{The third line of \eqn{eq:ColorDressedCompton} contains superclassical contributions that obey the BCJ relations, and classical contributions that violate the BCJ relations.
}
the BCJ amplitude relations.
As already alluded to, further contact terms could in principle be added to the amplitude~\eqref{eq:ColorDressedCompton} that might enhance some desirable properties.
However, there are no obvious additional contact terms that would fully restore the BCJ relations, either for the quantum Compton amplitude or in the classical limit.
Any attempt to restore the BCJ relations would also modify the factorization channels and thus implicitly modify the three-point amplitudes for spins $s \ge 3/2$, as already discussed in \rcite{Johansson:2019dnu}.  

As discussed, the constraints from the classical limit inform us about the abelian contact terms, which multiplies the anticommutator $\{T^{c_3},T^{c_4}\}$. One may wonder if a similar analysis would require new contact terms proportional to the commutator $[T^{c_3},T^{c_4}]$.
However, in our classical-limit analysis we find that such contact terms will not contribute to the classical amplitude, unless their numerical coefficients violate our assumptions and grow as a function of the spin eigenvalue $s$, see \sec{sec:contacttermbreaking}.

\section{Classical Compton amplitudes for $\sqrt{\text{Kerr}}$}
\label{sec:ClassCompAmps}

A major reason for studying the $\sqrt{\text{Kerr}}$ theory is that it exhibits interesting behavior in the classical limit, corresponding to a charged rotating ring (or disk)~\cite{Monteiro:2014cda, Arkani-Hamed:2019ymq,Guevara:2020xjx}, in close analogy with the ring singularity of a classical Kerr black hole. We thus proceed to analyze the classical limit of our Compton amplitude.

\subsection{Classical kinematics and variables}
\label{sec:classVars}

Classical Compton scattering corresponds to a process where the massive particle is only slightly deflected and the radiation (massless gauge bosons) is soft.
One can extract classical physics from the scalar quantum Compton amplitudes by parameterizing Compton momenta and kinematic variables such that, when $\hbar\to 0$, they behave as
\cite{Kosower:2018adc}
\be \label{eq:clMomScaling}\!\!
p^\mu := p_1^\mu \sim 1 \,, \quad~
q^\mu := (p_3 + p_4)^\mu \sim \hbar \,, \quad~
q_\perp^\mu := (p_4 - p_3)^\mu \sim \hbar \,, \quad~
\chi^\mu := \bra{3}\sigma^\mu|4] \sim \hbar \,,  
\ee
and the independent Lorentz invariants behave as 
\be \label{ClassicalLimitScaling1}
p^2 = m^2 \sim 1 \,, \qquad
q_\perp^2 = 2 p \cdot q = -q^2 \sim \hbar^2 \,, \qquad
p \cdot q_\perp \sim \hbar \,, \qquad
p \cdot \chi \sim \hbar \,.
\ee

In the classical limit, we can identify the physical parameters $\omega, \, \theta$ as follows:
\be
\omega = \frac{p \cdot q_\perp}{2m}\,, ~~~~~~ \cos\theta = 1+\frac{q^2}{2\omega^2} = 1+ \frac{2}{\xi}\,,
\ee
where $\omega$ is the energy of the radiation in the rest frame of the $\sqrt{\rm Kerr}$  object, and $\theta$ is the deflection angle of the radiation, due to the scattering with the massive object. The variable $\xi=(p \cdot q_\perp)^2/(m^2q^2)$ is known as the optical parameter.

The classical limit reduces the non-spinning color-dressed amplitude in \eqn{eq:ColorDressedCompton} to
\beal
\label{eq:scalarClCompton}
{\cal A}(\bm{1}^{s=0}\!,\bm{2}^{s=0}\!,3^-\!,4^+) & = \bra{3}1|4]^2
   \bigg( \frac{T^{c_3} T^{c_4}}{s_{12} t_{14}}
        + \frac{T^{c_4} T^{c_3}}{s_{12} t_{13}} \bigg) \\ &
 = (p\cdot\chi)^2  \bigg( \frac{[T^{c_3},T^{c_4}]}{q^2 (p \cdot q_\perp)}
 + \frac{1}{2}  \frac{\{T^{c_3},T^{c_4}\}}{(p \cdot q_\perp)^2} \bigg)
 + {\cal O}(\hbar) \,.
\eeal
The color factors are best expressed in terms of the commutator and anticommutator of the gauge-group generators, as the corresponding kinematic factors scale differently.
In the classical limit, the kinematic factor multiplying the commutator scales as ${\cal O}(\hbar^{-1})$, while the factor of the anticommutator scales as ${\cal O}(\hbar^0)$.
So in order to obtain a proper classical limit containing both term, we assume that the commutator and anticommutator themselves scale as
\begin{align} \label{ClassicalLimitScaling2}
[T^{c_3},T^{c_4}] \sim \hbar\,, \qquad
\{T^{c_3},T^{c_4}\} \sim 1\,.
\end{align}
The above limit corresponds to considering large representations of the gauge group, in which case the commutators become suppressed compared to the anticommutators. This is the appropriate limit for classical color charge (see \eg \rcite{delaCruz:2020bbn}).

Next we consider classical spin, which is a vector quantity $S^\mu=m a^\mu$.
We will mostly make use of the transverse spacelike vector~$a^\mu$, $p\cdot a=0$, and call it a ring radius in analogy with that of a Kerr BH, for which its magnitude $|a|=\sqrt{-a^2}$ corresponds to the size of the ring singularity.
In the classical limit, we assume that the ring radius is a macroscopic length, so in Planck units it must scale inversely with $\hbar$:
\be \label{ClassicalLimitScaling3}
a^\mu \sim \frac{1}{\hbar}\,.
\ee
This is another way of saying that $|a| \approx \hbar s/m$ is finite while $\hbar \to 0$, so $s\to\infty$, but without restoring expicit $\hbar$'s anywhere.

The classical helicity-conserving Compton amplitude can be written as an entire function in the following four spin-dependent,  helicity-independent and dimensionless variables 
\beal \label{eq:clComptonVars}
x & = a \cdot q_\perp \,, \qquad\;\,\qquad
y = a \cdot q\,, \\
z & = |a| \frac{p \cdot q_\perp}{m} \,, \qquad \quad
w = \frac{a \cdot \chi\;p \cdot q_\perp}{p\cdot\chi} \,.
\eeal
The $z$ variable is sometimes called the spheroidicity parameter~\cite{Dolan:2008kf}\footnote{In our convention, $z=2\omega|a|$ has an extra factor of two compared to \rcite{Dolan:2008kf}.} and is not to be confused with the SU(2) little-group wavefunction, for which we use the same letter.  

In addition to the four spin-dependent classical variables that can appear to any power, the classical Compton amplitude can be a non-trivial function of a single dimensionless kinematic variable, which we take to be the optical parameter $\xi$.
However, we need to account for the Gram-determinant relation, $G(p,q,q_\perp,a,\chi)=0$, which in the classical limit gives a relation between the optical parameter and the above spin-dependent variables
\be \label{eq:clGramDet}
\xi^{-1}=\frac{m^2 q^2}{(p \cdot q_\perp)^2} =\frac{(w - x)^2 - y^2}{z^2-w^2}\,.
\ee
In principle, the optical parameter $\xi$ can be eliminated, however we choose to include $\xi$ in our ansatz to avoid introducing spurious poles in the spin variables $x,y,z,w$. Thus we allow for inverse powers $\xi^{-I}$, with integer $I\ge0$, since the combination $z^{2I}\xi^{-I}=(-a^{2} q^{2})^I$ is local.

At intermediate steps we will encounter Levi-Civita tensors contracted with four of the five vectors $p^\mu$, $q^\mu$, $q_\perp^\mu$, $a^\mu$, $\chi^\mu$. Since the ring radius is a pseudo-vector, certain contractions involving it can be reduced to simple dot products.
For example, we have the following useful relations in the classical limit:\footnote{We denote $\epsilon(p, q, q_\perp,a):=\epsilon_{\mu\nu\rho\sigma}p^\mu q^\nu q_\perp^\rho a^\sigma$, and the Levi-Civita tensor is normalized as $\epsilon^{0123}=1$.
}
\beal \label{eq:clLeviCivitaIds}
\frac{i \epsilon(p, q, q_\perp,a)}{p \cdot q_\perp }& = w-x+\frac{w}{\xi} = x+\frac{1}{w}\Big(y^2-x^2 +\frac{z^2}{\xi}\Big) \,, \\
\frac{i \epsilon(p, q, \chi,a)}{p \cdot \chi } &= w-x\,,
\eeal
where we divided by appropriate dot products so as to make the right-hand sides have classical scaling and no helicity dependence.  

Taking into account all of the above classical variables we can now assemble a generic ansatz for a classical tree-level amplitude that is broad enough to capture every physical theory. A helicity-conserving Compton amplitude for a classical spinning non-abelian massive object should have the form 
\beal \label{GenericClassicalAmplitude}
{\cal A}(\bm{1},\bm{2},3^-\!,4^+) &
= [T^{c_3},T^{c_4}] \frac{(p \cdot \chi)^2}{q^2 (p \cdot q_\perp)} \Big(\genE(x,y,z,w) + \sum_{I=1}^{\infty} z^{2I}\xi^{-I} \genE_I(x,y,z,w)  \Big) \\ & 
+\frac{1}{2} \{T^{c_3},T^{c_4}\} \frac{(p \cdot \chi)^2 }{(p \cdot q_\perp)^2}\Big(\tilde \genE(x,y,z,w) + \sum_{I=1}^{\infty} z^{2I}\xi^{-I} \tilde \genE_I(x,y,z,w)  \Big) \,,
\eeal
where the $\genE, \tilde\genE, \genE_I, \tilde\genE_I$ are general functions that must be analytic at the origin (\ie have a Taylor series expansion around the spinless case). In fact, based on the structure of our quantum amplitudes, we expect that the functions must be analytic everywhere in the complex plane $\mathbb{C}^4$, hence they are entire functions. 

The requirement of locality (absence of spurious poles) gives the following refinements for the $\genE$'s
\be
\genE(x,y,z,w) = \genE(x,y,z,0) + w \genE'(x,y,z,0) + \frac{w^2}{2} \genE''(x,y,z,0) \,. 
\ee
and there is a analogous three-term expansion for $\tilde \genE$.
The reason that $w$ can appear at most quadratically is that it contains a spurious pole in the variable $p\cdot\chi$, which can be canceled by the same overall helicity factor of the amplitude~\eqref{GenericClassicalAmplitude}.

For the $\genE_I$'s (and $\tilde \genE_I$'s), the analogous locality constraint combined with the Gram determinant constraint gives the refinement\footnote{In principle, the second term could admit the form $w \genE'_I(x,y,z,0)\rightarrow \frac{w}{z} \genE'_I(x,y,z,0)$, since the non-locality cancels against the positive $z$ powers in~\eqn{GenericClassicalAmplitude}, however, such odd-$z$ powers can only appear in time-reversal asymmetric amplitudes (dissipation).}
\be
\genE_I(x,y,z,w) = \genE_I (x,y,z,0) + w \genE'_I(x,y,z,0) \,.
\ee
The reason we only expand to linear order in $w$ is that the Gram-determinant relation \eqref{eq:clGramDet} allows us to reduce any quadratic term $\sim w^2\xi^{-1}$.
Further locality constraints can be obtained by considering the absence of the simultaneous poles in $p\cdot q_\perp$ and $q^2$ in the first line of~\eqref{GenericClassicalAmplitude}. However, we will not need these here, as we will in practice get the physical pole terms correct from the quantum amplitude. 

Hermiticity and parity invariance of the quantum theory translates to further $\mathbb{Z}_2\times\mathbb{Z}_2$ constraints on the functional form, see \tab{tab:Parity}.
Exchange symmetry of the two massive particles (equivalent to time-reversal symmetry $\omega\rightarrow -\omega$) demands that the variable $z$ appears quadratically $z^2= 4 \omega^2 |a|^2$ in the amplitude. Exchange symmetry of the massless particles combined with parity demands that the variable $y$ also appears quadratically in \eqn{GenericClassicalAmplitude}.
Hence the entire functions satisfy
\be
\genE(x,\pm y, \pm z,w) = \genE(x,y,z,w) \,, \qquad \quad
\genE_{I}(x,\pm y, \pm z,w) = \genE_{I}(x,y,z,w) \,,
\ee
and likewise for the entire functions $\tilde \genE$ and $\tilde \genE_I$, which have identical expansion patterns.

Although in this paper we will not consider dissipative effects, in order to include such effects one may need to relax the exchange symmetry of the massive legs, since the incoming and outgoing $\sqrt{\rm Kerr}$ object may not be related by CPT symmetry.
This means that dissipative effects show up as odd functions in $z\propto |a|$ and correspond to non-hermitian interactions that violate time-reversal symmetry.

From the known literature~\cite{Guevara:2018wpp,Guevara:2019fsj,Huang:2019cja,Aoude:2021oqj}, the first few spin-multipole orders are known to exponentiate up to quadrupole order
\begin{align}
\genE(x,y,z,w)=\tilde \genE(x,y,z,w)=e^{x+w}+ {\cal O}(a^3)=e^{x}\Big(1+w+\frac{w^2}{2}\Big)+ {\cal O}(a^3)
\end{align}
and the remaining functions have currently no known contributions
\begin{align}
\genE_I(x,y,z,w)={\cal O}\big(a^{{\rm max}\{3,\, 2I\}}\big)= \tilde \genE_I(x,y,z,w) \,.
\end{align}

\subsection{Spin variables for quantum amplitudes}
\label{sec:spinvars}

Before we can discuss the classical limit of amplitudes involving spin, we need to introduce an alternative basis for quantum amplitudes. 
Starting from generic Lorentz generators~$M_{\mu\nu}$, we can obtain the Pauli-Lubanski pseudo-vector
\be
\mathbb{S}^\mu = \frac{1}{2m} \epsilon^{\mu\nu\rho\sigma} p_\nu M_{\rho\sigma} \,,
\ee
which is a spin operator that acts on Lorentz-group representations. The Lorentz generators are normalized as $[M^{\mu\nu},M^{\rho\sigma}] = i\big( \eta^{\mu\rho} M^{\sigma\nu}\!
- (\mu \leftrightarrow \nu)\big)  - (\rho \leftrightarrow \sigma)$. 

We will not make much use of $\mathbb{S}^\mu$, as we prefer to work with spin operators that act on the physical states, \ie little-group representations. First we introduce the appropriate spin expectation values, then we extract the needed operators.    
For spin 1/2, we can construct a ring-radius vector as
\begin{align} \label{ahalfspinVectors}
\ahalf^\mu&:= -\frac{1}{4 m^2} \Big( \langle{\bar{\bm{1}}}|\sigma^\mu|\bm{1}]+\langle\bm{1}|\sigma^\mu|{\bar{\bm{1}}}] \Big)= \frac{1}{m^2} \langle{\bar{\bm{1}}}|\mathbb{S}^\mu|\bm{1} \rangle= -\frac{1}{m^2} [{\bar{\bm{1}}}|\mathbb{S}^\mu|\bm{1}] \,,
\end{align}
with spinors for momentum $p=p_1$ defined as 
$|\bm{1}\rangle= |1^a\rangle z_a$, 
$|\bm{1}]= |1^a] z_a$,
$| {\bar{\bm{1}}}\rangle= |1^a\rangle \bar z_a$,  
$| {\bar{\bm{1}}}]= |1^a] \bar z_a$. The magnitude of this vector is
$|\ahalf|=\sqrt{-\ahalf^2}=\frac{1}{2m}|z|^2$, where $|z|^2=z_a \bar z^a = z_1 \bar z_2- z_2 \bar z_1$ gives the squared norm of the corresponding SU(2) wavefunction.  

The spin-$s$ ring-radius vector is defined by the expectation value
\be \label{RingRadiusSpinS}
a^\mu:=-(z_a \bar z^a)^{2s-1}\frac{s}{2m^2} \Big( \langle{\bar{\bm{1}}}|\sigma^\mu|\bm{1}]+\langle\bm{1}|\sigma^\mu|{\bar{\bm{1}}}] \Big)= \frac{1}{m^{2s+1}} \langle{\bar{\bm{1}}}|^{2s}  {\mathbb S}^\mu |\bm{1}\rangle^{2s} \,,
\ee
and it is related to the spin vector by the usual relation $S^\mu= m a^\mu$. The magnitude is $|a|=\sqrt{-a^2}=s|z|^{2s}/m$, which for properly normalized wavefunctions $|z|^2=1$ becomes the expected relation between the spin quantum number and the ring radius $|a|= s/m$. 

We can also define quantum-mechanical spin operators $\hat S^\mu$ and $\hat a^\mu$ acting on the space of spin-$s$ wavefunctions $(z)^{2s}:=z_a\otimes z_a \otimes \cdots \otimes z_a$, related to the spin vectors $S^\mu$ and $a^\mu$ through an expectation value, 
\begin{align}
S^\mu  = \big\langle \hat S^\mu \big\rangle  :=  (\bar z)^{2s} \cdot \hat S^\mu \cdot (z)^{2s}\,,~~~~~~~~
a^\mu  = \big\langle \hat a^\mu \big\rangle  :=  (\bar z)^{2s} \cdot \hat a^\mu \cdot (z)^{2s}\,,
\end{align}
where the spin operators can be extracted from \eqn{RingRadiusSpinS} by taking $2s$ derivatives with respect to both $z_a, \bar z_a$ variables,\footnote{Alternatively, it is not difficult to work out the Clebsch–Gordan coefficients needed for changing from spin-1/2 to spin-$s$ representation. This gives the following representation of the spin-operator:
\be
 (\hat S^\mu)_{A}^{\ B} = S^\mu \Big|_{\bar z_2^{2 s - k} \bar z_1^{k}\, z_1^{2 s - l} z_2^{l} ~\to~ \frac{i^{k+l} }{\sqrt{C^{2 s}_{k}C^{2 s}_{l}}} \delta_{k A}\delta^{l B}}\,, \nn
\ee
where $C^{n}_{k}$ are the binomial coefficients and $A,B,k,l=0,1,\ldots,2s$.
}
\be
(\hat S^\mu)_{a_1{\cdots}a_{2s}}{}^{b_1{\cdots}b_{2s}} ~ := ~ 
\frac{1}{(2s)!^2} \Big(\prod_{j=1}^{2s}
\frac{\partial~}{\partial \bar z^{a_j}}
\frac{\partial~}{\partial z_{b_j}}
\Big) S^\mu \,, \qquad \quad
\hat a^\mu= \frac{\hat S^\mu}{m} \,.
\ee
The spin operator $\hat S^\mu$ satisfies the appropriate quantum-mechanical properties and transversality,
\begin{align}
& [\hat S^\mu, \hat S^\nu] =
i\epsilon^{\mu\nu\rho} \hat S_\rho \,, \qquad \quad
\hat S^2 = -s (s+1) \mathbbm{1} \,, \qquad \quad
p_\mu \hat S^\mu=0 \,,
\end{align}
where the SU(2) structure constants are given by the Levi-Civita tensor $\epsilon^{\mu \nu \rho}=\frac{p_\sigma}{m} \epsilon^{\sigma \mu \nu \rho}$.
Note that $\hat S^\mu$ is a spacelike vector, hence $\hat S^2$ is negative in mostly-minus signature. 

In order to express scattering amplitudes in terms of the spin variables defined above, we begin by defining the Weyl spinors of particle~$\bm{2}$ through a Lorentz boost\footnote{The Lorentz boost guarantees that the $\ket{\bm{2}}$, $|\bm{2}]$ spinor automatically satisfy the Dirac equation for particle~$\bm{2}$, and obey the same normalization constraints as in~\eqn{normEqnP1}.
}
acting on the spinors of particle~$\bm{1}$, satisfying $\Lambda p_1 = p_1+q = -p_2$ and $\Lambda \bar {\bm \varepsilon}_1 =   {\bm \varepsilon}_2$,
\beal
&|\bm{2}\rangle = |2^a\rangle \bar z_{a}
:= \frac{1}{c}\Big(|\bar{\bm{1}}\rangle + \frac{1}{2m} |q|\bar{\bm{1}}]\Big) \,, \\
&|\bm{2}] = |2^a] \bar z_a
:= -\frac{1}{c}\Big(|\bar{\bm{1}}] + \frac{1}{2m} |q|\bar{\bm{1}}\rangle \Big) \,,
\eeal
such that the corresponding ``bra'' Weyl spinors are
\beal
\langle \bm{2}| & = \bar z_a \langle  2^a|
= \frac{1}{c} \Big( \langle\bar{\bm{1}}| - \frac{1}{2m} [\bar{\bm{1}}|q|\Big) \,, \\
[\bm{2}| & = \bar z_a [  2^a|
= -\frac{1}{c} \Big( [\bar{\bm{1}}| - \frac{1}{2m} \langle\bar{\bm{1}}|q|\Big) \,, 
\eeal
where the variable $c$ is a boost-dependent coefficient (see also \rcite{Aoude:2020onz})
\be
c=\sqrt{1 - \frac{q^2}{4m^2}} = \cosh\frac{\zeta}{2} \,.
\ee
Here $\zeta$ is the rapidity boost gained by the BH due to the Compton scattering (note that $q^2<0$), \eg an initial BH at rest recoils and gains the 3-momentum $|\bm{p}| = m \sinh\zeta$.
In the classical limit, the recoil is insignificant, $\zeta \sim \hbar$, and in all of our classical calculations $c=1$ is a good approximation.
We can also write the Lorentz boost explicitly in terms of $\zeta$,
\be
\Lambda= {\rm exp}\Big( \frac{i\zeta}{\sinh \zeta } \frac{q_\mu p_\nu}{m^2} M^{\mu \nu}\Big)\,.
\ee 
This boost can be used to relate the massive polarization vector $\bep_2^\mu$ to the conjugate of the polarization vector $\bep_1^\mu$,
\be
{\bm \varepsilon}_2^\mu = {\Lambda^\mu}_\nu \bar {\bm \varepsilon}_1^\nu = \bar{\bm \varepsilon}_1^\mu + \frac{q \cdot \bar{\bm \varepsilon}_1}{2 m^2 c} \Big( p_1^\mu + \frac{1}{2} q^\mu\Big) \,,
\ee
although this will not be needed for any calculation, as we will use spinor-helicity formulae.
The spin dependence of the Compton amplitude is encoded in the two spin-dependent and crossing-(anti)symmetric complex vectors,
\begin{subequations} \label{eq:rhoVectors} \begin{align}
\rho^\mu & :=\frac{1}{2} \Big(\bra{\bm{2}}\sigma^\mu|\bm{1}]
 + \bra{\bm{1}}\sigma^\mu|\bm{2}] \Big) \,,  \\ 
\bar\rho^\mu & :=\frac{1}{2} \Big(\bra{\bm{2}}\sigma^\mu|\bm{1}]
 - \bra{\bm{1}}\sigma^\mu|\bm{2}] \Big) \,.
\end{align} \end{subequations}
Given their spinorial weight, $\rho$ and $\bar{\rho}$ are natural spin-$1/2$ variables, however, one can span the spin-$s$ variables by taking products of the vectors. For integer spin, there is a quadratic map between the massive polarizations and $\rho\, ,\bar{\rho}$,
\be
\bep_1^\mu \bep_2^\nu = \frac{1}{2m^2} \Big( \rho^\mu \rho^\nu-\bar{\rho}^\mu\bar{\rho}^\nu -\eta^{\mu\nu} \rho^2 +i \epsilon^{\mu\nu\sigma\tau} \rho_\sigma \bar{\rho}_\tau \Big)\,,
\ee
and further relations can be found in \app{sec:appConventions}.

While the physical meaning of these vectors is less clear, their spin dependence can be made explicit with some algebra,
\beal
\label{eq:rhoSpinId}
\bar\rho^\mu &= -2m^2 c \, \ahalf^\mu + p^\mu \frac{q \cdot \ahalf }
  {c} \,, \\
\rho^\mu &=  |\ahalf| \frac{m}{c} (p_1^\mu-p_2^\mu)-
  \frac{i}{c} \epsilon^\mu(p,q, \ahalf) \,,
\eeal
where we again use a short-hand notation for the partially-contracted Levi-Civita tensor $\epsilon^\mu(p,q, \ahalf)=\epsilon^{\mu \nu \rho \sigma} p_\nu q_\rho \ahalf_\sigma$. We have also used the identity $|z|^2 = 2m |\ahalf|$, since the latter is the more convenient variable in the classical limit.

\subsection{Classical limit: infinite versus coherent spin}
\label{sec:clLim}

In \sec{sec:classVars}, we introduced the macroscopic spin vector $S^\mu=m a^\mu$, and the appropriate classical variables used to parameterize any classical Compton amplitude.
In order to map the quantum Compton amplitudes to these classical variables, we need to take a classical limit, where the ring-radius vector $a^\mu$ emerges from the expectation values of the corresponding quantum operator $\hat{a}^{\mu}$,
\begin{equation} \label{ClassicalLimitRule}
    \angle{\hat{a}^{(\mu_1} \hat{a}^{\mu_2} \cdots \hat{a}^{\mu_n)}} \to a^{\mu_1}a^{\mu_2}\cdots a^{\mu_n}.
\end{equation} 
This is equivalent to taking a limit where the quantum variance and higher central moments of the operator vanish,
\begin{equation}\label{eq:spinvariance}
\angle{\prod_{i=1}^n \hat{a}^{\mu_i}}-\prod_{i=1}^n \angle{\hat{a}^{\mu_i}} \overset{\hbar \to 0}{\longrightarrow}  0 \,.
\end{equation}
One can formulate a consistent classical limit for spin in various ways (see \eg~\rcites{Guevara:2019fsj,Arkani-Hamed:2019ymq,Aoude:2021oqj,Cangemi:2022bew,Cangemi:2022abk}).
In this paper, we will discuss two distinct approaches in parallel, and demonstrate that they produce the same classical Compton amplitudes. The two approaches differ in which quantum states are used, and what variables are scaled with respect to $\hbar$. They are
\begin{description}
\item[\namedlabel{largespinLim}{(I)}\,]  large quantum spin $s \rightarrow \infty$, with fixed wavefunction normalization $z_a \bar z^a = 1$;
    
\item[\namedlabel{coherentLim}{(II)}\,] coherent spin states, with wavefunction scaling $z_a, \bar z_a \sim \frac{1}{\sqrt{\hbar}}$ such that $z_a \bar z^a \sim \frac{1}{\hbar}$.
\end{description}

The limit~\ref{largespinLim} is practically implemented by re-expressing the spin-$s$ quantum amplitude in terms of the operator~$\hat{a}^\mu$, acting on the spin-$s$ representation wavefunction. To simplify the calculation, the rule \eqref{ClassicalLimitRule} can be used such that the variables that scale classically according to the $\hbar$-scalings \eqref{ClassicalLimitScaling1}, \eqref{ClassicalLimitScaling2}, \eqref{ClassicalLimitScaling3} can be pre-selected, and quantum ${\cal O}(\hbar)$ terms can be immediately dropped. Finally, the limit $s\rightarrow \infty$ can be taken to obtain the classical amplitude
\be
{\cal A}(\bm{1},\bm{2},3,4)
 = \lim_{s \to \infty}
   {\cal A}(\bm{1}^s\!,\bar{\bm{2}}^s\!,3,4) \,.
\ee 
One must be careful to identify the $s$-dependence appropriately when working at finite spin, in order to enforce the scaling $\hbar s \sim {\cal O}(1)$.\footnote{In particular, naively neglecting terms like $s \,  p \cdot q_{\perp}$ based on the scaling of the momentum $q^\mu \sim \hbar$ violates the classical limit.
Making use of the relation $|a| = s/m$, we find that such terms contribute classically as $s \, p \cdot q_{\perp} \xrightarrow[s\to\infty]{\hbar \to 0} m p \cdot q_{\perp} |a|$.
}

While this  approach requires some delicacy, it has practical advantages for the case where amplitudes are only known up to a fixed spin, since the $s\rightarrow \infty$ limit is delayed until the very last step and can be carried out with only partial knowledge of the $s$-dependence. 

The spin-$s$ amplitudes are polynomials of spin-$1/2$ quantum variables; variables that can be re-expressed exactly in terms of the ring-radius vector $\ahalf^\mu$ of a spin-1/2 state, given in \eqn{ahalfspinVectors}. This  conversion to $\ahalf$ can be done via the $\rho^\mu$, $\bar\rho^\mu$ vectors \eqref{eq:rhoVectors} and their identities \eqref{eq:rhoVectors}. 

\begin{table}[t]
    \centering
\begin{tabular}{|c||c|c|c|} 
 \hline 
 $$ & $\hbar^0$ & $\hbar^1$   & $\hbar^{\ge2}$  \\
 \hline
 \hline
$W_+/m^2$ & $-1$ & $\frac{1}{4m}q^2|\ahalf| $& ${\cal O}(\hbar^3)$  \\ [0.5ex]   
\hline
$W_-/m^2$ &  $-q \cdot \ahalf $ & $0$ & ${\cal O}(\hbar^3)$  \\ [0.5ex]    
\hline
$U/m^2$ & $1  +q_\perp\!\cdot \ahalf $ & $-\frac{1}{4m}q^2 |\ahalf|-\frac{1}{2m^2} p \cdot q_\perp q \cdot \ahalf  $ & ${\cal O}(\hbar^2)$  \\ [0.5ex]   
\hline
$V/m^2$ & $\frac{1}{m} |\ahalf| \, p \cdot q_\perp$ & $-\frac{1}{2m^2} i \epsilon(p,q,q_\perp, \ahalf)$ & ${\cal O}(\hbar^3)$  \\ [0.5ex]  
\hline
$\langle 3 | \rho | 4]/m^2$ &   $\frac{2}{m} |\ahalf| \,  p\cdot\chi $  & $ -\frac{1}{m^2}i\epsilon(p,q,\chi,\ahalf)$ & ${\cal O}(\hbar^3)$  \\ [0.5ex]  
\hline
$\langle 3 | \bar{\rho}| 4]/m^2$ &  $-  2 \chi \cdot \ahalf $ & $\frac{1}{m^2}p\cdot \chi \, q \cdot \ahalf $ & ${\cal O}(\hbar^2)$  \\ [0.5ex]   
\hline
\end{tabular}
\caption{In classical limit~\ref{largespinLim}, the six spin-dependent quantum variables are expanded using the spin-$1/2$ ring-radius vector $\ahalf^\mu$. The identity $2m |\ahalf|=1$ has been used for the leading contribution on the first and third line. The $\hbar$ scalings indicate the final behavior of the variables, relevant for the the macroscopic ring-radius vector. Note that $W_{\pm}=\frac{m}{2}(\langle \bm {12} \rangle\pm [\bm {12}])$.}
\label{CLITable}
\end{table}

In \tab{CLITable}, we summarize the map between the quantum variables and the leading few orders in the classical limit. The $\hbar$ scalings in \tab{CLITable} reflects the final scaling after the ring-radius vector has become macroscopic, since this is more useful for our purposes. Note that one might naively anticipate that the $\hbar^0$ column will give all relevant classical contributions; however, this is not true as quantum and superclassical terms have the potential to conspire.
While in principle all orders in $\hbar$ should be kept to maintain the relation $\hbar s \sim {\cal O}(1)$, in practice we find our classical amplitudes are insensitive to terms beyond $\hbar^1$ so that we can pre-select the relevant terms.

All variables in \tab{CLITable} are in principle linear in $\ahalf^\mu$; however, we have made use of the identity $2m|\ahalf|=z_a \bar z^a=1$ for two of the terms, in order to not artificially add divergences in the classical limit.\footnote{While the identity $2|\ahalf|=1$ can be freely used in finite spin quantum amplitudes, the infinite-spin limit requires that $|\ahalf|$ gets replaced by $|a|$ via \eqn{eq:spinrepchange}. This introduces spin-dependent combinatorical factors that alter the naive classical behavior and may introduce divergences in $s$. }    

Once the spin-$s$ amplitude is expressed in terms of the spin-$1/2$ ring-radius vector $\ahalf^\mu = \langle \hat a^\mu \rangle$, we need to re-write it into its correct quantum-mechanical representation. This is done using the spin-representation change formula for products of such objects,
\begin{equation} \label{eq:spinrepchange}
    \ahalf^{\mu_1} \ahalf^{\mu_2} \cdots \ahalf^{\mu_k} = \frac{(2s - k)!}{(2s)!} 
    \big\langle \hat a^{(\mu_1} \hat a^{\mu_2} \cdots \hat a^{\mu_k)}\big\rangle + {\cal O}(\hat{a}^2) \, , 
\end{equation}
which introduces combinatorial factors of $s$ that are sensitive to how many spin operators are present. The correction terms ${\cal O}(\hat{a}^2)$ correspond to contributions that have Casimir factors $\eta^{\mu_i \mu_j} \hat{a}^2\sim \eta^{\mu_i \mu_j}s(s+1)$. We will later argue that these terms are always suppressed in the large-$s$ limit, and can be dropped from any classical-limit calculation. Note that this does not mean that all Casimir factors are suppressed; here we mean only those that are coming from the above spin-representation change formula.

Note that in the letter \cite{Cangemi:2022bew}, we introduced a complimentary scaling of the spin variables, where we scaled the wavefunctions
\begin{equation}\label{eq:unphysScaling}
    z_a \,, \bar z_a \sim \frac{1}{\sqrt{\hbar}} \qquad \text{ while } \qquad z_a \bar z^a \sim 1 \, ,
\end{equation} instead of the quantum spin number $s \sim \hbar^{-1}$. The scaling \eqref{eq:unphysScaling} works at fixed quantum spin $s$ and can be implemented numerically. However, it is ultimately unphysical as it requires the wavefunctions $z_a$ and $\bar z_a$ to be large and independent. This can only be achieved if we analytically continue the ring radius to complex values, and thus $z_a$ and $ \bar z^a$ are no longer related by complex conjugation (hence SU(2) is generalized to ${\rm SL}(2)$). Nonetheless, this unphysical approach is still able to reproduce a remarkable portion of the classical information contained in the physical classical limit of \ref{largespinLim}. Notably, the final scaling of the variables in \tab{CLITable} can be read off immediately given that $\ahalf$ scales as $\hbar^{-1}$. However \eqn{eq:unphysScaling} misses any classical contributions of $|a|$ terms, as such terms are now subleading compared to $a^\mu$. Therefore, in this paper we implement limits \ref{largespinLim} and \ref{coherentLim}, although the scaling in \eqn{eq:unphysScaling} can reproduce all classical results in the upcoming sections up to missing $|a|$ terms. 

The coherent spin states construction \ref{coherentLim} involves a similar scaling of the wavefunctions as \eqn{eq:unphysScaling}
\begin{equation}
    z_a \,, \bar z_a \sim \frac{1}{\sqrt{\hbar}}\, , \qquad  \text{ while } \qquad z_a \bar z^a \sim \frac{1}{\hbar} \, ,
\end{equation}
but now the normalization of the wavefunctions is no longer constrained.
Despite the superclassical scaling of $|z|^2$, the coherent-state approach avoids divergences due to the fundamental differences in the setup of the scattering. As explained in detail in \rcite{Aoude:2021oqj}, one should scatter coherent spin states in place of the massive spin particles. Schematically, the coherent state is constructed from a sum over spin eigenstates $\ket{s,\{a\}}$,
\begin{equation}
\ket{\text{coherent}} = e^{-|z|^2/2} \sum_{s=0}^\infty
   \frac{(z_a)^{\otimes 2s}}{\sqrt{(2s)!}} \ket{s,\{a\}} \,,
\end{equation}
where here $\{a\}$ denotes the set of SU(2) little-group indices (not to be confused with the ring radius). The classical amplitude corresponds to the $\hbar\rightarrow 0$ limit of an infinite sum over diagonal finite-spin amplitudes,
\begin{align} \label{eq:coherentClAmp}
 {\cal A}(\bm{1},\bar{\bm{2}},3^\pm\!,4^\pm) &=\lim_{\hbar \to 0} e^{-|z|^2}\sum_{s=0}^{\infty} \frac{1}{(2s)!} {\cal A}(\bm{1}^s\!,\bar{\bm{2}}^s\!,3^{\pm}\!,4^{\pm}) \,.
\end{align} In principle, there could be non-diagonal contributions where the spins of external massive particles differ, $s_1 \neq s_2$. However, we will neglect such contributions in this paper, since we have not included off-diagonal interactions between physical fields in the Lagrangians in~\secs{sec:HigherSpinGauge}{sec:chiralform}.

In this approach the $\hbar \to 0$ limit must be taken after resumming the coherent states, such that there is no distinction\footnote{Coherent states and classical ring radius satisfy the identities
 $a^\mu = \bra{\text{coherent}}\hat{a}^\mu \ket{\text{coherent}} = \ahalf^\mu$.
}  between the spin-$1/2$ expectation value $\ahalf^\mu$ and the classical ring radius $a^\mu$, and there is no need to change representation \cite{Aoude:2021oqj}. Thus limit~\ref{coherentLim} generates different scalings for the expansion of the spin variables directly in terms of the classical ring radius $a^\mu$, as shown in \tab{table:CLIITable}.

\begin{table}[t]
\centering
\begin{tabular}{|c||c|c|c|c|} 
 \hline
& $\hbar^{-1}$ & $\hbar^0$ & $\hbar^1$ & next term  \\
 \hline
 \hline
$W_+/m^2$ & $-2 m |a|$ & $0$ & $\frac{1}{4m} |a| q^2$ & ${\cal O}(\hbar^3)$  \\ [0.5ex]   
\hline
$W_-/m^2$ & $0$ & $ -q {\cdot} a $ & $0$ & ${\cal O}(\hbar^2)$  \\ [0.5ex]    
\hline
$U/m^2$ & $2 m |a|$ & $ q_\perp {\cdot } a $ & $-\frac{1}{2m^2} p \cdot q_\perp  q {\cdot} a -\frac{1}{4m} |a|q^2 $ & ${\cal O}(\hbar^2)$  \\ [0.5ex]   
\hline
$V/m^2$ & $0$ & $ p \cdot q_\perp |a|/m $ & $-\frac{1}{2m^2} i \epsilon(p,q,q_\perp, a)$ & ${\cal O}(\hbar^2)$  \\ [0.5ex]  
\hline
$\langle 3 | \rho | 4]/m^2$ & $0$ & $2|a| p{\cdot}\chi/m $ & $-i\epsilon(p,q,\chi, a)/m^2$ & ${\cal O}(\hbar^2)$  \\ [0.5ex]   
\hline
$\langle 3 | \bar{\rho}| 4]/m^2$ & $0$& $-2 \chi {\cdot}a$ & $ p{\cdot}\chi \,  q {\cdot} a/m^2 $ & ${\cal O}(\hbar^2)$  \\ [0.5ex]   
\hline
\end{tabular}
\caption{Scaling of spin variables in the coherent-state approach. Note that $W_{\pm}=\frac{m}{2}(\langle \bm {12} \rangle\pm [\bm {12}])$.}
\label{table:CLIITable}
\end{table}

While in \tab{table:CLIITable} we have indicated the distinct $\hbar$ scaling of each term, we will find that the classically relevant terms remain the same as those indicated in \tab{CLITable}.
Note that in this approach possible $s$ divergences are traded traded for possible wavefunction divergences, however any spurious divergence automatically cancels out because of the overall normalization $e^{-|z|^2}$ in \eqn{eq:coherentClAmp}.

\subsection{Same-helicity intermezzo} \label{sec:samehel}

Let us first analyze the same-helicity amplitude \eqref{AmpGaugeAHH4idh}, whose spin dependence can be factored out fully from the helicity and pole structures. In the large-$s$ limit~\ref{largespinLim} the spin-dependent term can be expanded as
\be \label{SameHelClassical}
\frac{\angle{\bm{21}}^{2s}}{m^{2s}} \approx (1+q\cdot \ahalf)^{2s}= \sum_{k=0}^{2s} \frac{(2s)!}{(2s-k)! k!}(q\cdot \ahalf)^k
= \big\langle e^{q \cdot \hat a} \big\rangle + {\cal O}(\hat a^2) \,,
\ee
where $\approx$ implies we have only kept up to ${\cal O}(\hbar)$ in \tab{CLITable}. We first used the binomial expansion, and then the representation change formula in \eqn{eq:spinrepchange}. Equating the finite sum with the exponential operator is justified from the finite-spin representation of $\hat a$, and the difference will only involve contributions ${\cal O}(\hat a^2)$ that are proportional to powers of the Casimir $\hat a^2 \sim s(s+1)$. Another source of such Casimir contributions, in the above ${\cal O}(\hat a^2)$ term, come from their appearance in the spin-representation change formula~\eqref{eq:spinrepchange}. 

Now let us show that the ${\cal O}(\hat a^2)$ terms in \eqn{SameHelClassical} are suppressed in the classical limt. While in the large-wavefunction limit~\eqref{eq:unphysScaling} such terms are clearly subleading, in the proper large-spin classical limit \ref{largespinLim} it is not a priori clear whether they are classical or quantum since there are $s$-dependent combinatorial factors in \eqn{eq:spinrepchange}. Let us work out a few examples of such multipoles in the following table:
\begin{align}
\begin{tabular}{|c||c|c|c|} 
\hline
 & $e^{q \cdot \hat a}$ &$ {\cal O}(\hat a^2) $  &$ {\cal O}(\hat a^4) $ \\ [0.5ex]
\hline
\hline
dipole & $ \hat a \cdot q$ & $0$ & $0$ \\ [0.5ex]
\hline
quadrupole & $\frac{1}{2} (\hat a \cdot q)^2$ & $-  \frac{s}{4s_2}  \hat a^2 q^2 $ & $0$ \\ [0.5ex]
\hline
octapole & $\frac{1}{3!} (\hat a \cdot q)^3$ & 
$ -\frac{1}{4} \frac{1 - 3 s}{1 - 3s_2} \hat a^2 q^2 (\hat a \cdot q) $ 
& $0$ \\ [0.5ex]
\hline
hexadecapole  & $\frac{1}{4!} (\hat a \cdot q)^4$ & 
$-\frac{1}{4}\frac{2 - 3 s}{5 - 6s_2} \hat a^2 q^2 (\hat a \cdot q)^2$
& 
$\frac{s}{8} \frac{7 - 29s +  6 s^2 (2 + 3 s)}{12 (5 - 6 s_2)} \frac{\hat a^4 q^4}{s_2 (1/3 - s_2)} $
\\ [0.5ex]
\hline
dotriacontapole  & $\frac{1}{5!} (\hat a \cdot q)^5$ & 
$-\frac{1}{4}\frac{3 - 3 s}{27 - 18 s_2} \hat a^2 q^2 (\hat a \cdot q)^3$
&  
$\frac{1}{8}  \frac{14 - 63s + 83 s^2 - 30 s^4}{20 (3 - 2 s_2) (1 - 3 (1 - s_2) s_2)} \hat a^4 q^4 (\hat a \cdot q)  $
\\ [0.5ex]
\hline
\end{tabular}
\end{align}
where $s_2:= s(s+1)$ stands for the quantum-mechanical ``square'', and strictly speaking one needs to include the transversality projector in $q^2 \to (q \cdot q)- (q \cdot p)^2/m^2$, but the last term is suppressed by two powers of $\hbar$ and can be ignored. Note that all above operator products are symmetrized over all orderings, which we also assume throughout this paper.

We can work out the ${\cal O}(\hat a^2)$ terms to any order in the spin-multipole expansion,
\be \label{eq:allplusasq}
{\cal O}(\hat a^2)= -\hat a^2 q^2 \sum_{n=2}^{\infty}\frac{(q \cdot \hat a)^{n-2}}{(n - 2)!} \frac{n - 2 - 3 s}{(n - 2) (n + 1) - 12 s_2}   =  -\hat a^2 q^2  \frac{e^{q \cdot \hat a}}{4 s} + {\cal O}\Big(\frac{1}{s^2}\Big) \,,
\ee
which clearly vanish in the classical limit $s\to \infty$.
For the ${\cal O}(\hat{a}^4)$ terms, one can also work out the general behavior,
\be
{\cal O}(\hat{a}^4) = \hat a^4 q^4 \sum_{n=4}^{\infty}\frac{(q \cdot \hat a)^{n - 4}}{2 (n - 4)!}  \frac{\tau_{n,s}}{\theta_{n,s}}
= \hat a^4 q^4  \frac{e^{q \cdot \hat a}}{32 s^2} + {\cal O}\Big(\frac{1}{s^3}\Big)\,,
\ee
where 
\begin{align}
\tau_{n,s}= & \, (n - 4) (n - 2) (n + 1) (5 n - 11) +3 (n - 2) (129 - n - 10 n^2) s  \\
 &
 \null - 3 (642 - 253n  + 5 n^2) s^2
 +360 (n - 5) s^3
 -540 s^4\,, \nn \\*
\theta_{n,s}=& \, \big((n - 2) (n + 1) - 12 s_2 \big)\Big(\!(n {-} 4) (n {+} 1) (30 - 23 n + 5 n^2) - 5! (n - 3) (n - 2) s_2 + 6! s_2^2\Big). \nn
\end{align}
This also vanishes in the classical limit.

From the observed patterns, we can conjecture the leading-$s$ behavior of each ${\cal O}(\hat a^{2n})$ Casimir term,
\be
{\cal O}(\hat a^{2n}) = (-1)^n a^{2n} q^{2n} \frac{e^{q \cdot a}}{4^{n} n! s^n} + {\cal O}\bigg(\frac{1}{s^{n+1}}\bigg) \,.
\ee
Given that these corrections vanish increasingly fast as $\sim 1/s^n$ in the classical $s \to \infty$ limit, this suggests the Casimir terms generated in \eqn{eq:spinrepchange} are always quantum.

We are able to crosscheck that the ${\cal O}(\hat{a}^2)$ terms are indeed quantum by computing the classical amplitude via the coherent-spin approach \ref{coherentLim}. Considering again only the factor $\angle{\bm{21}}^{2s}$ and summing over all spins generates
\begin{align}
e^{- |z|^2} \sum_{s=0}^{\infty} \frac{1}{(2s)!}  \frac{\angle{\bm{21}}^{2s}}{m^{2s}}  = e^{- |z|^2 -\angle{\bm{12}}/m} = e^{q \cdot a} + {\cal O}(\hbar)\, ,
\end{align}
where the wavefunction normalization $|z|^2=2m|a|$ cancels out against a similar term in $\angle{\bm{12}}$, as seen in \tab{table:CLIITable}.

In summary, the classical same-helicity (helicity-violating) Compton amplitude takes the following simple form in the classical limit,
\be
{\cal A}(\bm{1},\bar{\bm{2}},3^+\!,4^+)= -2g^2
e^{q \cdot a}[34]^2 \bigg( \frac{[T^{c_3},T^{c_4}] }{q^2 (p \cdot q_\perp)} +\frac{1}{2} \frac{ \{T^{c_3},T^{c_4}\}}{(p \cdot q_\perp)^2} \bigg) \,,
\label{AmpGaugeAHH4ppClassical}
\ee
which means that it is given by the Newman-Janis shift of the non-spinning Compton amplitude for a charged scalar \cite{Arkani-Hamed:2019ymq,Aoude:2020onz}.

\subsection{Opposite-helicity analysis}
\label{sec:OppositeHelicity}

In contrast to the same-helicity case, the color structure and poles cannot be fully factored out in the opposite-helicity color-dressed amplitude \eqref{eq:ColorDressedCompton}.
This remains the case in the classical limit, such that the abelian and color-dressed amplitudes differ by more than just their pole structure~\eqref{GenericClassicalAmplitude}. 

\paragraph{Non-abelian amplitude.}
We will proceed by first considering the terms proportional to the commutator $[T^{c_3}, T^{c_4}]$ in \eqn{eq:ColorDressedCompton}, computing the classical spin-dependent functions $\genE$ and $\genE^{I}$ in the first line of \eqn{GenericClassicalAmplitude}.

After factoring out the scalar amplitude \eqref{eq:scalarClCompton}, the first spin-dependent function in \eqref{eq:ColorDressedCompton} can be expanded in the large-$s$ limit~\ref{largespinLim} as
\begin{align} \label{eq:firstStepClTerm}
(U+V )^{2s}  & \approx (1+ q_\perp \cdot \ahalf+|\ahalf| p\cdot q_\perp/m )^{2s} =\sum_{k=0}^{2s} \frac{(2s)!}{(2s-k)! k!}(q_\perp \cdot \ahalf+|\ahalf| p\cdot q_\perp/m)^k \nn \\*
 &= \Big\langle e^{q_\perp\!\cdot \hat a+ |\hat a| p\cdot q_\perp/m} \Big\rangle  + {\cal O}\bigg(\frac{1}{s}\bigg) \,.
\end{align}
Here $\approx$ implies that we only keep terms where each $q,q_\perp$ is matched with a $\ahalf$ factor, since other contributions are $\hbar$-suppressed, and the last step involves changing the representation of the spin operator \eqref{eq:spinrepchange}. 
The ${\cal O}\big(1/s\big)$ terms coming from the representation change of the $q_\perp \cdot \ahalf$ factors can be obtained by swapping $q\,\to\,q_\perp$ in the corresponding all-plus formula~\eqref{eq:allplusasq},
\begin{align}
{\cal O}\Big(\frac{1}{s}\Big) =  \Big(q^2+ \frac{(q_\perp\!\cdot p)^2}{m^2}\Big) \hat a^2 \frac{e^{q_\perp\!\cdot \hat a}}{4 s} + {\cal O}\Big(\frac{1}{s^2}\Big) \,.
\end{align}
Other such ${\cal O}(1/s)$ terms come from the representation change of the $|\ahalf|$ factor; however $|\hat a|$ is a mathematically ambiguous operator and its precise definition is not needed.\footnote{In the complete amplitude odd $|\hat a|$ terms are guaranteed to cancel out by time-reversal symmetry. The even factors can be defined as $|\hat a|^2= -\hat a^2$ and $|\hat a| f(\hat a)\, |\hat a| = -\hat a^2 f(\hat a) + {\cal O}(1/s)$ for some function $f$.}
In the following, we will neglect all ${\cal O}(1/s)$ contributions in the representation change formula \eqref{eq:spinrepchange}, since they are both irrelevant and uninteresting terms. 
Thus only the leading-order term contributes to the classical function $\genE(x,y,z,0)$ in \eqref{GenericClassicalAmplitude},
\be \label{eq:clNAbterm1}
\lim_{s\to \infty} (U+V )^{2s} =\lim_{s\to \infty}  \Big\langle e^{q_\perp\!\cdot \hat a+|\hat a| p\cdot q_\perp/m} \Big\rangle =  e^{x+z}\,,
\ee
where $x = q_\perp\!\cdot a$, $z = |a|\,p \cdot q_\perp/m$ and
$a^\mu =\langle \hat a^\mu \rangle$. 

In the the coherent-state approach \ref{coherentLim}, we first resum the $s$ dependence and then expand the variables according to \tab{table:CLIITable},
\begin{align} \label{eq:opphelterm1}
e^{- 2m|a|} \sum_{s=0}^{\infty} \frac{(U+V)^{2s}}{(2s)!}    &= e^{- 2m|a| + U+ V} = e^{x+z}+ {\cal O}(\hbar) \,,
\end{align} such that we confirm the result in \eqn{eq:clNAbterm1}.
Note that here we have written the SU(2) wavefunction normalization $|z|^2$ in terms of $|a|$ in order to not confuse it with our classical variable $z$, which uses the same letter.

Continuing with the large-spin analysis, we find the second spin-dependent term in the commutator sector of the color-dressed amplitude \eqref{eq:ColorDressedCompton} generates contributions to both $\genE(x,y,z,0)$ and $\genE'(x,y,z,0)$ in \eqn{GenericClassicalAmplitude}.
Namely, after factoring out the scalar amplitude, we get
\begin{align}\label{eq:opphelterm2}
\frac{ t_{14}\langle 3|\bar\rho{-}\rho|4] }{2 \langle 3|1|4]} P^{(2s)}_2 &
\approx -\Big(\frac{1}{2} +\frac{m \chi \cdot \ahalf}{2 p \cdot \chi |\ahalf|}\Big) \Big((1+ q_\perp \cdot \ahalf + |\ahalf| p \cdot q_\perp )^{2s}-(1+ q_\perp \cdot \ahalf - |\ahalf| p \cdot q_\perp )^{2s}\Big)
\nn \\ &\!\!\!
\stackrel{s\to \infty}{=}  -e^{q_\perp{\cdot}a}  {\rm sinh}\big(|a|  p \cdot q_\perp \big)- e^{q_\perp \cdot  a} \frac{\chi \cdot  a \, p \cdot q_\perp}{p \cdot \chi } {\rm sinhc}\big(|a|  p \cdot q_\perp \big)\,.
\nn \\
 &=-e^{x}  {\rm sinh} \, z  - w \,  e^{x} {\rm sinhc}\, z  \,,
\end{align}
where  $w = (\chi \cdot a)(p \cdot q_\perp)/(p \cdot \chi)$ and ${\rm sinhc}\, z = \frac{{\rm sinh} \,z}{z}$ .
In the first line we have dropped higher orders in $\hbar$, whilst in the limit $s \to \infty$ we used the simplified representation-change formula~\eqref{eq:spinrepchange} that is insensitive to the Casimirs ${\cal O}(\hat a^2)$.
An analogous, but simpler, calculation using coherent states confirms the result. Note that the odd powers in $z$ cancels between \eqns{eq:opphelterm1}{eq:opphelterm2}, as is guaranteed by the time-reversal symmetry of our Compton amplitude.
See \eqn{eq:KinIdentity} in \app{app:factorisation} for a brief discussion on the non-trivial $1 \leftrightarrow 2$ identity satisfied by the pole terms of our Compton amplitude.  

In the large-spin approach, the entire functions are generated when resumming the binomial with the relevant combinatorial factors dictated by the representation change formula in \eqn{eq:spinrepchange}.
In comparison, in the coherent-state approach, the entire functions are generated by the infinite sums in $s$.
These infinite sums can in general be rather difficult to perform. However, the special polynomials in our amplitudes ensure that sums encountered are remarkably simple. They are variations of the following two cases:
\begin{subequations} \begin{align}
\sum_{k=0}^{\infty}\frac{1}{k!} P^{(k)}_{2} &= \frac{e^{\vs_1}}{\vs_1 - \vs_2}+ (\vs_1 \leftrightarrow \vs_2) \,, \\
\sum_{k=0}^{\infty} \frac{1}{k!}P^{(k)}_{4} &= \frac{e^{\vs_1}}{(\vs_1 - \vs_2)(\vs_1 - \vs_3)(\vs_1 - \vs_4)}+ \text{cyc}(\vs_1\, , \vs_2\, , \vs_3\, , \vs_4) \,.
\end{align} \end{subequations}
This means that the coherent-spin approach is vastly simpler for our amplitudes than it could have been. 

Continuing the amplitude analysis, the third term in \eqn{eq:ColorDressedCompton}, proportional to $P_2^{2s-1}$, vanishes in the classical limit, as confirmed by both the large-spin limit and coherent-spin approach.
The last two terms in the color-dressed amplitude require a more careful treatment.
Individually, they contain divergences: in the large-$s$ limit they appear as divergences in $s$, but as superclassical $\hbar^{-1}$ terms in the coherent-state approach. 

Once we factor out the relevant scalar amplitude and the color factor $[T^{c_3},T^{c_4}]$, the relevant terms in the non-abelian amplitude have the following behavior: 
\beal \label{eq:divergingterms}
\text{term 4: } \qquad t_{13} t_{14} & \frac{\braket{\bm{1}3} \braket{3\bm{2}} [\bm{1}4] [4\bm{2}]}{m^{4s-2} \langle 3|1|4]^2}\braket{\bm{12}}[\bm{12}] P_{4}^{(2s-1)} &&\quad\sim\, {\cal O}(s) \,, \\
\text{term 5: } \quad\,\qquad t_{14} & \frac{\braket{\bm{1}3} \braket{3\bm{2}} [\bm{1}4] [4\bm{2}]}{m^{4s-2} \langle 3|1|4]^2} 2 V P^{(2s)}_{4}  &&\quad\sim\, {\cal O}(s) \,,
\eeal
where, in the second line, we use the following $\hbar$-expansion for the $V$ prefactor,
\be
    V = \frac{1}{2}p \cdot q_\perp - \frac{1}{2}i \epsilon(q_\perp,p,q,\ahalf) + {\cal O}(\hbar^3) \,.
\ee
Compared to \tab{CLITable}, we have used the identity $2m|\ahalf|=1$ in order to get the softer scaling $V \sim \hbar$, such that the powers of $q^\mu,q_\perp^\mu$ and powers of $\ahalf^\mu$ match in this expression. 
Remarkably, the divergences of the ${\cal O}(s)$ behavior cancel between the two terms, so that in the sum the following finite classical contribution is obtained: 
\beal \label{eq:clnonAbcancellation}
-t_{14}\frac{\braket{\bm{1}3} \braket{3\bm{2}} [\bm{1}4] [4\bm{2}]}{m^{4s-2} \langle 3|1|4]^2}   \Big[ &t_{13}\braket{\bm{12}}[\bm{12}] P_{4}^{(2s-1)} -  2 V P_4^{(2s)} \Big] \\* &
\stackrel{s \to \infty}{=} \frac{(w^2-z^2)}{2} \Big[  E\big(x,\,y,\,z \big) - \frac{2i\epsilon(q_\perp,p,q,\ahalf)}{p \cdot q_\perp}  \Eb \big(x,\,y,\,z \big) \Big] \,,
\eeal
where $\Ea$ and $\Eb$ are entire functions given by
\begin{align} \label{eq:clEfns}
\Ea(x,y,z)&=\frac{e^y - e^x \cosh z + (x - y) e^x \, {\rm sinhc}\,z}{(x - y)^2 - z^2}~+~ (y\to -y) \,, \\
\Eb(x,y,z)&=\frac{2 x \cosh y-2 x e^x  \cosh z +(x^2- y^2+z^2) e^x \,{\rm sinhc}\, z
 +(x^2+y^2- z^2)\,{\rm sinhc}\, y}{\big((x - y)^2 - z^2\big)\big((x + y)^2 - z^2\big)} \,. \nn
\end{align} 
The cancellation of the individual divergences also follows in the coherent-state picture, where the superclassical pieces cancel and the classical terms match \eqn{eq:clnonAbcancellation}.

The full non-abelian sector of the classical amplitude, generated independently by \ref{largespinLim} and \ref{coherentLim}, is given by
\beal\label{eq:clCommSector}\!\!
{\cal A}(\bm{1},\bar{\bm{2}},3^-\!,4^+)|_{[T^{c_3},T^{c_4}]} = 2g^2
\frac{(p \cdot \chi)^2 }{q^2 (p \cdot q_\perp)} %
\Big(&e^x \cosh z - w \, e^x {\rm sinhc}\, z + \frac{w^2\!- z^2}{2} \Ea(x,y,z) \\*
&- (w^2\!- z^2) \frac{i \epsilon(p,q,q_\perp,a)}{p \cdot q_\perp } {\Eb}(x,y,z) \Big) [T^{c_3}\!,T^{c_4}] \,,\!\!\!
\eeal
where we recall $x= a \cdot q_\perp$, $y=a \cdot q$, $z = |a|\,p \cdot q_\perp/m$,
$w=(\chi \cdot a)(p \cdot q_\perp)/(p\cdot\chi)$.
Note that the Levi-Civita term is not an independent variable and can be written as a function of $x,y,z$ using \eqn{eq:clLeviCivitaIds} and the classical-limit Gram determinant \eqref{eq:clGramDet} such that
\be
(w^2 - z^2) \frac{i \epsilon(p,q,q_\perp,a)}{p \cdot q_\perp } = x ( w^2\!+z^2) - w(x^2\!-y^2\!+z^2) \,.
\ee
One can read off the entire functions in \eqn{GenericClassicalAmplitude} from the above expression
\begin{alignat}{2}\label{eq:genEcommutator}
    &\genE(x,\!y,\!z,\!0) {=} e^x \cosh z{+}\frac{z^2}{2}\Ea(x,y,z){+}xz^2 \Eb(x,y,z)\,, \!\quad &&\genE''(x,\!y,\!z,\!0) {=} \frac{1}{2} \Ea(x,y,z){-}x\Eb(x,y,z)\,,\nn\\
    &\genE'(x,\!y,\!z,\!0) {=}{-}e^x {\rm sinhc}\,z{+}(x^2\!-y^2\!+z^2) \Eb(x,y,z)\,, \!\quad &&\genE_{I}(x,\!y,\!z,\!w){=}0 \,.
\end{alignat}
We will postpone further analysis of the expression until we have the full color-dressed classical amplitude.

\paragraph{Abelian amplitude.}
The abelian amplitude shares many terms with the non-abelian sector such that the first two terms of the classical abelian amplitude are identical to those in \eqn{eq:clCommSector} up to differing pole structure.
The difference lies in the treatment of the shared diverging term, \ie term 4 in \eqn{eq:divergingterms}.
In the abelian sector, the cubic Lagrangian does not generate a term that cures the divergence in ${\cal O}(s)$, suggesting the amplitude is only consistent if we add contact terms to cure the divergence.

Following the construction of contact terms explained in \sec{sec:abAmp}, we find the simplest consistent contact term to be
\begin{equation}
    C^{(s)} = -\frac{\braket{\bm{1}3} \braket{3\bm{2}} [\bm{1}4] [4\bm{2}]}{ 2 m^{4s-1}} \big(\braket{\bm{12}} + [\bm{12}]\big)\Big(P^{(2s)}_4 -P^{(2s-2)}_2 \Big) \,.
\end{equation}
It conspires with the divergent fourth term, such that a classically finite result is obtained:
\be \label{eq:abDivCancellation}
     - \frac{\braket{\bm{1}3} \braket{3\bm{2}} [\bm{1}4] [4\bm{2}]}{m^{4s-2}}\braket{\bm{12}}[\bm{12}] P_{4}^{(2s-1)} + C^{(s)} ~\stackrel{s\to \infty}{=}~ -
\frac{(p \cdot \chi)^2 }{(p \cdot q_\perp)^2} \frac{w^2 - z^2}{2}\Ea(x,y,z) \,,
\ee
where $\Ea$ is the same entire function \eqref{eq:clEfns} that appeared in the non-abelian sector.
The corresponding classical abelian amplitude is then
\be \label{eq:clAbSector}
{A}_{\text{U}(1)}(\bm{1},\bar{\bm{2}},3^-\!,4^+) = -
\frac{(p \cdot \chi)^2 }{(p \cdot q_\perp)^2}
\Big(e^x \cosh z - w\,e^x {\rm sinhc}\,z + \frac{w^2 - z^2}{2} \Ea(x,y,z)\Big) \,.
\ee
Once again the entire functions $\tilde{\genE},\tilde{\genE}',\tilde{\genE}''$ can be read off from the amplitude and correspond to the functions defined in \eqref{eq:genEcommutator} where $\Eb$ is set to zero.

As mentioned at the end of \sec{sec:abAmp}, the choice of quantum contact term $C^{(s)}$ is not unique, even after imposing classical consistency. An example of an alternative contact term, which is fully consistent with our constrains, is
\begin{equation}
    {C'}^{(s)} =-\frac{\braket{\bm{1}3} \braket{3\bm{2}} [\bm{1}4] [4\bm{2}]}{m^{4s-2}} \braket{\bm{12}} [\bm{12}] \Big(2P_4^{(2s-1)} + \frac{m}{2}\big(\braket{\bm{12}} +[\bm{12}]\big)P_4^{(2s-2)} \Big) \,.
\end{equation} However, ${C'}^{(s)}$ and all other such quantum contact terms contribute identically to $C^{(s)}$ in the classical limit, such that the abelian amplitude is uniquely fixed, given our chosen constraints below \eqn{eq:C(s)ansatz}. In \sec{sec:spinuniversality} and \sec{sec:contacttermbreaking}, we discuss the consequences of weakening certain constraints in our contact term construction and discuss how it introduces free parameters in the classical amplitude.


\paragraph{Color-dressed amplitude.}
We can now assemble the full, color-dressed classical Compton amplitude in the candidate $\sqrt{\text{Kerr} }$ theory. It is given by, to all orders in spin,
\begin{align} \label{eq:NAbClAmp}
{\cal A}(\bm{1},\bar{\bm{2}},3^-\!,4^+) \quad~& \nn \\
 = 2g^2 (p \cdot \chi)^2 \bigg\{\!
   \bigg( & \frac{[T^{c_3},T^{c_4}]}{q^2 (p \cdot q_\perp)}
        + \frac{1}{2} \frac{\{T^{c_3},T^{c_4}\}}{(p \cdot q_\perp)^2} \bigg)\!
   \Big( e^x \cosh z - w\,e^x {\rm sinhc}\,z
       + \frac{w^2\!-z^2}{2} \Ea(x,y,z)
   \Big) \nn \\
 - & \frac{[T^{c_3},T^{c_4}]}{q^2 (p \cdot q_\perp)}
   \Big(x ( w^2\!+z^2) - w(x^2\!-y^2\!+z^2) \Big)
   \Eb(x,y,z)
   \bigg\} \,.
\end{align} 
It is instructive to inspect the spin-multipole expansion at low orders. We first note the similarity in the expansion of the two entire functions $\Ea(x,y,z)$ and $\Eb(x,y,z)$,
\beal\label{eq:EfnSpinExpansion}
    \Ea(x,y,z) & = 1 + \frac{2}{3}x + \frac{1}{12} (3 x^2 + y^2 + z^2) + \frac{1}{30} x (2 x^2 + y^2 + 2 z^2) + {\cal O}(a^4) \,, \\
    \Eb(x,y,z) & = \frac{1}{6} + \frac{1}{12}x + \frac{1}{120} (3 x^2 + y^2 + z^2) + \frac{1}{360} x (2 x^2 + y^2 + 2 z^2) + {\cal O}(a^4) \,.
\eeal
The relationship between the monomials in the $\Ea$ and $\Eb$ expansions is captured by the differential equation
\be
\frac{\partial}{\partial \lambda} \lambda^3 \Eb(\lambda x,\lambda y,\lambda z)\Big|_{\lambda=1} =\frac{1}{2}\Ea(x,y,z) \,.
\ee

The classical amplitude \eqref{eq:NAbClAmp} expanded up to the octupole order is then 
\begin{align}
{\cal A}&(\bm{1},\bar{\bm{2}},3^-\!,4^+) = 2g^2 (p \cdot \chi)^2 \nn \\*
& \times \bigg\{\!
  \bigg(  \frac{[T^{c_3},T^{c_4}]}{q^2 (p \cdot q_\perp)}
        + \frac{1}{2} \frac{\{T^{c_3},T^{c_4}\}}{(p \cdot q_\perp)^2} \bigg)\!
  \Big( 1+ x-w +\frac{1}{2} (w - x)^2 +\frac{1}{6} (w - x) (2 w x - x^2 - z^2) \Big) \nn \\* & \qquad
-   \frac{ [T^{c_3},T^{c_4}]}{ \, 6 q^2 (p \cdot q_\perp)} 
  \Big( x(w^2  +  z^2) - w (x^2 - y^2 + z^2) \Big) \bigg\}
+ {\cal O}(a^4) \, .
\end{align}
The expansion up to quadrupole order matches that of the exponential $e^{x-w}$, consistent with \rcite{Aoude:2020mlg}.
This follows from our chosen constraints below \eqn{eq:C(s)ansatz}, since the finite-spin classical limit of the spin-$1$ amplitude has this expansion.
However, in principle, this property can be relaxed, see \sec{sec:spinuniversality} below for more details.
Starting at the octupole order, the spin multipoles deviate from the exponential pattern, which they need to, since factors of $w^3$ and higher powers would give rise to unwanted spurious poles.
Our classical amplitude is free from unphysical poles and it has the expected factorization properties, see \app{app:factorisation}.

As a final minor remark, we note that beyond quadrupole order the amplitude  
\eqref{eq:NAbClAmp} does not exhibit the spin shift symmetry $a^\mu \to a^\mu + q^\mu/q^2$ proposed in \rcites{Aoude:2022trd, Bern:2022kto}.
While the variables $x, w$ are invariant under the shift, the breaking is due to the dependence on $y^2$ and $z^2$ in \eqref{eq:NAbClAmp}.
Therefore, the classical amplitude \eqref{eq:NAbClAmp} differs from the analogous gauge-theory Compton amplitudes obtained in \rcite{Aoude:2022trd}.
The difference in the result is due to contact terms proportional to $z^2$, as discussed in \sec{sec:contacttermbreaking}.

\subsection{Observations on spin universality}
\label{sec:spinuniversality}

The three-point \AHH amplitudes exhibit a universal behavior in a fixed-spin classical limit, a property referred to in the literature as \textit{spin universality} \cite{Holstein:2008sw,Holstein:2008sx,Vaidya:2014kza, Aoude:2022trd, Cangemi:2022abk}. Indeed, the quantum amplitudes can be {\it exactly} expressed as an exponential of the spin operator sandwiched between spin-$s$ eigenstates~\cite{Cangemi:2022abk}, 
\be 
{\cal A}_{\rm AHH}(\bm{1}^s,\bar{\bm{2}}^s,3^{\pm})= {\cal A}^{(0)}_3\, \langle e^{\pm \hat a\cdot q} \rangle\,,
\ee
even before any limit $\langle e^{\hat a\cdot q} \rangle\rightarrow e^{a\cdot q}$ is taken. 

More generally, a fixed-spin classical amplitude is typically obtained from some $\hbar \to 0$ procedure while working with a fixed quantum spin number $s$, giving a pseudo-classical amplitude of the form
\begin{equation}\label{eq:pseudoClAmp}
{\cal A}_{\text{cl}}(\bm{1}^s\!,\bar{\bm{2}}^s\!,3,\ldots) = \sum_{k+l=0}^{2s} c^{(s,l)}_{\mu_1 \cdots \mu_k} |a|^l a^{\mu_1} \cdots a^{\mu_k} ~ + ~ {\cal O}(a^{2s+1}) \,.
\end{equation}
Here the tensors $c^{(s,l)}_{\mu_1 \dots \mu_k}$ contain the physical information that informs us of the $S^{k+l}$ multipoles at finite $s$, and ${\cal O}(a^{2s+1})$ denotes terms not accessible from a computation involving finite-spin representations.
Since a proper classical limit should correspond to $s \to \infty$, the classical amplitude can in principle be inferred from  $c^{(\infty,l)}_{\mu_1 \cdots \mu_k}$. However, in practice, the amplitudes might not be known beyond a certain fixed $s$, thus complicating matters.

Families of amplitudes that exhibit spin universality will have $S^{k+l}$-multipole coefficients that are independent of $s$,
\begin{equation}
c^{(s,l)}_{\mu_1 \cdots \mu_k} = c^{(\infty,l)}_{\mu_1 \cdots \mu_k}\,,
\end{equation}
and as such they can be determined by a calculation involving any quantum amplitude of spin $2s\ge k+l$. This is a very desirable property, since each multipole can be computed from a single quantum  amplitude, rather than from an infinite family of amplitudes.

Let us briefly discuss the $\sqrt{\text{Kerr}}$ family of same-helicity Compton amplitudes, which clearly exhibit a version of spin universality, similar but not identical to the above three-point case. The quantum same-helicity amplitude has the form~\eqref{SameHelClassical}, 
\begin{equation} \label{ExactAmplExp}
    {\cal A}(\bm{1}^s\!,\bar{\bm{2}}^s\!,3^+\!,4^+)= {\cal A}^{(0)}_4 \langle e^{\hat a\cdot q} \rangle + {\cal O}\Big( \frac{1}{s}q^2 \hat a^2\Big) + {\cal O}(\hbar ) \,.
\end{equation}
where ${\cal A}^{(0)}$ is the scalar amplitude, and the expectation value $\langle e^{\hat a\cdot q} \rangle$ consists of the exponentiated spin operator acting on the spin-$s$ wavefunctions. The correction terms ${\cal O}( q^2 \hat a^2/s)$ correspond to Casimirs that originate both from spin-representation change (see \eqn{eq:spinrepchange}) and finite-representation identities of the $\hat a$ operator, and the terms ${\cal O}(\hbar )$ come from small $q$ factors that are not balanced by corresponding $\hat a$ factors. If we ignore the two last terms in \eqn{ExactAmplExp}, we can read off the spin-multipole tensors in \eqn{eq:pseudoClAmp} from the exponentiated quantum multipole moments, such that $c^{(s, l)}_{\mu_1 \dots \mu_k} = \delta_{l0} \, q_{\mu_1}\dots q_{\mu_k}/k!$, and we define the pseudo-classical amplitude as
\begin{equation}
{\cal A}_{\text{cl}}(\bm{1}^s\!,\bar{\bm{2}}^s\!,3^+\!,4^+)=  {\cal A}^{(0)}_4  e^{a\cdot q}    ,
\end{equation}
which also happens to be the correct classical amplitude. That is, the $s$ dependence is completely absent on the right-hand side, which thus exhibits spin universality. While the terms that we neglected in  \eqn{ExactAmplExp} do not exhibit spin universality, this is not of any practical issue, since those terms are easy to identify and remove, by hand, from any finite-$s$ calculation. 

We can apply the same analysis to the general-spin opposite-helicity Compton amplitude. The sector of the non-abelian amplitude~\eqref{eq:ColorDressedCompton} proportional to the commutator $[T^{c_3},T^{c_3}]$ in the classical limit behaves in an analogous way to the same-helicity amplitudes. One can define the pseudo-classical amplitude by computing the fixed-spin quantum amplitudes and neglecting any Casimirs generated from the spin-representation change \eqref{eq:spinrepchange}. However, the individual terms of that amplitude do not necessarily exhibit it, for instance the third term, proportional to $P_2^{(2s-1)}$, has an explicit dependence on $s$
\begin{equation}
t_{13}t_{14}\frac{\braket{\bm{1}3} \braket{3\bm{2}} [\bm{1}4] [4\bm{2}]}{m^{4s} \langle 3|1|4]^2}   P_{2}^{(2s-1)} \cleq - \frac{2}{s}\Big( e^{x} (w^2 - z^2)\, {\rm sinhc} \, z \Big) \,,
\end{equation} 
where $\cleq$ indicates that we extracted the pseudo-classical terms. Thus the term seemingly breaks the spin universality. 
However, this $s$-dependence is cancelled against a similar contribution on the second line of \eqn{eq:nabGaugeAmp}, such that the non-abelian amplitude as a whole is spin universal. The observed spin universality is partially explained by the fact that, in the non-abelian color-stripped amplitude, only the superclassical terms $\sim \hbar^{-1}$ survive when multiplied by $[T^{c_3},T^{c_3}] \sim \hbar$, and thus these terms are chiefly pole-term contributions that inherit the spin universality of the three-point amplitude. However, as discussed in the next subsection, contact terms can also develop a superclassical scaling if they grow with $s$, and thus in principle alter the universality properties. 

In contrast, the abelian sector of \eqref{eq:ColorDressedCompton}, proportional to the anticommutator $\{T^{c_3},T^{c_4}\}$, does not exhibit spin universality, as there is residual spin-dependence even after neglecting the Casimirs in the representation change formula. The full color dressed amplitude has the following non-universal contribution  
\beal
{\cal A}_{\rm cl}(\bm{1}^s\!,\bar{\bm{2}}^s\!,3^-\!,4^+) &
 - {\cal A}(\bm{1},\bar{\bm{2}},3^-\!,4^+)  \\ &
 = -\frac{(p \cdot \chi)^2}{(p \cdot q_{\perp})^2}
   \frac{w^2 -z^2}{ 2s (2s-1) } \big( \cosh z + (x+1-2s)\mathrm{sinhc}\,z\big) {\{T^{c_3},T^{c_4}\}} \,,
\eeal
and since only ${\{T^{c_3},T^{c_4}\}}$ appears the breaking of spin universality is in the abelian sector. Note, here ${\cal A}$ is the proper classical amplitude and ${\cal A}_{\rm cl}$ is the pseudo-classical one, as is clear from the indicated $s$ dependence. 

The breaking of spin universality in the abelian sector can be traced back to a sole term in the abelian amplitude: the $P_2^{(2s-2)}$ contribution in the contact term $C^{(s)}$. If we remove it by modifying the contact term $C^{(s)}\to C^{(s)}-\Delta C^{(s)}$, where the offending term is
\begin{equation}
    \Delta C^{(s)} = -\frac{\braket{\bm{1}3} \braket{3\bm{2}} [\bm{1}4] [4\bm{2}]}{ 2 m^{4s-1}} \big(\braket{\bm{12}} + [\bm{12}]\big) P^{(2s-2)}_2 \, ,
\end{equation}
then the abelian amplitudes also exhibit spin universality. However, this modification violates one of our main constraints used in the construction of the contact term, listed in \sec{sec:abAmp}, since $\Delta C^{(3/2)} \propto P_2^{(1)} \neq 0$. 
Thus, with this modification,  the  spin-$3/2$ amplitude obtained in \rcite{Chiodaroli:2021eug} would receive corrections.
Such a modification is not necessarily in conflict with the constraints given by massive gauge invariance; however, it would potentially require worsening the high-energy behavior of the spin-$3/2$ quantum amplitude, which appears undesirable form a fundamental-theory perspective.  
It is also important to note that, for the purpose of classical scattering, the $\Delta C^{(s)}$ term is irrelevant, and whether or not to keep it depends on one's preferences. We choose to work with the original contact term \eqref{eq:abContact}, were $C^{(3/2)}=0$.

\subsection{Consequences for other contact-term choices}
\label{sec:contacttermbreaking}

So far we have restricted our analysis to the amplitudes and contact terms that satisfy the constraints listed in \sec{sec:abAmp}, which result in a unique color-dressed classical amplitude \eqref{eq:NAbClAmp}. As expected, if we break one or several constraints, the classical amplitude is no longer unique and depends on free parameters.

For instance, let us consider loosening the constraint on the spin quadrupole; that is, we no longer assume the quadrupole is given by the spin-1 quantum amplitude. We can modify the abelian contact term by $C^{(s)}\to C^{(s)}+\Delta C^{(s)}$, where 
\be
\Delta C^{(s)} = \delta \frac{\braket{\bm{1}3} \braket{3\bm{2}} [\bm{1}4] [4\bm{2}]}{2m^{4s}}W_+ \Big(P_4^{(2s)}- W_+^2\, P_4^{(2s-2)}- P_2^{(2s-2)}\Big) \,,
\ee
which is constructed to satisfy all other constraints listed in \sec{sec:abAmp}. In the classical limit, the free parameter $\delta$ survives and the classical abelian amplitude is deformed to
\be \label{oneparafam}\!\!
A_{\rm U(1)}(\bm{1},\bar{\bm{2}},3^-\!,4^+) = -
\frac{(p \cdot \chi)^2 }{(p \cdot q_\perp)^2}
\Big( e^x \cosh z - w \, e^x {\rm sinhc}\, z + (1 - \delta)\frac{w^2- z^2}{2} \Ea(x,y,z)\Big) \,.
\ee 
The one free parameter modifies the classical amplitude to all orders in spin, starting with the quadruple. To restore the spin-$1$ quadrupole, which is the expected result even if it is not currently proven from first principles, the parameter must be fixed to $\delta =0$. Exploring more complicated but similar modifications $\Delta C^{(s)}$, constructed from polynomials $P^{(k)}_{2}$ and $P^{(k)}_{4}$, suggests that removing the quadrupole constraint always lead to the one-parameter classical amplitude~\eqref{oneparafam}. 

In \sec{sec:nonabAmp}, we choose to not add any contact terms proportional to the commutator $[ T^{c_3},T^{c_4} ]$, since the candidates were irrelevant in the classical limit.
In order to contribute classically, such contact terms must have kinematic dependence that is superclassical  ${\cal O}(\hbar^{-1})$ and are ruled out by restrictions on helicity structure and spin dependence.
Let us now consider relaxing more constraints so that we do get non-abelian contact terms in the classical limit. 

If we constrain ourselves to the $\braket{\bm{1}3} \braket{3\bm{2}} [\bm{1}4] [4\bm{2}]$ helicity structure, contact terms that respect the symmetries of the amplitude, see \sec{sec:ChiralQuarticOps}, are forced to be subleading in the classical limit. Contact terms can only contribute classically if we allow for coefficients that explicitly grow with $s$, breaking one of the constraints introduced in \sec{sec:abAmp}. For example, the following terms grow linearly in $s$:
\beal
\Delta C^{(s)}& \sim s  \frac{\braket{\bm{1}3} \braket{3\bm{2}} [\bm{1}4] [4\bm{2}]}{m^{4s-3}} (t_{14}-t_{13})P^{(2s+1)}_4 [ T^{c_3},T^{c_4} ]\,, \\
\Delta C^{(s)}& \sim s  \frac{\braket{\bm{1}3} \braket{3\bm{2}} [\bm{1}4] [4\bm{2}]}{m^{4s-1}} V P^{(2s)}_4 [ T^{c_3},T^{c_4} ]\,,
\eeal
where we restored the non-abelian commutator for clarity. 
Both examples above give the same classical-limit expressions,
\be
 \Delta {C}^{(\infty)} \sim  -
(p \cdot \chi)^2\frac{(w^2 - z^2) }{(p \cdot q_\perp)^2}  \frac{z^2}{p \cdot q_{\perp}} \,{\Eb}(x,y,z)[ T^{c_3},T^{c_4} ] \,,
\ee
and will contribute from the $S^4$ spin-multipole order. 

There exist other relevant contact terms that do not have explicit $s$ prefactors, if we allow other helicity-dependent structures. Starting at $s=2$ the structure $\braket{\bm{1}3}^2  [4\bm{2}]^2- \braket{\bm{2}3}^2 [4\bm{1}]^2$ is compatible with massive gauge invariance, as shown in \eqn{eq:nonabquarticContact}. A viable contact term is then
\be
\Delta C^{(s)} \sim \frac{\braket{\bm{1}3}^2  [4\bm{2}]^2- \braket{\bm{2}3}^2 [4\bm{1}]^2}{m^{4s-3}} P^{(2s+1)}_4 [ T^{c_3},T^{c_4} ]\,.
\ee
The term contributes classically as
\be
\Delta C^{(\infty)} \sim -
(p \cdot \chi)^2\frac{w z^2 }{(p \cdot q_\perp)^3} \,{\Eb}(x,y,z)[ T^{c_3},T^{c_4} ] \,,
\ee
and will contribute first from the cubic order in spin. Note that this contact term is odd in $w$, as opposed to the $(w^2-z^2)$ prefactor present in all other contact terms and inherited from the helicity structure $\braket{\bm{1}3} \braket{3\bm{2}} [\bm{1}4] [4\bm{2}]$. When attempting to match to the spin-shift symmetric result in \rcite{Aoude:2020onz}, we note the need to add contact terms with more general helicity structures, given the mismatch is not proportional to $(w^2-z^2)$. However, such contact terms seems less natural from the quantum amplitude structure, obtained from the chiral Lagrangian \eqref{FinalRootKerrLagrangian}, and would require further justification for their inclusion.

\section{Conclusion}
\label{sec:outro}

The present paper builds upon recently introduced ideas for describing higher-spin amplitudes for Kerr black holes, such as massive higher-spin gauge symmetry~\cite{Cangemi:2022bew} and the chiral field approach~\cite{Ochirov:2022nqz}, and applies them to the gauge-theory case known as $\sqrt{\text{Kerr}}$~\cite{Arkani-Hamed:2019ymq}.

The classical $\sqrt{\text{Kerr}}$ object should correspond to a rotating charged ring (or disk) of matter~\cite{Monteiro:2014cda} interacting with an electromagnetic or non-abelian gauge field. For our purposes, it is simply an interesting toy model where we can learn about the same features that one encounters for a Kerr BH.
The connection between Kerr and $\sqrt{\text{Kerr}}$ comes from the classical double copy~\cite{Monteiro:2014cda}.
Its quantum counterpart at three points corresponds to the spin-$s$ AHH amplitudes~\cite{Arkani-Hamed:2017jhn}, while at the Compton level the correspondence is expected to be less immediate for higher-spin cases~\cite{Johansson:2019dnu,Chiodaroli:2021eug}.   

Construction of higher-spin gauge-theory Compton amplitudes has seen some recent activity~\cite{Aoude:2022trd,Bjerrum-Bohr:2023jau,Alessio:2023kgf,Bjerrum-Bohr:2023iey} as a stepping stone for obtaining gravitational amplitudes. While the gravitational Compton amplitudes can be inferred from general-relativity matching calculations using black-hole perturbation theory~\cite{Bautista:2023szu}, a suitable gauge-theory framework for $\sqrt{\text{Kerr}}$ has not been explored. Since it is currently not clear what is the precise physics that controls the contact ambiguities of the $\sqrt{\text{Kerr}}$ Compton amplitudes, we explore a range of possible approaches and constraints in this work. In the forthcoming paper~\cite{Upcoming2}, will report our detailed findings regarding gravitationally interacting massive higher-spin theories relevant for Kerr Compton amplitudes.

There are several complementary higher-spin ideas used in the current work: off-shell higher-spin gauge symmetry, its on-shell version in the form of Ward identities, and the chiral description.
A priori, massive fields are non-gauge systems with complicated second class constraints that are hard to control, and the main idea behind introducing the gauge description of higher-spin fields is to make the constraints manifest themselves as gauge invariance, which is conceptually easy to impose, see \eg \rcites{Zinoviev:2001dt,Zinoviev:2006im,Zinoviev:2008ck,Zinoviev:2009hu,Zinoviev:2010cr,Buchbinder:2012iz} and the recent \rcite{Lindwasser:2023dcv}.
The free Lagrangian ${\cal L}_2$ is invariant under the linear gauge symmetries $\delta_0$, and the interactions are constructed order by order in fields, starting from the cubic terms ${\cal L}_3$ and from the deformations $\delta_1$ of the gauge symmetries that are linear in fields.
Importantly, certain constraints on the number of derivatives have to be imposed on the deformations.
An observed guiding principle for targeting $\sqrt{\text{Kerr}}$ objects, as well as Kerr BHs, is to look for the interactions with the lowest possible number of derivatives and hence the best possible high-energy behavior. 

Using Ward identities on amputated correlation functions (off-shell amplitudes) has immediate practical advantages. The deformations $\delta_{k>1}$ of the free gauge symmetry $\delta_0$ are not needed explicitly, and the amputated correlation function evaluated on shell directly gives the amplitude, which is the main objective. In order to constrain the Compton amplitude, one can solve the reduced Ward identities at the cubic level, \ie only for the vertices that actually contribute to the exchange diagram.

A single double-traceless tensor $\Phi_{\mu_1\ldots\mu_s}$, also known as the Fronsdal field~\cite{Fronsdal:1978rb}, is sufficient for describing a single free massless spin-$s$ field; similarly, there is a part of $\delta_0$ which is the usual massless gauge symmetry.
It is then not surprising that, at least at the cubic order, the leading part of the vertex coincides with the one of a massless spin-$s$ field.
For the EM and gravitational interactions, the simplest vertex is known to have $2s-1$ and $2s-2$ derivatives, respectively, see \eg \rcites{Berends:1984rq,Bengtsson:1986kh,Metsaev:2005ar}.\footnote{Note that the are also electromagnetic and gravitational vertices with one and two derivatives, respectively \cite{Bengtsson:1986kh}, whose existence seem to also play a role for massive higher-spin fields.
}
It would be interesting if there is a more direct relation to higher-spin gravities (theories with massless higher-spin fields that are governed by infinite-dimensional gauge symmetries) beyond cubic vertices, which require infinite multiplets of massless higher-spin fields, e.g. via a higher-spin analog of the Higgs mechanism.

We employed the chiral approach for constructing interactions of massive higher-spin fields, proposed in \rcite{Ochirov:2022nqz}. The main idea is to rely on field variables $\Phi_{\alpha_1\ldots \alpha_{2s}}$ that are in $(2s,0)$ representation of the Lorentz algebra ${\rm sl}(2,\mathbb{C})$.
The central advantage is the relaxation of the transversality constraint; however, parity invariance is not automatic. At least for low spins, the chiral fields are directly related to the (spin)-tensor fields via certain parent actions~\cite{Chalmers:1997ui,Chalmers:2001cy}.
There exists other ideas for controlling the degrees of freedom for massive higher-spin fields, which would be interesting to apply in the future: the covariant ones \cite{Buchbinder:2005ua,Buchbinder:2007ix,Kaparulin:2012px,Kazinski:2005eb} and the light-cone gauge \cite{Metsaev:2005ar,Metsaev:2007rn,Metsaev:2022yvb}. 

Additional assumptions were used to narrow down the $\sqrt{\text{Keer}}$ Compton amplitude, including consistency with previously known patterns. The need for additional constraints does not immediately imply that the massive higher-spin gauge symmetry is less restrictive.
Indeed, gauge symmetry is a way to construct all consistent interactions, and extra assumptions related to minimality of interactions is likely needed in order to land on the correct theory describing the dynamics of Kerr BHs and $\sqrt{\text{Keer}}$ objects. Other compact objects can be modelled by turning on higher-derivative non-minimal interactions. It would be interesting to see if the minimality of the interactions imposed at higher orders will automatically fix the free coefficients at lower orders, starting from the quartic one, see \eg \rcite{Haddad:2023ylx} for an example of the cubic-quartic interplay.
For example, this is what happens in higher-spin gravities: while consistency at the cubic order leaves infinitely many parameters free, the higher-order consistency fixes them completely, see \eg \rcites{Metsaev:1991mt,Metsaev:1991nb,Ponomarev:2016lrm}.
The same effect can be observed in supergravities, but (gauged) supersymmetry still leaves infinitely many free parameters to be seen via counterterms.
Another interesting direction would be to relate our approach to recent versions of the worldline formalism with spin~\cite{Porto:2016pyg,Levi:2018nxp,Liu:2021zxr,Jakobsen:2021zvh,Goldberger:2022rqf,Ben-Shahar:2023djm}. For instance, one could explore realizations of massive gauge symmetry on the worldline, or see if connecting the two approaches can help nail down the BH amplitudes uniquely.

One effect that has not been considered in the paper at all, but can in principle be implemented is the spin-changing interactions.
Indeed, one can add various vertices of type $s{-}s'{-}1$ or $s{-}s'{-}2$, for electromagnetic and gravitational interactions, respectively.
On the same footing one can introduce mass-changing interactions.
Such interactions have recently be considered in \rcites{Bern:2023ity,Aoude:2023fdm,Jones:2023ugm}.
Provided both types of interactions are introduced and the masses are kept discrete (with arbitrarily small gaps), one gets a spectrum that is similar to string theory, \ie infinitely degenerate in spin with mass gaps of order ${\alpha'}^{-1}$.
Therefore, the complete effective theory of BH dynamics is a string-like theory in $4d$ whose spectrum contains massive higher-spin fields and a graviton.
Whether it is valid beyond the EFT regime can be an interesting question on its own.

\begin{acknowledgments}

We thank Francesco Alessio, Fabian Bautista, Maor Ben-Shahar, Zvi Bern, Lara Bohnenblust, Paolo Di Vecchia, Agata Grechko, Alfredo Guevara, Kays Haddad, Chris Kavanagh, Lionel Mason, Fei Teng, Radu Roiban, and Justin Vines for useful discussions related to this work.
This research is supported in part by the Knut and Alice Wallenberg Foundation under grants KAW 2018.0116 (From Scattering Amplitudes to Gravitational Waves) and KAW 2018.0162.
The work of M.C. is also supported by the Swedish Research Council under grant 2019-05283.
E.S. is a Research Associate of the Fund for Scientific Research (FNRS), Belgium. The work of E.S. was supported by the European Research Council (ERC) under the European Union’s Horizon 2020 research and innovation programme (grant agreement No 101002551). 

\end{acknowledgments}

\appendix
\section{Variables and conventions}
\label{sec:appConventions}

The massive Weyl spinors for particle $i = 1,2$ are defined as
\beal
\ket{{\bm i}} & = \ket{i^a} z_{i a} \,, \qquad \quad
\ket{\bar {\bm i}} = \ket{i^a} \bar z_{i a} \,, \\ 
|{\bm i}] & = |i^a] z_{i a} \,, \qquad\,\quad
|\bar {\bm i}] = |i^a] \bar z_{i a} \,, 
\eeal
where the unbarred spinors are incoming, and the barred spinors are outgoing.
The $z_a$ and $\bar z_a$ variables are ${\rm SU}(2)$ little-group spinors, they describe the physical (spin up or down) wavefunction for a massive particle.
The little-group indices are raised and lowered using the ${\rm SU}(2)$ Levi-Civita symbol, \eg $z^a=\epsilon^{ab} z_b$ and $z_a=\epsilon_{ab} z^b$, where $\epsilon^{12}=\epsilon_{21}=1$. Under complex conjugation, the variables behave as $(z_a)^* =\bar z^a $ and $(z^a)^* =-\bar z_a$.

The massive spinors obey the on-shell constraints, \ie the Dirac equation:
\be
p_i \cdot \sigma |{\bm i}] = m \ket{{\bm i}} \,, \qquad \quad
p_i \cdot \bar \sigma \ket{{\bm i}} = m |{\bm i}] \,.
\ee
They are normalized as 
\be \label{normEqnP1}
\braket{\bar{\bm i} {\bm i}} = [{\bm i} \bar{\bm i}] = m |z_i|^2 \,, \qquad \quad
\braket{{\bm i}{\bm i}} = [{\bm i}{\bm i}] = 0 \,,
\ee
and, in general, the spinor brackets are antisymmetric $\langle  {\bm i} {\bm j}\rangle=-\langle  {\bm j} {\bm i}\rangle$, $[{\bm i} {\bm j}]=-[ {\bm j} {\bm i}]$. 
The natural contraction of the SU(2) spinors is 
\be
|z_i|^2=z_{i a} \bar z_i^a =z_{i 1} \bar z_{i 2} - z_{i 2}  \bar z_{i 1} \,,
\ee
and we often assume the constraint $|z_i|^2=1$ to have properly normalized wavefunctions. Note that the wavefunctions can also be defined projectively as ``rays'', in which case $|z|^2\neq1$ but the overall scale is irrelevant. With either the constraint, or the projective interpretation, the $z_a$ coordinates parameterize the 3-sphere (SU(2) group manifold).    

Having introduced the massive spinor-helicity variables, we can now construct massive polarization vectors for particles $1$ and $2$ out of their massive spinors, 
\be
{\bm \varepsilon}_1^\mu = \frac{\langle\bm{1}|\sigma^\mu|\bm{1}]}{\sqrt{2}m} \,, \qquad \quad
{\bm \varepsilon_2^\mu} = \frac{\langle\bm{2}|\sigma^\mu|\bm{2}]}{\sqrt{2}m} \,,
\ee
such that they are null, ${\bm \varepsilon}_i^2=0$, and satisfy transversality $p_i \cdot {\bm \varepsilon}_i=0$. The individual little-group components can be extracted by taking derivatives with respect to the wavefunctions $z_i^a$, see \rcite{Chiodaroli:2021eug} for more details.

We will consider Compton scattering in an all-incoming convention. The incoming massive $\sqrt{\text{Kerr}}$ object (with momentum $p_1$ and spin $s$) and massless plane wave (with momentum $p_3$ and helicity $h_3$) scatter and produce outgoing $\sqrt{\text{Kerr}}$ object ($-p_2$ and $s$) and massless plane-wave ($-p_4$ and helicity $-h_4$), where
\be
p_1 + p_2 + p_3 + p_4 = 0 \,, \qquad \quad
p_1^2 = p_2^2 = m^2 \,, \qquad \quad
p_3^2 = p_4^2 =0 \,.
\ee

For convenience, we also make use of the following independent momenta
\be
p=p_1\,, \qquad \quad
q=p_3+p_4\,, \qquad \quad
q_\perp=p_4-p_3\,,
\ee
which satisfy
\be
p^2=m^2\,, \qquad \quad
q_\perp^2 = 2 p \cdot q=-q^2 \,, \qquad \quad
q_\perp\!\cdot q=0 \,.
\ee
For the opposite-helicity configuration $h_3=-1, h_4=1$, we choose the massless polarization vectors as
\be
\varepsilon_3^{-\mu}= \frac{\langle 3|\sigma^\mu|4]}{\sqrt{2}[34]} \,, \qquad \quad
\varepsilon_4^{+\mu}= \frac{\langle 3|\sigma^\mu|4]}{\sqrt{2}\langle 34 \rangle}\,,
\ee
which corresponds to picking the gauge where
$p_4 \cdot \varepsilon_3^{-} =0 = p_3 \cdot \varepsilon_4^{+}$,
although all the amplitudes written are gauge-invariant.
We can thus encode the helicity information in a complex vector $\chi^\mu := \langle 3|\sigma^\mu|4]$. 

\paragraph{Vector variables.}
Note that for the opposite-helicity Compton amplitude one can trade all spinorial variables for the following (over-)complete basis of six vector variables
\be
p^\mu\,,~ q^\mu\,,~ q_\perp^\mu\,,~ \chi^\mu\,,~ \rho^\mu\,,~ \bar\rho^\mu\,,~
\ee
defined in \eqns{eq:clMomScaling}{eq:rhoVectors}. Any quantum opposite-helicity Compton amplitude must be built out of Lorentz invariants formed by these six vectors. 
There are at most nine independent such dot products:
\begin{alignat}{3} \label{9dotproducts}
 & p \cdot q_\perp \,, \qquad &&
   q^2 \,, \qquad &&
   p \cdot \chi \,, \nn \\
 & p \cdot \rho \,, \qquad &&
   \rho \cdot q_\perp \,, \qquad &&
   \rho \cdot \chi \,, \\
 & p \cdot \bar\rho \,, \qquad &&
   \bar\rho \cdot q_\perp \,, \qquad &&
   \bar\rho \cdot \chi \,. \nn
\end{alignat}
The simple linear relations that eliminate the remaining dot products are
\beal
 & p^2=m^2\,, \qquad\,\quad
   q_\perp^2 = -q^2\,, \qquad \quad
   p \cdot q = -\frac{1}{2} q^2\,, \\
 & \bar\rho^2 = -\rho^2
\,, \qquad
   p \cdot \bar\rho = -\frac{1}{2} q \cdot \bar\rho\,, \\
 & q \cdot q_\perp = q \cdot \rho = q \cdot \chi= q_\perp\!\cdot \chi = \rho \cdot \bar\rho = \chi^2 = 0\,,
\eeal
as well as the simple quadratic relation
\begin{align}
     \rho^2 = \frac{(p \cdot \rho)^2-(p \cdot \bar{\rho})^2}{m^2}\,.
\end{align}

There are further more complicated non-linear relations, most of which follow from Gram determinant relations, since the six vectors span an overcomplete basis in four dimensions.
The following quadratic identity, in principle, allows to eliminate the spinless variable $p\cdot\chi$:
\be \label{ExtraNonLinearConstraint}
p\cdot\chi\  = \frac{2 p\cdot\rho \, \rho\cdot\chi + \bar \rho\cdot q_\perp \rho\cdot\chi - 
  \rho\cdot q_\perp \,\bar \rho\cdot\chi}{2 \rho^2}\,.
\ee
However, since it comes at the expense of a spurious non-locality, this relation must be used with some care. 
Likewise, the Mandelstam variables $q^2$ and $p \cdot q_\perp$ can be eliminated at the expense of spurious non-localities:
\beal
q^2 &= \frac{1}{m^2}\frac{4\big(\rho \cdot p \, p \cdot \chi - m^2 \rho \cdot\chi\big)^2 - (p \cdot \chi)^2 (q \cdot \bar\rho)^2}{ (\rho \cdot \chi)^2 - (\bar \rho\cdot \chi)^2} \,, \\
(p\cdot q_\perp)^2 &= \frac{-4 p\cdot \chi \,
    q_\perp\cdot \rho \big(p\cdot \chi \, q_\perp\cdot \rho - 
     2 p\cdot q_\perp\,\rho\cdot \chi\big) + 
  q^4 (\bar\rho\cdot \chi)^2}{4 (\rho\cdot \chi)^2} \,.
\eeal

In principle, we could also consider Levi-Civita terms built out of the six vectors, however, they they all reduce to combinations of the nine dot products~\eqref{9dotproducts}. 
This can be deduced from the three dualization identities
\beal
    i\epsilon^{\mu \nu \rho \sigma} q_{\rho} \chi_\sigma &=q_\perp^\mu \chi^\nu- q_\perp^\nu \chi^\mu\,, \\
    i\epsilon^{\mu \nu \rho \sigma} q_{\perp \rho} \chi_\sigma  &=q^\mu \chi^\nu- q^\nu \chi^\mu \,, \\
    i \epsilon^{\mu \nu \rho \sigma}  p_{1 \nu} \rho_{\rho} \bar{\rho}_\sigma &= \rho^\mu 
    \,(p_1 \cdot \rho)-\bar{\rho}^\mu \, (p_1 \cdot \bar{\rho})-p_1^\mu \rho^2.
\eeal

Let us check that one can re-express the spinorial Compton variables into the above vector variables; we have the helicity-independent variables
\beal
\label{eq:ComptonSpinVars}
W_+ &= 
\frac{m}{2}\left( \braket{\bm{12}}+[\bm{12}] \right)  
= -p\cdot \rho\,, \\
W_- &= 
\frac{m}{2}\left( \braket{\bm{12}}-[\bm{12}] \right) 
=-p \cdot \bar{\rho}\,, \\
U &= \frac{1}{2 }\big(\langle\bm{1}|4|\bm{2}] - \langle\bm{2}|4|\bm{1}]\big)- m[\bm{12}] =  p \cdot \rho  - \frac{1}{2}\bar \rho \cdot q_\perp\,, \\
V &= \frac{1}{2}\big(\langle\bm{1}|4|\bm{2}] + \langle\bm{2}|4|\bm{1}]\big) = \frac{1}{2}\rho \cdot q_\perp\,.
\eeal

\paragraph{Compton building blocks.}
For the matrix elements considered in \eqn{ContactTermsF4pt}, we have the following expressions:
\beal
\label{ContactTermsF4pt2}
\bra{\Phi} \FF_1 \ket{\Phi}\big|^{s=0}_{(2,3^-\!,4^+\!,1)} &
 = \bra{3}1|4]^2
 = (p\cdot\chi)^2 \,,
\\
\bra{\Phi} \FF_2 \ket{\Phi}\big|^{s=1/2}_{(2,3^-\!,4^+\!,1)} & = \frac{1}{2} \bra{3}1|4]
   \big( \braket{\bm{2}3} [4\bm{1}] + [\bm{2}4] \braket{3\bm{1}} \big)
 = \frac{1}{2} (p\cdot\chi) (\rho\cdot\chi) \,,
\\
\bra{\Phi} \FF_3 \ket{\Phi}\big|^{s=1/2}_{(2,3^-\!,4^+\!,1)} & = \frac{1}{2} \bra{3}1|4]
   \big( \braket{\bm{2}3} [4\bm{1}] - [\bm{2}4] \braket{3\bm{1}} \big)
 = \frac{1}{2} (p\cdot\chi) (\bar{\rho}\cdot\chi) \,,
\\
\bra{\Phi} \FF_4 \ket{\Phi}\big||^{s=1}_{(2,3^-\!,4^+\!,1)} & = \frac{1}{2}
   \big(\braket{\bm{2}3}^2 [4\bm{1}]^2 - [\bm{2}4]^2 \braket{3\bm{1}}^2\big)
 = \frac{1}{2} (\rho\cdot\chi)(\bar{\rho}\cdot\chi) \,,
\\
\bra{\Phi} \FF_5 \ket{\Phi}\big|^{s=1}_{(2,3^-\!,4^+\!,1)} & = \frac{1}{2}
   \big(\braket{\bm{2}3}^2 [4\bm{1}]^2 + [\bm{2}4]^2 \braket{3\bm{1}}^2\big)
 = \frac{1}{4} \big((\rho\cdot\chi)^2\!+ (\bar{\rho}\cdot\chi)^2\big) \,,
\\
\bra{\Phi} \FF_6 \ket{\Phi}\big|^{s=1}_{(2,3^-\!,4^+\!,1)} &
 = \frac{1}{2} \braket{\bm{2}3} \braket{3\bm{1}} [\bm{2}4] [4\bm{1}]
 = \frac{1}{8} \big((\rho\cdot\chi)^2\!- (\bar{\rho}\cdot\chi)^2\big) \,.
\eeal

\begin{table}[t]
\begin{center}
\begin{tabular}{|c||c|c|c|c|c|c|c|c|c|c|} 
\hline
$k \backslash s$ & 0 & $\frac{1}{2}$ & 1 & $\frac{3}{2}$ & 2 & $\frac{5}{2}$ & 3 &$\frac{7}{2}$& 4  \\ [0.5ex]
\hline
\hline
0& 1 & 6 & 20 & 46 & 87 & 146 & 226 & 330 & 461 \\ [0.5ex]
\hline
1 & 2 & 10 & 28 & 58 & 102 & 162 & 240 & 338 & 458 \\ [0.5ex]
\hline
2 & 3 & 14 & 37 & 74 & 127 & 198 & 289 & 402 & 539 \\ [0.5ex]
\hline
3 & 4 & 18 & 46 & 90 & 152 & 234 & 338 & 466 & 620 \\ [0.5ex]
\hline
4 & 5 & 22 & 55 & 106 & 177 & 270 & 387 & 530 & 701\\ [0.5ex]
\hline 
\end{tabular}
\caption{\label{Table5:OperatorCounts}The number of independent four-point operators for the opposite-helicity Compton amplitude. The operators are classified by spin~$s$ and the power~$k$ of Mandelstam variables. The $C$, $P$ and $T$ symmetries are not yet imposed, which would reduce the counts.}
\end{center}
\end{table}

Using the above vector variables, together with the linear and non-linear constraints, one can build up independent matrix elements for the Compton contact terms, which are in one-to-one correspondence with independent local operators. 
We find that the number of independent four-point operators for the opposite-helicity Compton amplitude are given by the formula
\be
N_{s,k} =(2 s + 1) (4 s^2 + 2 s + 1) (k + 1) - N^{\rm GD}_{s,k} \,,
\ee
where $k$ is the power of the Mandelstam variables $\sim D_i \cdot D_j$ (where no chiral indices are contracted with the fields). 

The function $N^{\rm GD}_{s,k}$ is the subtraction needed from the naive count, and it is due to Gram determinant relations as well as the non-linear constraint \eqn{ExtraNonLinearConstraint},
\be
 N^{\rm GD}_{s,k}=
\left\{
\begin{matrix}
s (2 s-1)^2 \,, & k = 0  \\
\frac{2}{3} s (2 s - 1) (4 s - 1) + 4 k s^2  (2 s + 1) \,, ~~~~~ & k > 0
\end{matrix} \right\}
\ee
and we use the rule that Mandelstam variables are reduced to the lowest power possible using the Gram-determinant relations.
Note that these operators are counted without imposing any crossing or CPT symmetries, and thus they include abelian, non-abelian and non-hermitian terms (dissipative). See Tables~\ref{Table5:OperatorCounts} and \ref{Table6:OperatorCounts}  for the explicit counts at low spin $s$ and low powers $k$ of Mandelstam-variable factors.

\begin{table}[t]
\begin{center}
\begin{tabular}{|c||c|c|c|c|c|c|c|c|c|c|} 
\hline
$s$ & 0 & $\frac{1}{2}$ & 1 & $\frac{3}{2}$ & 2 & $\frac{5}{2}$ & 3 &$\frac{7}{2}$& 4  \\ [0.5ex]
\hline
\hline
& 0 & 0 & 3 & 12 & 30 & 60 & 105 & 168 & 252  \\ [0.5ex]
\hline
\end{tabular}
\caption{\label{Table6:OperatorCounts}Counting independent operators for the interesting case where the two Mandelstam variables as well as $\chi \cdot p$ are absent. These are given by the tetrahedral numbers $(2s - 1) 2s (2s + 1)/6$ times a factor of 3, because of the three possible quadratic monomials of $\rho \cdot \chi$, $\bar \rho \cdot \chi$.}
\end{center}
\end{table}

Under the two $\mathbb{Z}_2$ symmetries (combinations of crossing symmetry and complex conjugation, as in the parity constraint~\eqref{ParityConstraintOrdered}) of the amplitude, these variables transform as
\begin{subequations} \begin{align}
\bm{1} \leftrightarrow \bm{2}~:~&
\begin{cases} \quad
U \to -U \,, \qquad\,\quad V \to V \,,~\qquad
W_\pm \to -W_\pm \,, \\
p\cdot\chi \to -p\cdot\chi \,, \quad
\rho\cdot\chi \to \rho\cdot\chi \,, \quad
\bar{\rho}\cdot\chi \to - \bar{\rho}\cdot\chi \,,
\end{cases} \\
3 \leftrightarrow 4 \,+ \,{\rm c.c.} :~&
\begin{cases} \quad
U \to U \,, \qquad \qquad V \to -V \,, \qquad\!~
W_\pm \to \pm W_\pm \,, \\
p\cdot\chi \to -p\cdot\chi \,, \quad
\rho\cdot\chi \to -\rho\cdot\chi \,, \quad
\bar{\rho}\cdot\chi \to \bar{\rho}\cdot\chi \,.
\end{cases}
\end{align} \end{subequations}
The polynomials $P_2^{(k)}$ and $P_4^{(k)}$ are invariant under the latter symmetry, while they pick up a factor of $(-1)^{k+1}$ under the former. 
For bosonic states $\Phi^s$, the Compton amplitude is invariant under both symmetries, and for fermionic states it is odd under the first symmetry and even under the second.
In terms of our polynomials, this translates to $\{P^{(k={\rm odd})}_{2}\,,~ U P^{(k={\rm even})}_{2} \}$ depending only on powers of $\{U^2,V^2\}$, and likewise $\{ P^{(s={\rm integer)}}_4\,,~ U P^{(s=\text{half-integer})}_4 \}$ depending only on powers of $\{U^2,V^2,W_\pm^2, U W_{+}\}$, since these variable combinations are invariant under the symmetries.

\section{Chiral contact-term freedom}
\label{app:ChiralQuarticContact}

Here we present a way to classify the terms that can be freely added to the basic chiral Lagrangian $\braket{D_\mu \Phi|D^\mu \Phi} - m^2 \braket{\Phi|\Phi}$.
As we have seen in \sec{sec:ChiralCubicLag}, such non-minimal terms are required to restore parity.
Moreover, they are needed to describe compact objects that are more complicated than black holes.

By definition, the set of non-trivial contact terms is the quotient by field redefinitions of all possible contact terms that do not vanish on shell.
In practice, one often wishes to impose certain (anti)-symmetry with respect to the exchange of the massive legs due to the reality condition~\eqref{QuasiReality}, but let us not take it into account for the moment.
In order to write down the on-shell non-trivial terms, we need to know the on-shell jet space (in the mathematical sense) of the gauge fields and the chiral massive fields.
The former amount to the gluon's (anti)-self-dual field strengths $F_-^{\ga\gb}$ and $F_+^{\gad\gbd}$.  
It is quite easy to show that all of their derivatives that are distinct on shell can be parameterized by
\be
F^{\ga(n+2),\gbd(n)} := D^{\ga\gbd} \cdots D^{\ga\gbd} F_-^{\ga\ga}\,, \qquad
F^{\ga(n),\gbd(n+2)} := D^{\ga\gbd} \cdots D^{\ga\gbd} F_+^{\gbd\gbd}\,, \qquad 
n \ge 0 \,,
\label{FieldStrengthDerivatives}
\ee
where all like indices are understood as symmetrized. 
Meanwhile, the on-shell derivatives of the massive field are in one-to-one correspondence with the following set of spin-tensors:
\be
\Phi^{\ga(2s-2k+n),\gbd(n)}
:= D^{\ga(n-k)\gc(k)\gbd(n)}\Phi^{\ga(2s-k)}{}_{\gc(k)} \,, \qquad
n \ge 0 \,, \qquad 0 \leq k \leq n \,.
\label{ChiralFieldDerivatives}
\ee
Any contact term is a linear combination of singlets in the tensor product of the appropriate number of the above spin-tensors.
Total derivatives can be modded out by requiring one external leg not to have derivatives.	

Let us denote the irreducible representation corresponding to a symmetric spin-tensor $T^{\ga(m),\gbd(n)}$ as $(m,n)$ and discuss the relevant lower-point vertices.
For instance, the set of three-point operators of the type $(D...D\Phi)^\star (D...D\Phi) \Fp$ is spanned by singlets in
\be
(2s-2k+n,n) \otimes (2s-2k'+n',n') \otimes (0,2) \,.
\ee
Since all undotted indices must be contracted between the first and the second entry, we have $2s-2k+n=2s-2k'+n'$.
The two dotted indices of the third (field-strength) entry, on the other hand, may be contracted either with both (matter) entries or with one of them.
This means that either $n=n'$ or $n=n' \pm 2$,
and hence $k=k'$ or $k=k' \pm 1$, respectively.
Note that the AHH amplitudes~\eqref{AmpGaugeAHH3} did not require us to introduce any operators of this kind.
Instead, the actual $2s$ three-point operators in the Lagrangian~\eqref{ActionGaugeAHH} belong to the type $(D...D\Phi)^\star (D...D\Phi) \Fm$, for which we consider
\be
(2s-2k+n,n) \otimes (2s-2k'+n',n') \otimes (2,0) \,.
\ee
The singlets require $n=n'$, therefore $k=k'$ or $k=k' \pm 1$,
where only the first option is realized in \eqn{ActionGaugeAHH}.
Moreover, only the $k=n$ operators were shown to contribute, \ie for each dotted derivative index of $(D...D\Phi)$ one undotted index is contracted between a derivative and $\Phi$.
This is clearly a small subspace of possible cubic vertices.

More generally, we know from purely on-shell considerations~\cite{Arkani-Hamed:2017jhn} that there are $2s$ independent three-point couplings contributing to the minus-helicity amplitude (excluding the covariant-derivative coupling constant).
On the other hand, if we assume the maximum number of derivatives is $2(2s-1)$, as suggested by the cubic Lagrangian~\eqref{ActionGaugeAHH}, we observe $s(6s-1)$ off-shell operators of the type $(D...D\Phi)^\star (D...D\Phi) \Fm$.
This overcounting can be explained by the possibility to generate trivial interactions via field redefinitions $\Phi \to \Phi+(D...DF_-)(D...D\Phi)$, which we have not tried to rule out here. 
In other words, the present analysis gives us a significantly \emph{overcomplete} set of off-shell operators.
At each multiplicity~$n$, however, we may make a suitable choice,
such as the three-point action~\eqref{ActionGaugeAHH}, which will automatically include certain higher-order contributions, but the same analysis at the next order will once again give us access to the entire space of $(n+1)$-point couplings.

At the quartic level, there are three types of contact terms that are given by the singlets in the following tensor products:
\be
(2s-2k+n,n) \otimes (2s-2k'+n',n') \otimes (m+2,m)\otimes (2,0)
\label{Omm}
\ee
for terms of the type $(D...D\Phi)^\star (D...D\Phi) (D...D\Fm) \Fm$,
\be
(2s-2k+n,n) \otimes (2s-2k'+n',n') \otimes (m+2,m)\otimes (0,2)
\label{Omp}
\ee
for $(D...D\Phi)^\star (D...D\Phi) (D...D\Fm) \Fp$, and finally
\be
(2s-2k+n,n) \otimes (2s-2k'+n',n') \otimes (m,m+2)\otimes (0,2)
\label{Opp}
\ee
$(D...D\Phi)^\star (D...D\Phi) (D...D\Fp) \Fp$.

The cubic Lagrangian~\eqref{ActionGaugeAHH} suggests that only a limited number of possible contact terms should be relevant if we are interested in modeling black holes, for which a certain minimality concept is believed to exist.
In particular, \cite{Ochirov:2022nqz} explains that no terms of the third type $(D...D\Phi)^\star (D...D\Phi) (D...D\Fp) \Fp$ are needed at all.
In fact, the spin-1 action~\eqref{Spin1ActionChiralFull} contains only certain terms of the first type $(D...D\Phi)^\star (D...D\Phi) (D...D\Fm) \Fm$ and none of the second type $(D...D\Phi)^\star (D...D\Phi) (D...D\Fm) \Fp$.
More generally, we need terms of the first type to restore parity
and cannot entirely rule out the terms of the second type.

Let us illustrate this general method on a few lower-spin cases.
To bound this problem, we need to formulate a power-counting assumption applicable in the chiral formalism.
Note that we cannot use our intuition from the non-chiral actions, as even the three-point assumption~\ref{PowerCounting} of having at most $2s{-}1$ derivatives is explicitly violated by the cubic chiral Lagrangian~\eqref{ActionGaugeAHH}, which contains $4s{-}1$ derivatives, including those in the field strength.
Taking this $4s{-}1$ counting as a given, we can minimally extend it to four points by assuming that
\begin{description}
\item[\namedlabel{PowerCounting3}{(PC3)}] the number of derivatives
in higher-point vertices contributing to an amplitude should not exceed the momentum counting of the cubic diagrams in this amplitude.
\end{description}

This assumption is consistent with the all-plus amplitudes, such as \eqref{AmpGaugeAHH4pp}, that follow strictly from the minimal-coupling part of the Lagrangian~\eqref{ActionGaugeAHH} and thus require that all additional terms involve at least one $F^-$.
For instance, the cubic vertex for positive helicity has $1$ derivative, so the cubic diagrams contributing to ${\cal A}(\bm{1}^s\!,\bar{\bm{2}}^s\!,3^+\!,4^+)$ have $1+1-2=0$ derivatives, if we take into account $-2$ derivatives from the propagator.
Since we insist that all higher-point vertices depend on the four-potential $A_\mu$ through the field-strength, such a counting forbids any vertices that only involve $F^+$.

Applying this to the opposite-helicity case, we observe that ${\cal A}(\bm{1}^s\!,\bar{\bm{2}}^s\!,3^-\!,4^+)$ contains only up to $1+(4s-1)-2=4s-2$ derivatives, judging by the trivalent diagrams.
Since two derivatives are contained within the two corresponding field strengths $F^-$ and $F^+$, there are a maximum of $n+n'+m=4s-4$ derivatives that need to be addressed when classifying singlets in the tensor-product space~\eqref{Omp}.
The counts produced by our code are given in \tab{tab:4ptCounts}.
As explained above, these counts only give an upper bound for the number of independent operators under the described assumptions.
\begin{table}[t]
\centering
\begin{tabular}{|c||c|c|c|c|c|c|c|c|} 
 \hline
$s$ & 0 & \!$\,^1\!/\!_2$\! & $1$ & $\,^3\!/\!_2$ & $2$ & $\,^5\!/\!_2$ & $3$ \\
 \hline
count & 0 & 0 & 0 & 15 & 84 & 279 & 714 \\
 \hline
\end{tabular}
\caption{Number of singlets in the four-point chiral operator product space~\eqref{Omp}}
\label{tab:4ptCounts}
\end{table}

\section{Details on factorization poles}
\label{app:factorisation}

In this appendix, we give some useful formulae for checking the factorization channels of the $\sqrt{\text{Kerr}}$ Compton amplitudes, both in the quantum and classical cases. 

First we note that the residues of the physical poles in the abelian Compton amplitude~\eqref{abGTCompton} can be shown to match those of the corresponding AHH amplitude~\cite{Arkani-Hamed:2017jhn},
\be \label{FactorizationAHH}
\Big(\frac{\bra{3}1|4]^2 (U + V)^{2s}}{m^{4s}  t_{13} t_{14}}
- \frac{\braket{\bm{1}3} \bra{3}1|4] [4\bm{2}] }{m^{4s} t_{13}} P_2^{(2s)}\Big)\Big|_{\rm poles} = \frac{\big(\braket{\bm{1}3}[4\bm{2}]+\braket{\bm{2}3}[4\bm{1}]\big)^{2s}}{t_{13} t_{14} \bra{3}1|4]^{2s-2}}\Big|_{\rm poles}\,,
\ee
where the two massive poles are located at $t_{13}=(p_1+p_3)^2-m^2=0$ and at $t_{14}=0$.
While the AHH amplitude on the right-hand side also contains so-called spurious poles, $\bra{3}1|4]=0$, the left-hand side is manifestly free of them.
The absence of spurious poles is, of course, a consequence of working with amplitudes that originate from a local Lagrangian~\eqref{FinalRootKerrLagrangian}.

Note that there is an obvious asymmetry between the appearance of the two denominators, $t_{13}$ versus $t_{14}$, on on the left-hand-side of~\eqn{FactorizationAHH}, but not on the right-hand side. However, this is an artifact of our chosen compact presentation of the amplitude~\eqref{abGTCompton}.
The following non-trivial identity holds for generic kinematics and spin:
\be \label{eq:KinIdentity}\!\!
\frac{\bra{3}1|4]^2 (U + V)^{2s}}{m^{4s}  t_{13} t_{14}}
- \frac{\braket{\bm{1}3} \bra{3}1|4] [4\bm{2}] }{m^{4s} t_{13}} P_2^{(2s)}
= \frac{\bra{3}1|4]^2 (U - V)^{2s}}{m^{4s}  t_{13} t_{14}}
- \frac{\braket{\bm{2}3} \bra{3}1|4] [4\bm{1}] }{m^{4s} t_{14}} P_2^{(2s)} \,.
\ee
Thus, in principle, one could average over these two presentations of the pole terms, and obtain a manifestly symmetric presentation, at the price of a slightly longer formula.  

For the non-abelian color-ordered amplitude~\eqref{eq:nabGaugeAmp}, the physical pole structure again match those of the corresponding AHH amplitude~\cite{Arkani-Hamed:2017jhn},
\begin{align}
& \bigg[\frac{\bra{3}1|4]^2 (U + V)^{2s}}{m^{4s} s_{12} t_{14}}
- \frac{\braket{\bm{1}3} \bra{3}1|4] [4\bm{2}] }{m^{4s} s_{12}} P_2^{(2s)}
 + \frac{\braket{\bm{1}3} \braket{3\bm{2}} [\bm{1}4] [4\bm{2}]}{m^{4s} s_{12}} t_{13} P_2^{(2s-1)} \\* & \quad
 - \frac{\braket{\bm{1}3} \braket{3\bm{2}} [\bm{1}4] [4\bm{2}]}{m^{4s-2} s_{12}} 
 \big(t_{13} \braket{\bm{12}} [\bm{12}] P_4^{(2s -1)}\!+ 2 V P_4^{(2s)}\big) \bigg]\Big|_{\rm poles} = \frac{\big(\braket{\bm{1}3}[4\bm{2}]+\braket{\bm{2}3}[4\bm{1}]\big)^{2s}}{s_{12} t_{14} \bra{3}1|4]^{2s-2}}\Big|_{\rm poles}, \nn
\end{align}
where the $t_{14}=0$ residue matching follows from the abelian case  \eqref{FactorizationAHH}, but the $s_{12}=0$ residue matching is more non-trivial as it requires several many terms. 

It is interesting to also exhibit the factorization properties of the classical all-spin Compton amplitude \eqref{eq:NAbClAmp}.
On the massless factorization pole the classical Gram determinant \eqref{eq:clGramDet} reduces to 
\begin{equation}
0 = (w-x-y)(w-x+y) \,,
\end{equation}
where the two solutions correspond to the two choices of helicity for the propagating gluon.
On the massless pole, the only surviving term in the classical amplitude \eqref{eq:NAbClAmp} comes from the interplay between the functions $\Ea$ and $\Eb$,
\begin{align}
\frac{w^2{-}z^2}{2} \Big( & \Ea(x,y,z)
 + \frac{i \epsilon(q_\perp,p,q,a)}{p \cdot q_\perp}
   \Eb(x,y,z)\Big) \xrightarrow[q^2 \to 0]{} 
   \frac{w^2{-}z^2}{2}
   \Big( \Ea(x,y,z) + (w{-}x) \Eb(x,y,z)\Big) \nn \\* &
 = \frac{x^2 - y^2 - z^2}{2}
   \Big( \Ea(x,y,z) \pm y \Eb(x,y,z)\Big) 
 = e^{\pm y} + {\cal O}(e^{x+z},e^{x-z}) \,,
\end{align}
where the terms proportional to $ {\cal O}(e^{x+z},e^{x-z}) $ cancel against the first two terms in \eqn{eq:NAbClAmp}. The remaining exponential factor corresponds to the product of the classical three point amplitudes, $e^{\pm y} = e^{\pm k_3 \cdot a} \times e^{\pm k_4 \cdot a}$.

On the massive pole $t_{14}$, the cut kinematics enforces all momentum dot products to be at least of order ${\cal O}(\hbar^2)$:
\be
p_1 \cdot k_3 + p_1 \cdot k_4 + k_3 \cdot k_4 = 0
\qquad \Rightarrow \qquad
p_1 \cdot k_3 = - \frac{1}{2} q^2 = {\cal O}(\hbar^2) \,.
\ee
Therefore, we should take $w, z \to 0$ and $p \cdot q_\perp = \frac{1}{2} q^2$, such that the classical amplitude \eqref{eq:NAbClAmp} on the cut simplifies to
\beal
t_{14} {\cal A}(\bm{1},\bm{2},3^-\!,4^+) \big|_{t_{14}\to 0} &
 = -\frac{(p \cdot \chi)^2}{q^2}
   \left([T^{c_3},T^{c_4}]  -  \frac{1}{4}\{T^{c_3},T^{c_4}\}  \right) e^x \\ &
 = (p \cdot \varepsilon_3^-) (p \cdot \varepsilon_4^+)
   \left( [T^{c_3},T^{c_4}]  -  \frac{1}{4}\{T^{c_3},T^{c_4}\}  \right) e^x \,.
\eeal
This is once again the expected product of three-point amplitudes $e^x = e^{k_4 \cdot a}\times e^{-k_3 \cdot a}$, where the mismatch in reference vector for $a^\mu$ is irrelevant on the cut.

\bibliographystyle{JHEP}
\bibliography{references}

\end{document}